\documentclass[acmsmall,screen]{acmart}

\newif\ifarxiv
\arxivtrue

\setcopyright{cc}
\setcctype{by}
\acmDOI{10.1145/3763151}
\acmYear{2025}
\acmJournal{PACMPL}
\acmVolume{9}
\acmNumber{OOPSLA2}
\acmArticle{373}
\acmMonth{10}

\ifarxiv
\else
\received{2025-03-26}
\received[accepted]{2025-08-12}
\fi

\usepackage{sl}
\usepackage[utf8]{inputenc}
\usepackage{bussproofs}
\usepackage{enumitem}
\usepackage{amsmath}
\usepackage{wrapfig}
\usepackage{mathtools}
\usepackage{pifont}

\usepackage{tikz}
\usetikzlibrary{trees}
\usetikzlibrary{patterns}
\usetikzlibrary{shapes.misc, shapes.geometric}
\usetikzlibrary{positioning}
\usetikzlibrary{decorations.pathmorphing, decorations.pathreplacing, decorations.shapes}
\usetikzlibrary{calc}
\usetikzlibrary{arrows, arrows.meta}
\usetikzlibrary{matrix}
\usetikzlibrary{cd}

\definecolor{mblue}{HTML}{ced5ea}
\definecolor{lblue}{HTML}{e9ebf5}
\definecolor{hblue}{HTML}{b5c7e8}
\definecolor{softgreen}{HTML}{3dbd1a}

\usepackage{gnuplottex}

\usepackage{tabularx}
\usepackage{subcaption}

\usepackage{soul}
\let\strikethrough\st

\usepackage{graphicx}
\graphicspath{ {./images/} }

\usepackage{macros}
\usepackage{c_macros}
\usepackage{js_macros}

\usepackage{xspace}

\usepackage{mdframed}

\usepackage{tcolorbox}
\tcbuselibrary{theorems}

\theoremstyle{definition}
\newtheorem{instance-definition}{Instance Definition}
\newtheorem{instance-definition-example}{Instance Definition Example}
\newtheorem{instance-property}{Instance Property}

\newcommand{\cmark}{\textcolor{softgreen}{\ding{51}}}

\renewcommand{\iff}{\Leftrightarrow}
\renewcommand{\implies}{\Rightarrow}

\begin{document}

\title[Compositional Symbolic Execution for the Next 700 Memory Models\ifarxiv{} (Extended Version)\else\fi]{Compositional Symbolic Execution \\ for the Next 700 Memory Models\ifarxiv{} (Extended Version)\else\fi}

\author{Andreas Lööw}
\orcid{0000-0002-9564-4663}
\affiliation{%
  \institution{Imperial College London}
  \city{London}
  \country{United Kingdom}}
\email{a.loow@ic.ac.uk}

\author{Seung Hoon Park}
\orcid{0000-0001-7165-6857}
\affiliation{%
  \institution{Imperial College London}
  \city{London}
  \country{United Kingdom}}
\email{s.park23@ic.ac.uk}

\author{Daniele Nantes-Sobrinho}
\orcid{0000-0002-1959-8730}
\affiliation{%
  \institution{Imperial College London}
  \city{London}
  \country{United Kingdom}}
\email{d.nantes-sobrinho@ic.ac.uk}

\author{Sacha-Élie Ayoun}
\orcid{0000-0001-9419-5387}
\affiliation{%
  \institution{Imperial College London}
  \city{London}
  \country{United Kingdom}}
\email{s.ayoun17@ic.ac.uk}

\author{Opale Sjöstedt}
\orcid{0009-0003-7545-1383}
\affiliation{%
  \institution{Imperial College London}
  \city{London}
  \country{United Kingdom}}
\email{opale.sjostedt23@ic.ac.uk}

\author{Philippa Gardner}
\orcid{0000-0002-4187-0585}
\affiliation{%
  \institution{Imperial College London}
  \city{London}
  \country{United Kingdom}}
\email{p.gardner@ic.ac.uk}



\begin{abstract}
Multiple successful compositional symbolic execution (CSE) tools and platforms exploit separation logic (SL) for compositional verification and/or incorrectness separation logic (ISL) for compositional bug-finding, including VeriFast, Viper, Gillian, CN, and Infer-Pulse. Previous work on the Gillian platform, the only CSE platform that is parametric on the memory model, meaning that it can be instantiated to different memory models, suggests that the ability to use custom memory models allows for more flexibility in supporting analysis of a wide range of programming languages, for implementing custom automation, and for improving performance. However, the literature lacks a satisfactory formal foundation for~memory-model-parametric~CSE~platforms.

In this paper, inspired by Gillian, we provide a new formal foundation for memory-model-parametric CSE platforms. Our foundation advances the state of the art in four ways. First, we mechanise our foundation (in the interactive theorem prover Rocq). Second, we validate our foundation by instantiating it to a broad range of memory models, including models for C and CHERI. Third, whereas previous memory-model-parametric work has only covered SL analyses, we cover both SL and ISL analyses. Fourth, our foundation is based on standard definitions of SL and ISL (including definitions of function specification validity, to ensure sound interoperation with other tools and platforms also based on~standard~definitions).
\end{abstract}

\maketitle
\renewcommand{\shortauthors}{A. Lööw, S. H. Park, D. Nantes-Sobrinho, S.-É. Ayoun, O. Sjöstedt, and P. Gardner}


\section{Introduction}%
\label{sec:intro}

Multiple successful analysis tools and platforms  provide {compositional verification} and/or {compositional bug-finding} for heap-manipulating programs, including VeriFast~\cite{verifast}, Viper~\cite{Muller16}, Gillian~\cite{gillianpldi,gilliancav,cse1}, CN~\cite{Pulte23}, Infer-Pulse~\cite{Le22}, by animating their automated and semi-automated reasoning using {\em compositional symbolic execution (CSE)} grounded on ideas from separation logic (SL)~\cite{seplogic,reyseplogic} and/or incorrectness separation logic (ISL)~\cite{isl} (usually bottoming out in a call to an underlying SMT solver, such as Z3~\cite{z3}). In this case, compositional (or functionally compositional) reasoning  means that it is scalable in that the analysis works on functions in isolation, at any point in the codebase, and then records the results in simple function specifications that can be used in broader calling contexts. 
SL provides a good grounding for sound verification (proving the absence of bugs) through compositional \emph{over-approximate (OX) reasoning}; ISL provides a good grounding for sound bug-finding (proving the presence~of~bugs) through compositional \emph{under-approximate (UX) reasoning}. Important examples of CSE analyses based on the two logics include OX function-specification verification (implemented in, e.g., VeriFast, Viper, Gillian, and CN) and UX true bug-finding based on bi-abduction~\cite{Le22} (implemented in, e.g., Infer-Pulse~and~Gillian).

A key challenge with the design of  CSE platforms that aim for wide applicability is to manage the diverse range of memory models employed across different applications.\footnote{To avoid confusion: whereas some authors reserve the term memory model for weak-memory concurrency, we use the term broadly in this work (the many different ways of representing, updating, and analysing the heap).}
This need for different memory models arises from multiple sources. First, of course, different programming languages are defined over different language memory models. Second, {different analyses are defined over different types of memory ghost state}, e.g., ghost state for different kinds of ownership disciplines, such as exclusive ownership vs. fractional ownership, or negative-information ghost state used in UX analyses to ensure UX compositionality.  Third, even choosing  the programming language and the analysis still does not determine the memory model: there is \emph{no one-size-fits-all} memory models because the different axes of the memory-model design space 
are often-times in antagonistic relationship with each other: e.g., the choice of what part of the language to be analysed, accuracy and abstraction level of the model, implementation effort of the model, performance of the model, and automation/annotation burden associated with the analysis of the model. We  cannot move freely along different axes in the design space and therefore must solve a difficult trade-off problem when selecting a memory model to use: e.g., a complex memory model might give better performance than a simple memory model but will require more implementation effort.

The analysis platform Gillian stands out as the only CSE platform that faces this memory-model challenge head-on. Gillian is the only CSE platform that is \emph{parametric} on the memory model, meaning that no memory model is hard-coded into the platform and instead the platform can be instantiated to different memory models, depending  on which model has the best position in the model design trade-off space for a situation at hand. All other CSE platforms are \emph{monomorphic} on the {memory model} in that they only support one fixed memory model that has been hard-coded into the platform. It is therefore awkward, or impossible, to 
 use the memory model that is most appropriate for a situation at hand since it must be encoded into the fixed memory model~of~the~platform.
 
\paragraph{Literature gap.} No previous work provides a satisfactory formal foundation for CSE platforms that are parametric on the memory model; in particular,  {a formal foundation for Gillian's approach to memory-model parametricity is missing}. 
There was some initial work on the foundations of Gillian~\cite{gillianpldi,gilliancav}, which outlined mathematical definitions and gave a sketch of a soundness proof for parts of the CSE engine of Gillian. This work, however, suffers from \emph{four weaknesses}:
\begin{enumerate}
\item it was not mechanised;
\item it did not prove that any of their memory-model instances were sound, and thus  did not validate their definitions and conjectures;
\item it only covered SL-based analyses, not ISL-based analyses;
\item it did not use standard SL definitions, such as the definition of function specification validity, thus making the engine awkward to interoperate with~other~analysis~tools~and~platforms.
\end{enumerate}
Other previous work on the foundations of CSE have only covered tools and platforms that are monomorphic on the memory model~\cite{Jacobs15,Zimmerman24,Dardinier25,cse1}. (We discuss related work in more detail in \S\ref{sec:related-work}.)

\paragraph{Contribution.} In this paper, we contribute a new CSE theory that addresses all four weaknesses in the current formal foundations of CSE platforms that are parametric on the memory model. Our new CSE theory is inspired by the design of Gillian but is independent of its particular implementation; we have designed our theory to be a CSE analogue of separation logics that are parametric on the memory model, such as abstract separation logic~\cite{calcagno:lics:2007} and subsequent generic/modular/parametric separation-logic frameworks like the views framework~\cite{dyoung:popl:2013,cisl} and, perhaps the most well-known example, the Iris framework~\cite{jung:jfunc:2018}.

An important strength of our CSE theory is that it is remarkably simple; in fact, its definitions and metatheory are not much more complex than existing monomorphic CSE theories. \emph{This suggests that, while there are clear advantages, there are no clear disadvantages in making CSE platforms parametric on the memory model.}

Technically, our new CSE theory, which we have mechanised in Rocq~\cite{Rocq}, consists of:
\begin{itemize}
\item a definition of ``memory model'' in the CSE setting, including, two sets of OX and UX soundness requirements on memory models;
\item a formal semantics for a CSE engine that is parametric on the memory model;
\item two soundness results for the engine: if a memory model satisfies our OX/UX soundness requirements, then the engine is OX/UX sound when instantiated with the memory model.
\end{itemize}

\begin{table}[t]
\centering
\caption{Summary of our memory-model instances. The columns ``OX'' and ``UX'' specify whether the model is OX sound and/or UX sound; ``Rocq kLoc'' specifies the size of the Rocq proof script file for the model; ``Origin'' specifies which CSE platforms have implemented the model. The star (``*'') in the Rocq column for the OOP model specifies that we did not mechanise the model because of its large overlap with the block-offset model, and the double star (``**'') for the CHERI model specifies that the mechanised proof only covers OX soundness (UX soundness is left for future work because of the size of the task).}
\begin{tabular}{lccccc}
 \toprule
 Memory-model name & \S & OX & UX & Rocq kLoC & Origin \\
 \midrule
 Linear model (running example in paper) & \ref{sec:mem-linear} & \cmark & \cmark & 1 & \\
 Linear model with unique-match branching & \ref{sec:mem-linear} & \cmark & \cmark & 1 & \\
 Linear model with cut branching & \ref{sec:mem-linear} &  & \cmark & $\approx 0.5$ & \\
 Linear model without negative information & \ref{sec:mem-linear} & \cmark &  & $\approx 0.5$ & \\
 Fractional ownership model & \ref{sec:mem-frac} & \cmark & \cmark & 2 & \\
 Block-offset model for C & \ref{sec:mem-block-offset} & \cmark & \cmark & 4 & Gillian \\
 Model for OOP languages (e.g., JavaScript) & \ref{sec:mem-oop} & \cmark & \cmark & 0* & Gillian \\
 CHERI-assembly model & \ref{sec:mem-cheri} & \cmark & \cmark & 19** & New model \\
 VeriFast-and-Viper-inspired model for C & \ref{sec:mem-verifast} & \cmark &  & $\approx 0.5$ & VeriFast and Viper \\
 \bottomrule
\end{tabular}%
\label{tab:instances}
\end{table}

To validate our CSE theory and show that it has broadly applicability, we instantiate our theory to a broad collection of memory models ranging from models for low-level languages like assembly and C to high-level languages like JavaScript, as summarised in Tab.~\ref{tab:instances}. In more detail: in \S\ref{sec:mem-linear} and \S\ref{sec:mem-frac}, we cover the standard models used in theoretical investigations into SL and ISL, which we call linear memory models. In \S\ref{sec:mem-linear}, we show that multiple variants of these linear memory models fit into our theory, including OX-and-UX sound models, OX-only models, and UX-only models. In \S\ref{sec:mem-frac}, to show that different ownership disciplines fit into our theory, we instantiate our theory with a linear memory model implementing fractional ownership rather than standard exclusive ownership. In \S\ref{sec:mem-block-offset}, we shift the discussion towards more realistic memory models, starting of the discussion with a memory model inspired by the memory model of the CompCert compiler. The memory model is implemented in Gillian and has been used in Gillian-based teaching. In \S\ref{sec:mem-oop}, to show that memory models for high-level languages like object-oriented languages also fit our theory, we discuss the JavaScript memory model implemented in Gillian. In \S\ref{sec:mem-cheri}, we discuss a new memory model for CHERI, which is the largest model we have mechanised. As this model is new and, hence, has not been implemented in any CSE platform, the model shows that new models can be designed using only our CSE theory as guidance. Lastly, in \S\ref{sec:mem-verifast}, to show broad CSE platform coverage, we discuss the hard-coded memory model of VeriFast and Viper.

Our main technical contributions can be summarised as follows:
\begin{itemize}
\item We provide the first foundation of CSE platforms that are parametric on the memory model~(\S\ref{sec:language-symbolic}) that: (1) is mechanised, (2) is validated, (3) covers both SL- and ISL-based analyses, and (4)~is~interoperable.
\item We demonstrate that two important analyses can be soundly hosted on top of our memory-model-parametric CSE engine~(\S\ref{sec:language-symbolic}):  namely, OX function-specification verification;  and UX true bug-finding based on bi-abduction.
\item We discuss instantiations of our CSE theory~(\S\ref{sec:more-memory-models}), as summarised in Tab.~\ref{tab:instances}.
\item We make available all source code and proofs of our Rocq mechanisation of our CSE theory and its instantiations in the artefact of this paper (see our data-availability statement).
\end{itemize}

\paragraph{Scope limitations and caveats} For this paper, we only consider \emph{sequential} memory models not \emph{concurrent} memory models. We, however, believe our work is a useful starting point for future work on symbolic execution of different concurrent memory models. Additionally, we work with a simple demonstrator programming language, specifically, a memory-model-parametric variant of a standard (sequential) imperative language. In other words, to focus the discussion on our core contribution, which is memory-model parametricity, we do not vary other parts of the language.

\section{Overview}%
\label{sec:overview}

\tcbset{
  takeawaybox/.style={
    colframe=black!70,
    boxrule=1pt,
    arc=1mm,
    left=1mm, right=1mm, top=0.5mm, bottom=0.5mm,
    width=\linewidth,
  }
}

In this section, we highlight the main \emph{takeaways} of our new CSE theory. To be able to do so, we give a compressed overview of our theory. The \emph{core contribution of our theory} is that it is parametric on a set of parameters that factors out its memory-model dependent part; the focus in this section is therefore these parameters.

\subsection{Background: Monomorphic CSE Theory}%

\newcommand{\sbaction}{\baction}

Before introducing our new memory-model-parametric CSE theory, we give a summary of traditional memory-model-monomorphic CSE theory. The judgements we use in the summary are simplified judgements from our theory.

\paragraph{Engine architecture} In this paper, we work specifically with CSE engines implementing the consume-produce engine architecture. This is the architecture implemented by Gillian and other similar modern CSE engines like VeriFast and Viper. In this architecture, two operations called \mac and \produce are used to implement the execution of commands/constructs based on assertions (separation-logic points-to assertions, etc.), which are the commands/constructs making the engine a compositional engine, such as using function specifications to reason about function calls. In short, the \mac operation takes as input an assertion and removes (``consumes'') the corresponding symbolic state from the engine's current symbolic state and the \produce operation also takes as input an assertion but instead adds (``produces'') the corresponding symbolic state to the current symbolic state. For example, to execute a function call using a function specification, the precondition of the specification is first consumed and its~postcondition~is~then~produced.

\paragraph{Judgements} Assuming we work with (a monomorphised version of) our demonstrator language, a monomorphic CSE theory needs to specify at~least~the~following~judgements:
\begin{itemize}
\item $\st, \cmd \baction \st'$ -- judgement for the \emph{concrete} semantics of the language, where $\cmd$ denotes a language command, $\st$ is a concrete input state, and $\st'$ is a concrete output state.
\item $\st \models A$ -- judgement for the satisfaction relation for assertions (used to, e.g., define the semantics of function specifications).
\item $\sst, \cmd \sbaction \sst'$ -- judgement for the CSE engine, i.e., the \emph{symbolic} semantics of the language, where $\sst$ and $\sst'$ are symbolic states.
\item $\st \models \sst$ -- judgement for the satisfaction relation for symbolic states, i.e., the relation between concrete and symbolic states.
\end{itemize}
The two important operations $\mac$ and $\produce$ are part of the definition of the symbolic engine, i.e., $\sst, \cmd \sbaction \sst'$.

\paragraph{Definition of soundness} There are two standard soundness statements that relate the concrete and symbolic execution judgements: OX soundness and UX soundness~\cite{symb:exec:survey} (although different papers use different terminology). OX soundness is the following relation between the~two~types~of~execution:
\[
\st, \cmd \baction \st' \land \st \models \sst \implies \exists \sst', \sst, \cmd \sbaction \sst' \land \st' \models \sst'.
\]
Intuitively, the relation enforces that all states reachable by concrete execution are reachable by symbolic execution, i.e., symbolic reachability overapproximates concrete~reachability. UX soundness enforces the opposite relation, where, note, the universal quantification is over~the~final~states:
\[
\sst, \cmd \sbaction \sst' \land \st' \models \sst' \implies \exists \st, \st, \cmd \baction \st' \land \st \models \sst.
\]

In more analysis application-oriented terms: OX soundness is a good foundation for verification and UX soundness is a good foundation for bug-finding. E.g., for verification: it follows from OX soundness that if we have proved that a behaviour (such as a bug) is unreachable using symbolic execution, then the behaviour is also unreachable~by~concrete~execution.

\subsection{The Step to Parametric CSE Theory}

\begin{table}[t]
\centering
\begin{minipage}{0.39\textwidth}
\LARGE
\begin{tikzpicture}[
 node distance=0.5cm and 0.5cm,
 every node/.style={draw, rectangle, align=center, minimum width=2.5cm, minimum height=1cm, thick},
 every path/.style={thick, ->, >=latex},
 input/.style={fill=cyan!20,densely dotted},
 output/.style={fill=gray!20},
 scale=0.6,transform shape
]

\node (cmm) [input] {CMM};
\node (rm) [input, above=of cmm] {RM};
\node (smm) [input, above=of rm] {SMM};
\node (cse) [output, above=of smm] {CSE engine};

\node (smm_ox) [input, left=of cse] {OX REL};
\node (cse_ox) [output, above=of smm_ox] {CSE engine \\ OX sound};

\node (smm_ux) [input, right=of cse] {UX REL};
\node (cse_ux) [output, above=of smm_ux] {CSE engine \\ UX sound};

\draw[<-] (cmm) -- (rm);
\draw[<-] (rm) -- (smm);

\draw[<-] (smm) -- (cse);
\draw[<-] (smm) -- (smm_ox);
\draw[<-] (smm) -- (smm_ux);

\draw[<-] (smm_ox) -- (cse_ox);
\draw[<-] (cse) -- (cse_ox);

\draw[<-] (smm_ux) -- (cse_ux);
\draw[<-] (cse) -- (cse_ux);
\end{tikzpicture}

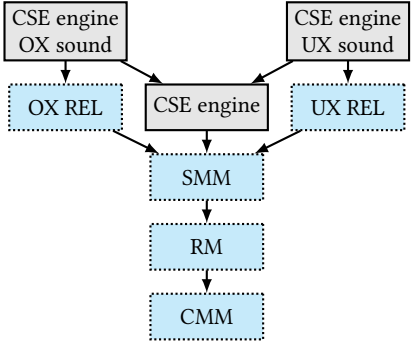
\captionof{figure}{Dependency structure of our theory.}
\label{fig:overview}
\end{minipage}
\hspace{3mm}
\begin{minipage}{0.57\textwidth}
\caption{Summary of required IDefs.}
\begin{tabularx}{\textwidth}{llX}
 \toprule
 \# & Abs. & Description \\
 \midrule
 \ref{idef:cmm} & CMM & Memory data type, empty memory, and composition operator \\
 \ref{idef:cmm-semantics} & CMM & Concrete semantics of memory actions \\
 \ref{idef:res-srel} & RM & Resource satisfaction relation \\
 \ref{idef:smm} & SMM & Memory data type and empty memory \\
 \ref{idef:smm-srel} & SMM & Memory satisfaction relation \\
 \ref{idef:smm-semantics} & SMM & Symbolic semantics of memory actions \\
 \ref{idef:smm-cp} & SMM & Semantics of $\resconsume$ and $\resproduce$ \\
 \bottomrule
\end{tabularx}%
\label{tab:idefs}
\end{minipage}
\centering
\vspace{1em}
\caption{Summary of required IProps., where the ``Deps.'' column specifies the IDefs. dependencies.}
\begin{tabularx}{0.88\textwidth}{lllX}
 \toprule
 \# & Abs. & Deps. & Description \\
 \midrule
 \ref{iprop:cmm-pcm} & CMM & \ref{idef:cmm} & Memory forms a partial commutative monoid (PCM) \\
 \ref{iprop:cmm-frame} & CMM & \ref{idef:cmm} and \ref{idef:cmm-semantics} & Memory actions satisfy OX/UX frame properties \\
 \ref{iprop:mem-sound} & REL & All except \ref{idef:res-srel} and \ref{idef:smm-cp} & OX/UX soundness of symbolic memory actions \\
 \ref{iprop:cp-sound} & REL & All except \ref{idef:cmm-semantics} and \ref{idef:smm-semantics} & OX/UX soundness of $\resconsume$ and $\resproduce$ \\
 \bottomrule
\end{tabularx}%
\label{tab:iprops}
\vspace{-1em}
\end{table}

We now discuss how our CSE theory parameterises the judgements and soundness statements introduced above. We differentiate between two types of parameters, which we also refer to as \emph{instance data}: \emph{instance definitions} (abbreviation: ``IDefs'') and \emph{instance properties} (abbreviation: ``IProps''). Additionally, we group the parameters into the following \emph{four abstractions}:
\begin{enumerate}
\item concrete memory model (CMM) -- the parameters of $\st, \cmd \baction \st'$ (concrete semantics);
\item resource model (RM) -- the parameters of $\st \models A$ (satisfaction relation for assertions);
\item symbolic memory model (SMM) -- the parameters of $\sst, \cmd \sbaction \sst'$ (symbolic semantics/engine) and $\st \models \sst$ (satisfaction relation for symbolic states);
\item OX and UX soundness relations (RELs) between CMMs, RMs, and SMMs -- the parameters of the soundness proofs of the engine.
\end{enumerate}

Fig.~\ref{fig:overview} depicts the dependency structure of the four abstractions (blue boxes) and the engine definition and its soundness proofs (grey boxes). All IDefs. and IProps. parameters are summarised, respectively, in Tab.~\ref{tab:idefs} and Tab.~\ref{tab:iprops}. We now introduce the parameters in more detail.

\paragraph{Concrete memory model and resource model} Because a CSE engine and a program logic for the same language have the same trusted computing base, the parameters of the concrete language and the assertion language of our CSE theory are the same as for comparable memory-model-parametric separation logics such as abstract separation logic~\cite{calcagno:lics:2007} (we are here speaking in terms of big-picture ideas, of course parameter details differ between different program logics). Therefore, we discuss the two abstractions concrete memory model~and~resource~model~together.

A concrete memory model, i.e., the parameters of $\st, \cmd \baction \st'$ (concrete language), specifies: the data type of memory (IDef.~\ref{idef:cmm}) and the memory actions and their semantics (IDef.~\ref{idef:cmm-semantics}). Examples of common memory actions include memory read, memory write, allocation, etc. Analogous to the setup in program logics, we require that the data type of memory comes with a composition operator that forms a PCM together with the data type (IProp.~\ref{iprop:cmm-pcm}) such that we can build the standard separation-logic infrastructure on top of the language. Additionally, to ensure that the concrete language satisfies the standard separation-logic frame properties, we require that the memory actions satisfy frame properties (IProp.~\ref{iprop:cmm-frame}) that we have derived from the standard properties.

A resource model, i.e., the parameters of $\st \models A$ (satisfaction relation for assertions), specifies: the resource assertions for the memory-model instance and their satisfaction relation (IDef.~\ref{idef:res-srel}).\footnote{Our assertions are deeply embedded because our CSE engine must be able to pattern match over their structure. Note that some presentations of memory-model-parametric separation logics (e.g., the abstract separation logic paper~\cite{calcagno:lics:2007}) shallowly embed their assertions, i.e., do not include an explicit satisfaction relation and instead define assertions~to~be~sets~of~state.} These resource assertions are the assertions that differ between memory models, the remaining assertion language is fixed. E.g., one common type of resource assertion is points-to assertions for heap cells (usually denoted $\lexp_1 \mapsto \lexp_2$). An alternative approach, followed by the original work on Gillian~\cite{gilliancav}, is to define the meaning of assertions in terms of $\mac$ and $\produce$ instead of a traditional satisfaction relation. This approach requires less instance data but makes the theory awkward to connect to other formalisms since the meaning of assertions is nonstandard.

\begin{tcolorbox}[takeawaybox]
\textbf{Takeaway.} Concrete memory models and resource models should look familiar; they are analogous to instance data also required by memory-model-parametric separation logics.
\end{tcolorbox}

\paragraph{Symbolic memory model} A symbolic memory model, i.e., the parameters of $\sst, \cmd \sbaction \sst'$ (symbolic semantics/engine), specifies: the data type of symbolic memory (IDef.~\ref{idef:smm}) and the satisfaction relation for symbolic memory (IDef.~\ref{idef:smm-srel}); the symbolic semantics of memory actions over the memory (IDef.~\ref{idef:smm-semantics}); additionally, the \mac and \produce operations of the engine are parametrised by $\resconsume$ and $\resproduce$ operations for the resource assertions of the memory model (IDef.~\ref{idef:smm-cp}).

Note that, in contrast to concrete memories, we do not require that symbolic memories form PCMs or satisfy any frame properties. This is unlike the original work on Gillian~\cite{gilliancav}, which defined concrete and symbolic memory models uniformly and therefore required the same instance data for both, i.e., required more instance data than us. Our work shows that the IDefs. and IProps. of a parametric CSE theory can be stated such that symbolic PCM and frame data is not needed for either the definition of the theory's engine or for its soundness proofs; instead, in the theory, all PCM reasoning and frame reasoning can be carried out at the concrete level.

\begin{tcolorbox}[takeawaybox]
\textbf{Takeaway.} Symbolic memory models and concrete memory models have different parameters and should therefore not be treated uniformly.
\end{tcolorbox}

\paragraph{OX and UX soundness relations} Our soundness relations, i.e., the parameters of the soundness proofs, tie together concrete memory models, resource models, and symbolic memory models. The OX soundness of our CSE engine (Thm.~\ref{thm:ox-sound}) follows from a series of OX IProps., while UX soundness (Thm.~\ref{thm:ux-sound}) follows from a series of UX IProps. Specifically, we define what it means for symbolic memory actions (IProp.~\ref{iprop:mem-sound}) and the $\resconsume$ and $\resproduce$ operations (IProp.~\ref{iprop:cp-sound}) to be OX sound and require this as instance data. See again Tab.~\ref{tab:iprops}, which specifies which IDefs. each IProp. ties together. We give analogous IProps.~for~UX~soundness.

\begin{tcolorbox}[takeawaybox]
\textbf{Takeaway.} From the Gillian implementation we know that the \emph{definition} of a CSE engine can be built on top of a symbolic memory model, this paper shows that the same is true of the \emph{soundness proof} of the engine; i.e., the soundness proof of the engine can be built on top of the soundness requirements for symbolic memory models as we define~them~in~this~paper~(i.e.,~our~IProps.).
\end{tcolorbox}

\begin{tcolorbox}[takeawaybox]
\textbf{Takeaway.} Additionally, our work shows that it is possible to make a clear separation between OX and UX soundness requirements and lift the two to the full engine independently of each other. In other words, OX soundness and UX soundness are independent of each other.
\end{tcolorbox}

\subsection{Structure of Rest of Paper}

The rest of the paper is structured as follows. In \S\ref{sec:language-concrete} to \S\ref{sec:language-symbolic}, we formally and incrementally present the parameters of our CSE theory. Throughout this presentation, we use a simple linear memory model as a running instantiation example. In \S\ref{sec:more-memory-models}, having introduced the parameters, we discuss other examples of memory models that fit~our~CSE~theory.

\section{Programming Language}%
\label{sec:language-concrete}

We introduce the \emph{syntax} and \emph{concrete semantics} of our demonstrator programming language, which is parametrised by a \emph{concrete memory model}, specifically, concrete memory data type (IDef.~\ref{idef:cmm}), memory actions (IDef.~\ref{idef:cmm-semantics}), and their associated IProps.

\paragraph{Syntax}

The syntax of the language is standard except that it is equipped with a \emph{memory-action command} $\passign{\vec{\pvar x}}{\act(\vec{\pexp})}$, where $\act \in \strings$, which is given a semantics using the memory-model IDefs. introduced below. The full definition of the syntax is as follows:
\[
\begin{array}{r@{~}c@{~}l}
 \gv \in \vals & ::= & \nil \mid \bv \in \bools \mid \nv \in \nats \mid q \in \rationals \mid \sv \in \strings \mid [\lst{v}] \quad \quad \pvar x, \pvar y, \pvar z, \ldots \in \pvars \\
  \pexp \in \pexps & ::= & \gv \mid \pvar x \mid \neg \pexp \mid
                           \pexp = \pexp \mid 
                           \pexp \land \pexp \mid
                           \pexp + \pexp \mid \pexp - \pexp \mid \pexp~/~\pexp \mid
                           \pexp < \pexp \mid \dots \\ 
  \cmd  \in \cmds & ::= &\pskip \mid \passign{\pvar{x}}{\pexp} \mid \pifelse{\pexp}{\cmd}{\cmd} \mid
                  \cmd; \cmd \mid
                  \pfuncall{\pvar{y}}{\fid}{\lst{\pexp}} \mid \passign{\vec{\pvar x}}{\act(\vec{\pexp})}
\end{array}
\]
where the vector notation (e.g. $\lst{v}$) denotes a list, $\vals$ the set of values ($\rationals$ is the positive rationals), $\pvars$ the set of program variables, $\pexps$ the set of expressions, and~$\cmds$~the~set~of~commands.

\paragraph{Concrete semantics} To define the semantics of the language and to ensure that the language can be used in compositional reasoning, we require the following instance data:\footnote{We use the $X \rightharpoonup Y$ to denote partial functions from $X$ to $Y$ and $X \rightharpoonup_{\mathit{fin}} Y$ to denote partial~functions~with~finite~support.}
\begin{instance-definition}
\label{idef:cmm}
We require a tuple $(\cmemss, \cwf, \emptycmem, \memcomp)$, where $\cmemss$ is a set of memories, $\cwf \subseteq \cmemss$ is a well-formedness predicate, $\emptycmem \in \cmemss$ the empty-memory element, and $\memcomp : (\cmemss, \cmemss) \rightharpoonup \cmemss$ is a memory composition operator.\footnote{It would also be possible to incorporate the $\cwf$ predicate into the concrete memory type itself using subtyping or dependent types. We chose to keep it separate so that our meta-theory is simply typed. This is a presentational choice; the condition is the same with both choices.} The empty memory must be well-formed and composition must maintain well-formedness.
\end{instance-definition}

\begin{instance-definition-example}\label{iex:lin_cmm}
In our running example linear memory model, $\cmemss$ is $\Nat \rightharpoonup_{\mathit{fin}} (\Val \uplus \{ \cfreed \})$, where the symbol $\cfreed$ records that a memory cell has been freed.
Tracking freed memory cells is a standard technique used in compositional UX reasoning~\cite{isl} to ensure that the memory model satisfies UX frame (IProp.~\ref{iprop:cmm-frame}). All memories are well-formed, i.e., $\cwf = \cmemss$. The empty memory $\emptycmem$ is the empty function and the composition of two memories $\cmem$ and $\cmem'$ is their disjoint union $\cmem \uplus \cmem'$ (i.e., their union defined only for nonoverlapping memories).
\end{instance-definition-example}

\begin{instance-property}
\label{iprop:cmm-pcm}
The components $(\cmemss, \emptycmem, \memcomp)$ form a PCM.
\end{instance-property}

\noindent
We now discuss our big-step operational semantics for the language, with judgement
\[
\csesemtransabstract{\st}{\scmd}{\st'}{\fictx}{\outcome}
\]
reading ``the execution of command $\scmd$ with function implementation context $\gamma$ in state $\st$ results in a state $\st'$ with outcome~$o$''. A program state is a pair $\cst = (\sto, \hp)$ comprising a variable store $\sto : \pvars \rightharpoonup_{\mathit{fin}} \vals$ and a memory $\hp \in \cmemss$. Outcomes are defined as $o ::= \oxok \mid \oxerr \mid \oxm$, denoting, respectively, a successful execution, a fault due to a language error, and a fault due to a missing resource error.
We must distinguish between the two kinds of faults as the missing-resource errors have a different role to play in compositional reasoning (see IProp.~\ref{iprop:cmm-frame} below) and bi-abduction (see \S\ref{sec:analyses}). Function implementation contexts $\fictx$ provide the function definitions used in function~calls.

The only interesting case in the definition of the semantics is the case for memory actions, which is given by instance data. We only discuss this case; for other cases, see \ifarxiv App.~\ref{app:concrete}\else the~extended~paper~\cite{arxiv}\fi.

\begin{instance-definition}
\label{idef:cmm-semantics}
Memory actions are defined by a relation, written  $\cmem.\act(\vec{v}) \rightsquigarrow o : (\cmem', \vec{v'})$, which executes an action $\alpha$ on memory $\cmem$ with parameters $\vec{v}$, and returns an outcome $o$, memory~$\cmem'$, and return values $\vec{v'}$. All memory actions must preserve well-formedness ($\cwf$).
\end{instance-definition}

\begin{instance-definition-example}\label{iex:lin_csem}
Our linear memory model has four memory actions: \texttt{lookup}, \texttt{mutate}, \texttt{new}, and \texttt{free}. We give the $\oxok$ and $\omiss$ rules for defining the  \texttt{lookup} action; the full set of rules is in \ifarxiv App.~\ref{app:memory-models}\else the~extended~paper~\cite{arxiv}\fi:
\begin{mathpar}
\small
\infer{\cmem.\mathtt{lookup}([n]) \rightsquigarrow \oxok : (\cmem, [v])}
      {\hp(n) = v}
\and
\infer{\cmem.\mathtt{lookup}([n]) \rightsquigarrow \omiss : (\cmem, [``\mathsf{MissingCell}", n])}
      {n \notin \dom (\hp)}
\end{mathpar}
\end{instance-definition-example}

The following two rules lift the semantics of memory actions to the command level:
\begin{mathpar}
\small
\infer{
    \csesemtrans{\sto, \cmem}{\vec{\pvar{x}} := \act(\lst\pexp)}{\sto [\vec{\pvar{x}} \storearrow \vec{v}' ], \cmem'}{\fictx}{\oxok}
  }{
    \begin{array}{c}
    \esem{\vec{\pexp}}{\sto} = \vec{v} \quad \cmem.\act(\vec{v}) \rightsquigarrow \oxok : (\cmem', \vec{v}') \quad |\vec{v}'| = |\vec{\pvar{x}}|
    \end{array}
  }
\and
\infer{
    \csesemtrans{\sto, \cmem}{\vec{\pvar{x}} := \act(\lst\pexp)}{\sto[\pvar{err} \mapsto \vec{v}], \cmem'}{\fictx}{\outcome}
  }{
    \begin{array}{c}
    \esem{\vec{\pexp}}{\sto} = \vec{v} \quad \cmem.\act(\vec{v}) \rightsquigarrow \outcome : (\cmem', \vec{v}) \quad \outcome \not= \oxok
    \end{array}
  }
%
%
\end{mathpar}
\noindent where $\esem{\pexp}{\sto}$ denotes the standard evaluation of an expression $\pexp$ with respect to a store $\sto$, resulting either in a value or a dedicated symbol $\undefd \notin \vals$ denoting an evaluation error.

\paragraph{OX and UX frame}

As is standard in compositional reasoning based on SL and ISL, we rely on the fact that the concrete semantics of the language satisfies frame properties. To ensure that we have these properties, we require that the concrete semantics of memory actions satisfy the standard OX and/or UX frame properties (which in turn straightforwardly lift to the full~concrete~semantics):
\begin{instance-property}\label{iprop:cmm-frame}
For $\cwf(\cmem)$ and $\cwf(\cmem_f)$:
\[
\begin{array}{lll}
(\text{OX}) & \text{If } & (\cmem \memcomp \cmem_f).\act(\vec{v}) \rightsquigarrow \result : (\cmem', \vec{v}')  \\
&\text{then } & \exists \cmem'', \vec{v}'', \result'.~
\cmem.\act(\vec{v}) \rightsquigarrow \result' : (\cmem'', \vec{v}'') \text{ and} \\
& & \qquad
 (\result' \neq \omiss
 \implies
 (\result' = \result \text{ and } \vec{v}'' = \vec{v}' \text{ and } \cmem' = \cmem'' \memcomp \cmem_f))
\\[1mm]
(\text{UX}) &\text{If } & \cmem.\act(\vec{v}) \rightsquigarrow \result : (\cmem', \vec{v}') \text{ and }
\result \neq \omiss \text{ and }
\cmem' \memcomp \cmem_f \text{ is defined}  \\
&\text{then } & 
(\cmem \memcomp \cmem_f).\act(\vec{v}) \rightsquigarrow \result : (\cmem' \memcomp \cmem_f, \vec{v}')
\end{array}
\]
\end{instance-property}

The OX property is more subtle than the UX property since to capture that we can extend analysis results to larger states we must say, perhaps counterintuitively, that \emph{removing} a ``frame'' $\cmem_f$ from $\cmem \cdot \cmem_f$ results in either a $\omiss$ outcome or the same behaviour as executing from the full state, rather than more straightforwardly stating something about \emph{adding} more state; see \citet{Yang02} for an in-depth discussion of the OX property. Note that while both properties capture that the analysis results can be extended to larger states, the frame $\cmem_f$ is added to the initial state $\cmem$ for OX reasoning (as, recall the definition in \S\ref{sec:overview}, OX soundness universally quantifies over all initial states) and to the final state $\cmem'$ for UX reasoning (as UX soundness instead universally quantifies over all final states).


\section{Assertions and Function Specifications}%
\label{sec:specifications}

We introduce our assertion language and its satisfaction relation, parametric on a \emph{resource model} comprising the resource assertions and satisfaction relation described in IDef.~\ref{idef:res-srel}. The assertions provide the pre- and postconditions of SL and ISL function specifications and are also used in assertion-based constructs of our CSE engine such as folding and unfolding of user-defined \mbox{predicates}.

\paragraph{Assertion syntax}

We define assertions, $\Assert$, assuming  a set of logical variables, $x, y, z, \in \LVar$, distinct from program variables, and a set of logical expressions, $\lexp \in \LExp$, which extends program expressions $\PExp$ to include these logical variables and two new expressions $\lexp \in \vals$ and $\lexp \in \type$ where $\type ::= \nils \mid \bools \mid \nats \mid \dots$, meaning, that the expression $\lexp$ successfully evaluates to a value and successfully evaluates to a value of type $\type$, respectively. The syntax of assertions~is~defined~by:
\[
\begin{array}{r@{~}c@{~}l}
A \in \Assert &\defeq& \lexp \mid \AssTrue \mid { A_1 \Rightarrow A_2} \mid A_1 \vee A_2 \mid \exists x \ldotp A \mid \emp \mid A_1 \lstar A_2 \mid \resource(\vec{\lexp_1}; \vec{\lexp_2}) \mid \pred(\vec{\lexp_1}; \vec{\lexp_2})
\end{array}
\]
for $\lvar{x} \in \LVar$, $\lexp \in \LExp$, $\vec{\lexp_1}, \vec{\lexp_2} \in \vec{\LExp}$, $\resource \in \strings$, and $\pred \in \strings$.
Our assertions comprise Boolean assertions $\lexp$, several first-order connectives and quantifiers, the empty-memory assertion $\emp$, assertions built using the separating conjunction $\lstar$, resource assertions $\resource(\vec{\lexp_1}; \vec{\lexp_2})$, and user-defined predicate assertions $\pred(\vec{\lexp_1}; \vec{\lexp_2})$. The parameters of resource and user-defined predicate assertions are split into \emph{in-parameters} and \emph{out-parameters} for automation purposes: in our CSE engine, the \mac operation requires the in-parameters \emph{to be known} before consumption and \emph{learns} the out-parameters during consumption; see \citet{matchplanning} for~further~details.

\paragraph{Satisfaction relation}

To define the satisfaction relation for assertions, we introduce logical interpretations, $\subst : \LVar \rightharpoonup_{\mathit{fin}} \Val$, and the  evaluation of logical expressions, $\esem{\lexp}{\subst,\sto}$, which extends program expression evaluation to interpret logical variables using $\theta$. The satisfaction relation, 
 $\subst, \st \models A$, is defined in \ifarxiv App.~\ref{app:specifications}\else the~extended~paper~\cite{arxiv}\fi; the interesting cases are:
\[
\begin{array}{@{}l@{~}l@{~}c@{~\ }l}
 \subst, (\sto, \hp)  \models
 & A_1 \lstar A_2 &\Leftrightarrow&
  \exists \hp_1, \hp_2 \ldotp \hp = \hp_1 \memcomp \hp_2 \text{ and  } \subst, (\sto, \hp_1) \models A_1 \text{ and } \subst, (\sto, \hp_2) \models A_2 \\
\subst, (\sto, \hp)  \models & \resource(\vec{\lexp_1}; \vec{\lexp_2}) &\Leftrightarrow&
  \esem{\vec{\lexp_1}}{\subst,\sto} = \vec{v_1} \text{ and }\esem{\vec{\lexp_2}}{\subst,\sto} = \vec{v_2} \text{ and } \hp \models_{\resources} \resource(\vec{v_1}; \vec{v_2}) \text{ defined in IDef.~\ref{idef:res-srel}.}
\end{array}
\]
As is standard for parametric separation logics, the semantics of $\lstar$ is defined with respect to the composition operator $\memcomp$ from the memory PCM instance data (IDef.~\ref{idef:cmm}). The semantics of resource assertions $\resource(\vec{\lexp_1}; \vec{\lexp_2})$ is also defined by instance data:
\begin{instance-definition}\label{idef:res-srel}
A set of resource assertions of the form $\resource(\vec{v_1}; \vec{v_2})$, and a satisfaction relation for the resource assertions, $\hp \models_{\resources} \resource(\vec{v_1}; \vec{v_2})$.
\end{instance-definition}

\begin{instance-definition-example}{}{}\label{iexa:assertion}
In our linear memory model, there are two types of resources: the positive cell assertion, $\lexp_1 \mapsto \lexp_2$ (in prettified syntax) with in-parameter $\lexp_1$ and out-parameter $\lexp_2$, and the negative cell assertion, $\lexp \mapsto \cfreed$ with in-parameter $\lexp$. Their~satisfaction~relation~is~as~follows:
\[
\begin{array}{@{}l@{~}l@{~}c@{~\ }l}
 \hp  \resmodels &
 n \mapsto v & \Leftrightarrow & \hp = \{n \mapsto v\} \\
 \hp  \resmodels & n \mapsto \cfreed & \Leftrightarrow & \hp = \{n \mapsto \cfreed \}
 \end{array}
\]
\end{instance-definition-example}

\paragraph{Program logics and function specifications}%

On top of our assertion language, we define the standard SL and ISL function triples (and quadruples, to describe both successful and erroneous outcomes). Our theory additionally supports exact separation logic triples~\cite{esl}, which are triples that are valid both in the SL and ISL senses (i.e., they can be used for both types of reasoning). As the SL, ISL, and ESL definitions are standard, we only give the formal definitions in \ifarxiv App.~\ref{app:specifications}\else the~extended~paper~\cite{arxiv}\fi. It is important that the definitions are indeed standard, otherwise CSE implementations based on our theory would not be formally interoperable with other analysis implementations based on SL and ISL (i.e., one could not formally exchange specifications between the implementations).

We additionally need the following definitions to eventually state our engine soundness theorems. A function specification context,~$\fsctx$, is a function from function identifiers to sets of function specifications. We say a function specification context $\fsctx$ is valid w.r.t. a function implementation context $\fictx$, denoted $\models (\fictx, \fsctx)$, when all function specifications in $\fsctx$ are valid w.r.t. $\fictx$. Formal definitions are, again, in \ifarxiv App.~\ref{app:specifications}\else the~extended~paper~\cite{arxiv}\fi.


\section{CSE Engine}%
\label{sec:language-symbolic}

We introduce our CSE engine and prove its OX and UX soundness theorems. Our discussion is focused on the parameters of the relevant theory that we say form a \emph{symbolic memory model} and our \emph{soundness relations}. The relevant parameters are: the symbolic memory data type~(IDef.~\ref{idef:smm}), its satisfaction relation (IDef.~\ref{idef:smm-srel}), symbolic memory actions (IDef.~\ref{idef:smm-semantics}), and consume and produce operations (IDef.~\ref{idef:smm-cp}) and their associated IProps. A larger excerpt of the formal rules of the engine is given in \ifarxiv App.~\ref{app:symbolic}\else the~extended~paper~\cite{arxiv}\fi; and the full set of rules is available in our~Rocq~mechanisation. 

\subsection{Symbolic States}

The symbolic states of our engine, denoted $\sst$, are built out of logical expressions $\LExp$ with the extra condition that none of the logical expressions have program variables. We use hat-notation to distinguish symbolic definitions,  such as $\sst$ for symbolic state compared with~$\st$~for~concrete~state.

The most interesting component of symbolic states is their symbolic memory component, which is given by instance data:
\begin{instance-definition}\label{idef:smm}
A pair $(\smemss, \emptysmem)$, where $\smemss$ is a symbolic memory and $\emptysmem \in \smemss$ is the empty memory.
\end{instance-definition}

\begin{instance-definition-example}\label{iex:lin_smm}
Our linear memory model comprises $\smemss$ equalling $ \LExp \rightharpoonup_{\mathtt{fin}} (\LExp \uplus \{ \cfreed\})$ and $\emptysmem$ equalling the empty function. 
\end{instance-definition-example}

We are now ready to give the full definition of symbolic state: a symbolic state $\sst$ is a tuple $(\ssto, \smem, \sps, \spc)$ where: $\ssto : \PVar \rightharpoonup_{\mathtt{fin}} \LExp$ is a symbolic store; $\smem \in \smemss$ is a symbolic memory from IDef.~\ref{idef:smm}; $\sps$ is a multiset of symbolic user-defined predicates, where a symbolic predicate has the form $\pred(\vec{\sexp}_1; \vec{\sexp}_2)$ where $\pred \in \strings$ is a predicate name and $\vec{\sexp}_1 \in \vec{\LExp}$  are the in-parameters and   $\vec{\sexp}_2 \in \vec{\LExp}$ the out-parameters; and $\spc \in \LExp$ is a path condition that captures constraints imposed during execution.  We use the syntax $\sst.\fieldsst{mem}$, $\sst.\fieldsst{pc}$ etc. to access components of symbolic state and the syntax $\sst[\sstupdate{mem}{\smem'}]$ etc. to update components of symbolic state. We use $\lv{\sst}$ to refer to the logical variables~of~a~symbolic~state~$\sst$.

We define the semantic meaning of symbolic states using a satisfaction relation between concrete and symbolic states of the form $\subst, (\sto, \cmem) \models \sst$ where  $\theta : \LVar \rightarrow \Val$ is a logical interpretation, $s: \PVar \rightarrow  \Val$ is a variable store and $\mu \in \cmemss$ a concrete memory. The  satisfaction relation for symbolic states depends on the  satisfaction relations for symbolic stores $\subst, \sto \models_\text{Sto} \ssto$, symbolic memories $\subst, \cmem \models_\text{Mem} \smem$, and symbolic predicates $\subst, \cmem \models_\text{Pred} \sps$: the satisfaction relations for symbolic stores and symbolic predicates are as expected, see \ifarxiv App.~\ref{app:symbolic}\else the~extended~paper~\cite{arxiv}\fi; the satisfaction relation for symbolic memories is given by instance data:
\begin{instance-definition}\label{idef:smm-srel}
A satisfaction relation for symbolic memory of the form $\subst, \cmem \models_\text{Mem} \smem$, with the property that $\subst, \cmem \models_\text{Mem} \emptysmem \iff \cmem = \emptycmem$.
\end{instance-definition}

\begin{instance-definition-example}\label{iex:lin_sat}
In our linear memory model, the  satisfaction relation for symbolic memory is defined as follows, where for succinct presentation we say $\esem{\cfreed}{\subst} = \cfreed$:
\[
\subst, \hp \models_\text{Mem} \{ \sexp_a^1 \mapsto \sexp_e^1, \dots, \sexp_a^n \mapsto \sexp_e^n \} \Leftrightarrow \cmem = \uplus_{i = 1}^{n} \{ \esem{\sexp_a^i}{\subst} \mapsto \esem{\sexp_e^i}{\subst} \}
\]

\end{instance-definition-example}

\noindent
The satisfaction relation $\subst, (\sto, \cmem) \models (\ssto, \smem, \sps, \spc)$ for symbolic state  is defined by:
\[
\exists \cmem_1, \cmem_2.~\cmem = \cmem_1 \memcomp \cmem_2 \text{ and } \subst, \sto \models_\text{Sto} \ssto \text{ and } \subst, \cmem_1 \models_\text{Mem} \smem \text{ and } \subst, \cmem_2 \models_\text{Pred} \sps \text{ and } \esem{\spc}{\subst} = \true
\]
We say that a symbolic state $\sst$ is satisfiable, denoted $\sat(\sst)$, when $\exists \theta, \sto, \cmem.~\theta, (\sto, \cmem) \smodels \sst$, and say that it implies an expression, denoted $\sst \smodels \sexp$, when $\forall \theta, \sto, \cmem.~\theta, (\sto, \cmem) \smodels \sst \implies \esem{\sexp}{\theta} = \true$.

\subsection{Engine Judgement}

The symbolic semantics of our CSE engine is given by a judgement of the form:
\[
\cseenginetrans{m}{\fsctx}{\oracle}{\sst}{\scmd}{\outcome}{\oracle'}{\sst'}
\]
where: oracles $\oracle, \oracle'$ of type $\nats \to \nats$ are used to model angelic nondeterminism,\footnote{See, e.g., \citet{Owens16} for further discussion on oracles. In short, each number $\oracle(0), \oracle(1), \dots$ represents an answer to a choice and the oracle is shifted once every time a choice is made (such that the oracle can always be queried by looking at $\oracle(0)$), ultimately resulting in an output oracle $\oracle' = \lambda{}n.~\oracle(n + m)$ where $m$ is the number of choices made. The oracle semantics we use is an intentionally simplistic model to avoid cluttering our formalism. In particular, the same angelic choice is taken in each demonic branch. See \citet{Dardinier25} (using multi-relations) or \citet{Jacobs15} (using monads) for more expressive formalisms for nondeterminism.} e.g. when there are multiple applicable function specifications to choose from for a function call; the mode $m$,  either OX, UX, or EX (for exact), 
enables  the engine to switch its behaviour depending on what type of soundness we need, e.g. what kind  of function specifications  to use in  function calls,
and the set of \emph{outcomes}, $o ::= \oxok \mid \oxerr \mid \oxm \mid \oxabort$,  extends the outcomes of the concrete language with $\oxabort$,  e.g. when a chosen function specification~is~not~applicable.

\subsection{Memory-action Command}

The symbolic semantics of memory-action commands, analogous to the concrete semantics, is parametric on an instance-given symbolic action execution relation.
\begin{instance-definition}\label{idef:smm-semantics}
A relation $\smem.\act(\vec{\sexp}) \rightsquigarrow o : (\smem', \spc', \vec{\sexp}')$, which executes an action $\alpha$ on memory $\smem$ with arguments $\vec{\sexp}$, and returns an outcome $o$, memory~$\smem'$, path condition~$\spc'$, and~return~values~$\vec{\sexp}'$. If $o \in \{\omiss, \oxabort\}$, then the symbolic memory must remain unchanged, i.e., $\smem = \smem'$.
\end{instance-definition}

\begin{instance-definition-example}\label{iex:lin_ssem}
To illustrate, we give some of the symbolic rules for the \texttt{lookup} action of our linear memory model (for the rest, see \ifarxiv App.~\ref{app:memory-models}\else the~extended~paper~\cite{arxiv}\fi), specifically, the success and missing resource rules:
\begin{mathpar}
\small
\inferrule
 {\smem(\sexp_l') = \sexp \and \spc' = (\sexp_l = \sexp_l')}
 {\smem.\mathtt{lookup}([\sexp_l]) \rightsquigarrow \oxok : (\smem, \spc', [\sexp])}
\and
%
%
%
\inferrule
 {\spc' = (\sexp_l \in \nats \land \sexp_l \not\in \domain(\smem))}
 {\smem.\mathtt{lookup}([\sexp_l]) \rightsquigarrow \omiss : (\smem, \spc', [\mathstr{\mathsf{MissingCell}}, \sexp_l])}
\end{mathpar}
where, note, the first rule branches over all possible addresses $\sexp_l'$ where $\sexp_l = \sexp_l'$ holds. We discuss other ``branching strategies'' in \S\ref{sec:mem-linear}, where we discuss variants of the memory model.

\end{instance-definition-example}

\noindent
The following symbolic rules lift, again analogously to the concrete semantics, the memory actions to the full semantics, where $\seval{\pexp}{\ssto}$ denotes symbolic expression evaluation, which evaluates a program expression $\pexp$ w.r.t.~a symbolic store $\ssto$:
\begin{mathpar}
\small
  \inferrule
  {
    \cseeval{\vec{\pexp}}{\ssto}{}{\vec{\sexp}}{} \and
    \smem.\act(\vec{\sexp}) \rightsquigarrow \oxok : (\smem', \spc', \vec{\sexp}') \and
    |\vec{\sexp}'| = |\vec{\pvar x}| \\\\
    \ssto' = \ssto [\vec{\pvar{x}} \storearrow \vec{\sexp}' ] \and
    \spc'' = (\ssto(\vec{\pvar{x}}), \vec{\sexp}, \vec{\sexp}' \in \Val \land \spc' \land \spc)
  }
  {
     \cseenginetrans{m}{\fsctx}{\oracle}{\sthreadp{ \ssto }{ \smem, \sps, \spc}}{\vec{\pvar x} := \act(\lst\pexp)}{\oxok}{\oracle}{\sthreadp{\ssto'}{ \smem', \sps, \spc'' }}
  } \hspace{0.5cm}
  \inferrule
  {
    \cseeval{\vec{\pexp}}{\ssto}{}{\vec{\sexp}}{} \and
    \smem.\act(\vec{\sexp}) \rightsquigarrow \outcome : (\smem', \spc', \vec{\sexp}') \and
    \outcome \neq \oxok \\\\
    \ssto' = \ssto [\pvar{err} \storearrow \vec{\sexp}' ] \and
    \spc'' = (\vec{\sexp} \in \Val \land \spc' \land \spc)
  }{
    \cseenginetrans{m}{\fsctx}{\oracle}{\sthreadp{ \ssto }{ \smem, \sps, \spc}}{\vec{\pvar x} := \act(\lst\pexp)}{\oxerr}{\oracle}{\sthreadp{\ssto'}{ \smem', \sps, \spc'' }}
  }
\end{mathpar}

\noindent
%

\noindent
We require the following property to be able to prove that our CSE engine, specifically, the memory-action commands, satisfy our two soundness theorems:
\begin{instance-property}\label{iprop:mem-sound}
The symbolic memory-action semantics must be OX/UX sound w.r.t. the concrete memory-action semantics, i.e., the two semantics must satisfy the OX/UX soundness definitions introduced in \S\ref{sec:overview} but with the command-level concrete semantics replaced by the memory-action-level concrete semantics, the satisfaction relation $\smodels$ replaced by the satisfaction relation $\memmodels$, etc. The full formal properties are given in \ifarxiv App.~\ref{app:symbolic}\else the~extended~paper~\cite{arxiv}\fi.
\end{instance-property}

\subsection{Consume and Produce Operations and Assertion-based Commands}

We now discuss the definition, soundness, and usage of our memory-model-parametric \mac and \produce operations that form the basis of our CSE engine's consume-produce architecture.

\paragraph{Definition}

The judgements of \mac and \produce are as follows. First, we introduce symbolic substitutions, $\ssubst : \LVar \rightharpoonup_{\mathtt{fin}} \LExp$, which the two operations use to instantiate free logical variables. Now, the judgements of \mac and \produce are:
\[
\mac(m, O, A, \ssubst, \sst) \rightsquigarrow (\outcome, O', \ssubst', \sst') \qquad \text{ and } \qquad \produce(A, \ssubst, \sst) \rightsquigarrow \sst'
\]
where the judgement for \mac states that the consumption of assertion $A$ in mode $m$ (which decides how Boolean information is consumed) with an oracle $O$ and an initial symbolic substitution $\ssubst$ from state $\sst$ results in outcome $\outcome$ ($\oxok$ or $\oxabort$), an updated oracle $O'$, symbolic state $\sst'$ where the symbolic state corresponding to $A$ has been removed, and extended symbolic substitution $\ssubst'$ now containing mappings for all free logical variables of $A$; and the judgement for \produce states that the production of $A$ with symbolic substitution $\ssubst$ in state $\sst$ results in state $\sst'$ where the symbolic state corresponding to $A$ has been added.

We only discuss how resource assertions $\resource(\vec{\lexp_1}; \vec{\lexp_2})$ are consumed and produced (remaining rules are inspired by the rules of \citet{cse1},\footnote{Our operations support the subset of assertions usually supported by \mac and \produce operations. That is, both operations support Boolean assertions, existential quantification (for \mac only in OX mode, we do not know of a use-case in UX mode), the empty-memory assertion, separating-conjunction assertions, resource assertions, and user-defined-predicate assertions, and, additionally, \produce supports disjunction assertions.} which in turn are inspired by Gillian). The \mac and \produce operations are parametric on two resource-level consume and produce operations, which we call, respectively, \emph{resource consume} $\resconsume$ and \emph{resource produce} $\resproduce$:
\begin{instance-definition}\label{idef:smm-cp}
Two operations $\resconsume$ and $\resproduce$ of the form explained below:
\[
\resconsume(m, O, r, \sexpin, \smem) \rightsquigarrow (\outcome, O', \sexpout, (\smem', \spc_i, \spc)) \quad \text{ and } \quad \resproduce(r, \sexpin, \sexpout, \smem) \rightsquigarrow (\smem', \spc)
\]

\end{instance-definition}

\begin{instance-definition-example}\label{iex:lin_cp}
For our linear memory model, the following are two of the rules for $\resconsume$ and $\resproduce$ (again, the rest of the rules are in \ifarxiv App.~\ref{app:memory-models}\else the~extended~paper~\cite{arxiv}\fi):
\begin{mathpar}
\small
\inferrule{\smem=\smem_f \uplus \{\sexp_1 \mapsto \sexp_2\}}
{\resconsume(m, O, \mapsto,[\sexp],\smem)
\rightsquigarrow (ok, O, {\sexp_2},(\smem_f, \true, \sexp=\sexp_1))}
%
%
\and
\inferrule{\smem'=\smem \uplus \{\sexp_1\mapsto \sexp_2\}}
{\resproduce(\mapsto, [\sexp_1],[\sexp_2],\smem)\rightsquigarrow (\smem', \true)}
\end{mathpar}
\end{instance-definition-example}

\begin{wrapfigure}{r}{0.59\textwidth}
\begin{center}
  \small

\begin{mathparpagebreakable}
\inferrule
{
\resconsume(m, O, r , \ssubst(\vec{\lexp}_\text{in}), \sst.\fieldsst{mem})\rightsquigarrow (\outcome, O', \sexpout, (\smem', \spc_i, \spc))\\\\
\sst \models \spc_i \and
\sst'=\sst[\sstupdate{mem}{\smem'},\sstupdate{pc}{\spc\wedge \sst.\fieldsst{pc}} ] \\\\
\textcolor{gray}{~\text{rest of the rule omitted}}
}
{\mac(m, O, r(\vec{\lexp}_\text{in}; \vec{\lexp}_\text{out}), \hat{\theta}, \sst) \rightsquigarrow (\outcome, O', \hat{\theta}', \sst'')}
\end{mathparpagebreakable}
\end{center}
\vspace{-1em}
\caption{Implementation of $\mac$.}
\vspace{-1em}
\label{fig:cons_impl}
\end{wrapfigure}

\noindent
The $\resconsume$ and $\resproduce$ operations, analogous to memory actions, are lifted into \mac and \produce. We explain the \mac case, the \produce case is similar. See the $\mac$ rule in Fig.~\ref{fig:cons_impl}, which illustrates the most interesting parts of how $\resconsume$ is lifted into \mac. There are two conditions that $\resconsume$ outputs: $\spc_i$, which must be implied by the initial state, and $\spc$, which is appended to the path condition of the updated state. The two conditions are used to implement different types of branching, which we illustrate by example when discussing variants of the linear memory model in~\S\ref{sec:mem-linear}.

\paragraph{Soundness} We introduce OX and UX soundness properties that formalise that the \mac and \produce operations ``correctly'' consume and produce their input assertions. Our soundness properties are based on the soundness properties for \mac and \produce introduced by \citet{cse1}, which we have refactored into four properties: (1) \mac OX soundness, (2) \produce OX soundness, (3) \mac UX soundness, (4) \produce UX soundness. The full properties are available in \ifarxiv App.~\ref{app:symbolic}\else the extended paper~\cite{arxiv}\fi; in short, the properties relate the behaviour of \mac and \produce to the satisfaction relation of the assertion language. For example, UX soundness of \mac requires that the composition of the models of the input assertion and the output symbolic state forms a model of the input~symbolic~state:
\[
\begin{array}{ll}
\text{If }& \mac(m, O, A, \ssubst, \sst) \rightsquigarrow  (ok, O', \ssubst', \sst_f)\\
&  \text{and } \subst, (\sto, \hp_A) \models \ssubst'(A) \text{ and } \subst,  (\sto, \hp_f) \models \sst_f \text{ and } (\hp_A\cdot \hp_f) \text{ is defined} \\
\text{then } & \subst, (\sto, \hp_A\cdot \hp_f)\models \sst
\end{array}
\]

\noindent
To ensure that our soundness properties of \mac and \produce hold, we require that resource-only variants of the properties to hold for the $\resconsume$ and $\resproduce$ operations.
\begin{instance-property}\label{iprop:cp-sound} The $\resconsume$ and $\resproduce$ operations must be OX/UX sound. Again, the full properties are stated in \ifarxiv App.~\ref{app:symbolic}\else the extended paper~\cite{arxiv}\fi. To exemplify the resource-only variant of the properties, we give the resource-only variant of the UX soundness property for \mac (the reader should compare the property with the property above):
\[
\begin{array}{ll}
\text{If}   &\resconsume(m, O, \resource, \sexpin, \smem)     \rightsquigarrow (ok, O', \sexpout,(\smem_f,\spc_{\mathit{fi}}, \spc_f)) \text{ and } \esem{\sexpin}{\subst} = \vin \text{ and }  \esem{\sexpout}{\subst}= \vout
   \\
& \text{and } \esem{\spc_f}{\subst}=\true \text{ and }  \subst, \hp_f\memmodels \smem_f \text{ and } \hp_r \resmodels \resource(\vin; \vout)\text{ and } (\hp_r\cdot \hp_f) \text{ is defined } \\
\text{then}&  \esem{\spc_{\mathit{fi}}}{\subst}=\true \text{ and }  \subst, (\hp_r\cdot \hp_f) \memmodels \smem
\end{array}
\]
\end{instance-property}

\noindent
With the above definitions in place, we have been able to prove the following:
\begin{lemma}\label{lem:cp}
Given OX/UX sound $\resconsume$ and $\resproduce$ operations, our \mac and \produce operations are OX/UX sound.
\end{lemma}

\paragraph{Usage} We have mechanised and proved sound the standard consume-produce definitions of the function-call command and fold/unfold commands for user-defined predicates, the full definitions are available in \ifarxiv App.~\ref{app:symbolic}\else the extended paper~\cite{arxiv}\fi{} for completeness. The function-call command we have proved OX and UX sound, whereas the fold/unfold commands we have proved OX sound.\footnote{\citet{cse1} claim that the fold command is UX sound if folding is restricted to strictly exact predicates (an assertion $A$ is strictly exact iff $\subst, (\sto, \hp) \models A \land \subst, (\sto, \hp') \models A \implies \hp' = \hp$~\cite[p.~149]{Yangphd}), but during our mechanisation work we found a counterexample to this claim. We have not investigated a new condition to make the command UX sound.} The successful proofs of these commands show that our consume-produce properties are sufficient for their core use cases. Additionally, as we discuss in \S\ref{sec:analyses}, the analyses we have built on top of our engine show that our consume-produce properties are also sufficient for those analyses.

\subsection{Engine Soundness}%
\label{sec:soundness}

Our OX soundness and UX soundness theorems are as follows:\footnote{Here, we have simplified away some uninteresting details of the statements, see the Rocq mechanisation~for the~full~details.}
\begin{theorem}[OX soundness]\label{thm:ox-sound}
Let $m \in \{ \macOX, \macEX \}$ and assume $\models (\fictx, \fsctx)$, when the CSE engine is instantiated with OX sound memory actions, $\resconsume$, and $\resproduce$, then the following holds:
\[
\begin{array}{ll}
\text{If } & \csesemtransabstract{\st}{\cmd}{\st'}{\fictx}{\outcome} \text{ and } \subst, \st \models \sst \text{ and} \\
& (\forall o, \oracle', \sst'.~
 \cseenginetrans{m}{\fsctx}{\oracle}{\sst}{\cmd}{\outcome}{\oracle'}{\sst'} \text{ and } \sat(\sst') \implies \\
& \qquad \outcome \neq \oxabort \text{ and  } (\outcome = \oxm \implies \sst'.\fieldsst{preds} = \emptyset)) \\
\text{then } & \exsts{\oracle', \sst', \subst'}
\cseenginetrans{m}{\fsctx}{\oracle}{\sst}{\cmd}{\outcome}{\oracle'}{\sst'} \text{ and }
\subst'|_{\lv{\sst}} = \subst \text{ and }
\subst', \st' \models \sst'
\end{array}
\]
\end{theorem}
\begin{theorem}[UX soundness]\label{thm:ux-sound}
Let $m \in \{ \macUX, \macEX \}$ and assume $\models (\fictx, \fsctx)$, when the CSE engine is instantiated with UX sound memory actions, $\resconsume$, and $\resproduce$, then the following holds:
\[
\begin{array}{ll}
\text{If } & \cseenginetrans{m}{\fsctx}{\oracle}{\sst}{\cmd}{\outcome}{\oracle'}{\sst'}
\text{ and } \outcome \neq \oxabort
\text{ and } (\outcome = \oxm \implies \sst'.\fieldsst{preds} = \emptyset)
\text{ and } \subst, \st' \models \sst' \\
\text{then } & \exsts{\st} \csesemtransabstract{\st}{\cmd}{\st'}{\fictx}{\outcome} \text{ and }\subst, \st \models \sst
\end{array}
\]
\end{theorem}

Both theorems have restrictions on $\oxabort$ and $\oxm$ outcomes. For the OX theorem, the condition should be read as follows: no reachable satisfiable state has an outcome $\oxabort$ or an outcome $\oxm$ unless there are no symbolic predicates in the state. For both theorems, the soundness of $\oxm$ outcomes cannot be guaranteed in the presence of symbolic predicates because the source of $\oxm$ outcomes, memory actions (IDef.~\ref{idef:smm-semantics}), do not take symbolic predicates into consideration (doing so would require implementing an automated complete unfolding procedure, which none of the existing CSE tools or platforms implement). To exemplify, say we are working with our running example linear memory model and have defined the following user-defined predicate: $\mathsf{foo}(;\!) \{1 \mapsto 1\}$. First, let $\cmd \defeq \pvar{x} := \mathtt{lookup}(1)$, $\sst \defeq (\emptyset, \emptysmem, \{\mathsf{foo}(;\!)\}, \true)$, and $\st \defeq (\emptyset, \{1 \mapsto 1\})$. Now, note that $\emptyset, \st \models \sst$, concrete execution of $\cmd$ from $\st$ only results in an $\oxok$ outcome, and symbolic execution of $\cmd$ from $\sst$ only results in a $\oxm$ outcome. This breaks both OX and UX soundness: there is no corresponding execution for the concrete execution and vice versa.

\subsection{Analyses}%
\label{sec:analyses}

We now discuss two analyses we have built on top of our CSE engine and proved sound: a \emph{function specification verification analysis} to exemplify an OX analysis application and a \emph{true bug-finding analysis based on bi-abduction} to exemplify an UX analysis application. We additionally discuss the trusted computing base (TCB) of analysis results when using our engine.

\paragraph{The two analyses} The development and verification of our OX analysis application was relatively uneventful: our work validates that our adapted consume-produce properties are sufficient for this OX application, but the analysis itself is standard in the consume-produce literature, and we did not run into any particular problems with porting its proof to our parametric setting. We therefore only discuss our UX application here; see \ifarxiv App.~\ref{app:symbolic}\else the extended paper~\cite{arxiv}\fi{} for our OX application.

Bi-abduction is a technique that facilitates automatic ISL specification synthesis by incrementally discovering the resources needed to execute a given piece of code starting from an empty pre-condition/symbolic state. It was first introduced in the OX setting~\cite{calcagno:popl:2009,calcagno:jacm:2011}, forming the basis of the Infer tool~\cite{calcagno:nasa:2011}. It was later ported to the OX consume-produce setting in the JaVerT 2.0 project~\cite{javert2}, by re-imagining bi-abduction as \emph{fixes-from-missing-resource-errors}. With the introduction of incorrectness separation logic~\cite{isl}, the original bi-abduction algorithm was ported to the UX setting of true bug-finding, underpinning the Infer-Pulse tool~\cite{Le22}. Following this, \citet{cse1} showed that the fixes-from-missing-resource-errors approach is UX sound in the setting of linear memory. Here, we generalise \citet{cse1}'s UX result to our memory-model-parametric setting.

We have built a bi-abductive engine with judgement $\cseenginetrans{\biexmode}{\fsctx}{\oracle}{\sst}{\scmd}{\outcome}{\oracle'}{(\sst', A)}$ on top of our engine with judgement $\cseenginetrans{\macUX}{\fsctx}{\oracle}{\sst}{\scmd}{\outcome}{\oracle'}{\sst'}$. The $A$ in the judgement is an assertion representing an \emph{anti-heap}, which captures the missing resources needed to execute the command $\scmd$ in~the~following~sense:
\begin{theorem}[CSE with Bi-Abduction: Soundness]\label{thm:bi-abduction}
\[
\begin{array}{ll}
\text{If } & \models (\fictx, \fsctx) \text{ and }
\cseenginetrans{\biexmode}{\fsctx}{\oracle}{\sst}{\scmd}{\outcome}{\oracle'}{(\sst', A)} \text{ and }
\subst, \st' \models \sst' \\
\text{then } &
\exists \sto, \cmem, \cmem_\text{fix}.~
\subst, (\sto, \cmem) \models \sst[\sstupdate{pc}{\sst'.\fieldsst{pc}}] \text{ and }
\subst, (\sto, \cmem_\text{fix}) \models A \text{ and }
\csesemtransabstract{(s, \cmem \memcomp \cmem_\text{fix})}{\scmd}{\st'}{\fictx}{\outcome}
\end{array}
\]
\end{theorem}

\noindent
In short, the bi-abductive engine works by catching missing-resource errors and abort errors during execution, which are given to a memory-model-dependent operation $\fix$ that takes, as input, the current symbolic state and constructs one or more assertions representing the resources needed for continued execution, which in turn are produced into the current symbolic state and appended to the anti-heap. The soundness of the engine~(Thm.~\ref{thm:bi-abduction}) follows from the UX soundness of our CSE engine (Thm.~\ref{thm:ux-sound}) and \produce~(Lem.~\ref{lem:cp}). To prove soundness, we had to adapt the soundness statement and proofs from previous work~\cite{javert2,cse1}, which relied on having a symbolic composition operator and a symbolic frame property, which we do not require as parameters.

\paragraph{Trusted computing base} As one would expect from a mechanised theory, our CSE theory comes with a strong TCB story: the TCB of analysis results includes only the semantics of assertions (to express pre- and postconditions) and the concrete semantics of the language. The TCB can be further reduced by considering only first-order assertions, as then the assertion language can be removed from the TCB (this is analogous to sound program logics, see, e.g., Iris'~adequacy~theorem~\cite{jung:jfunc:2018}).

The fact that the concrete memory model is part of the TCB (as it is part of the concrete semantics), means that one must choose the model carefully. Appropriate TCB models have a direct correspondence to a memory model specified by a language standard or the like. Because of space constraints, we do not discuss this point further but show in our Rocq development how analysis artefacts like ghost state annotations (e.g., like the annotations used by our fractional ownership model introduced in \S\ref{sec:mem-frac}) can easily be removed from the TCB by a~standard~simulation~argument.


\section{Memory-model Instances}%
\label{sec:more-memory-models}

Now having introduced the IDefs. and IProps. required to instantiate our CSE theory, we discuss additional examples of memory-model instances that fit into our CSE theory, as~summarised~in~Tab.~\ref{tab:instances}.

We emphasise that the primary purpose of our discussion in this section is to show that a wide array of memory models fit into our CSE theory. We do not have sufficient page budget to formally introduce all required IDefs. (i.e, IDefs.~\ref{idef:cmm}--\ref{idef:smm-cp}) and discuss proofs of the required IProps. (i.e., \mbox{IProp. \ref{iprop:cmm-pcm}--\ref{iprop:cp-sound}}) for each memory instance. Therefore, our discussion is informal and focused on what we have found to be the primary difficulty to get right in designing memory models: the memory data type (IDef.~\ref{idef:cmm}), such that compositional memory actions (IDef.~\ref{idef:cmm-semantics}) that work over SL/ISL-style partial state can be defined etc. (We give additional definitions in \ifarxiv App.~\ref{app:memory-models}\else the extended paper~\cite{arxiv}\fi{} and all definitions are available in our Rocq development.) With the right data type in place, we have found other major instance data, such as symbolic memory actions (IDef.~\ref{idef:smm-semantics}) and $\resconsume$ and $\resproduce$ operations (IDef.~\ref{idef:smm-cp}), to be relatively straightforward to define. In particular, for simpler memory models, symbolic components can be designed by ``symbolically lifting'' of the corresponding concrete component, in particular, the symbolic data type and memory actions. For an example, compare the concrete data type (IDef.~\ref{idef:cmm}) and symbolic data type (IDef.~\ref{idef:smm}) of our running example memory model and note how the symbolic data type is structurally identical to the concrete data type but abstracts both~$\nats$~and~$\vals$~to~$\lexps$.

\subsection{Linear Memory Models and Memory-model Design Considerations}%
\label{sec:mem-linear}

We have mechanised and proved sound multiple variants of our running example linear memory model (see again Tab.~\ref{tab:instances}). As discussed in the introduction (\S\ref{sec:intro}), there are multiple degrees of freedom available when designing a symbolic memory model. Simple linear memory models provide a good stage to illustrate this; as we have already introduced the various IDefs. of our running example linear memory model, we in this section discuss design considerations relating to OX vs. UX analysis, such as what types of ``branching strategies'' are allowed by different reasoning modes.

\paragraph{Operational meaning of OX vs. UX} The different requirements arising from OX and UX soundness can be exploited in memory-model design. Intuitively,  OX analyses like verification must consider all execution paths whereas  UX analyses like bug-finding only need to consider paths with bugs. More precisely: operationally, OX soundness allows for dropping information along execution paths but not dropping paths, whereas UX soundness allows for dropping paths but not information.

\paragraph{Path maintenance, illustrated through branching strategies} When updating and removing parts of memory, there are multiple ways to handle ``branching'', i.e., situations where there are multiple potential parts of memory to update/remove, sometimes referred to as ``matches''. To exemplify, we discuss branching in the context of $\resconsume$ for linear memory models (that is, variants of IDef.~\ref{idef:smm-cp}). For the discussion, it is important to have Fig.~\ref{fig:cons_impl} fresh in mind. Say we have $\smem = \{ 1 \mapsto 1, 2 \mapsto 1 \}$ and $\spc = 1 \le \sym{x} \le 2$ and are about to consume a resource assertion $x \mapsto 1$ knowing $x = \sym{x}$. Now consider the following three branching strategies, where $(\sexp_1, \oracle') \in \oracle(\dom(\smem))$ denotes that we angelically pick an element from $\dom(\smem)$:
\begin{mathparpagebreakable}
\small
\inferrule{\smem=\smem_f \uplus \{\sexp_1 \mapsto \sexp_2\}}
{\resconsume(m, O, \mapsto,[\sexp],\smem)
\rightsquigarrow \\\\ (ok, O, {[\sexp_2]},(\smem_f, \true, \sexp=\sexp_1))}
\and
\inferrule{
(\sexp_1, \oracle') \in \oracle(\dom(\smem))\\\\
\smem=\smem_f \uplus \{\sexp_1 \mapsto \sexp_2\}}
{\resconsume(m, O, \mapsto,[\sexp],\smem)
\rightsquigarrow \\\\ (ok, O', {[\sexp_2]},(\smem_f, \sexp=\sexp_1, \sexp=\sexp_1))}
\and
\inferrule{
(\sexp_1, \oracle') \in \oracle(\dom(\smem))\\\\
\smem=\smem_f \uplus \{\sexp_1 \mapsto \sexp_2\}}
{\resconsume(m, O, \mapsto,[\sexp],\smem)
\rightsquigarrow \\\\ (ok, O', {[\sexp_2]},(\smem_f, \true, \sexp=\sexp_1))}
\end{mathparpagebreakable}
The \emph{left rule} belongs to our running example linear memory model and the two other rules to variant models we have defined. The left rule branches over all possible matches; in our example, we get two branches: one branch with $\smem_f = \{ 2 \mapsto 1 \}$ and $\spc = 1 \le \sym{x} \le 2 \land \sym{x} = 1$ and one branch with $\smem_f = \{ 1 \mapsto 1 \}$ and $\spc = 1 \le \sym{x} \le 2 \land \sym{x} = 2$. The \emph{middle rule} implements unique-match branching since the rule is only applicable when there is a unique match. The rule is not applicable to our example as neither $\sym{x} = 1$ nor $\sym{x} = 2$ is implied by the current symbolic state. We have proved a linear memory model implementing this type of branching to be both OX sound and UX sound. Interestingly, the same rule but with conclusion $\dots \rightsquigarrow (ok, O', {[\sexp_2]},(\smem_f, \sexp=\sexp_1, \true))$  is OX sound but not UX sound because our CSE engine use the entire symbolic input state in the implication check -- meaning that the implication might not hold w.r.t. the smaller output state. (With a stricter implication check requiring that the path condition, rather than the full symbolic state, implies the matching condition (in the example: $\sexp=\sexp_1$), the rule would be UX sound.) Lastly, the \emph{right rule} implements, what we call, cut branching because the rule simply angelically picks one branch without checking if there are more matches. We have proved a linear memory model implementing this type of branching to be UX sound but not OX sound; the model is not OX sound because in OX reasoning we are not allowed to drop matches.

\begin{wrapfigure}{r}{0.37\textwidth}
\begin{mathparpagebreakable}
\small
\inferrule
 {\smem(\sexp_l') = \sexp_\textit{old} \and \smem' = \smem[\sexp_l' \mapsto \sexp] \\\\
  \spc' = (\sexp_l = \sexp_l' \land \sexp_\textit{old} \in \Val)}
 {\acttrans{\smem}{mutate}{[\sexp_l, \sexp]}{\oxok}{\smem', \spc', []}}
\end{mathparpagebreakable}
\caption{Successful \texttt{mutate} rule.}%
\label{fig:linear-mutate}
\end{wrapfigure}

\paragraph{Information maintenance} Beyond variants of our running example linear memory model with different branching strategies, we have also mechanised and proven sound a memory-model instance for the traditional linear memory model from the OX literature, with concrete data type $\Nat \rightharpoonup_{\mathit{fin}} \Val$. This is an OX-only model: the model does not keep track of freed cells, using $\cfreed$, and can therefore not be proven UX sound (because it does not satisfy UX frame). Since it is an OX-only model, dropping information is allowed. As a simple illustration, consider Fig.~\ref{fig:linear-mutate}, containing the successful \texttt{mutate} rule of our running example linear memory model. Note that the rule ensures that the information that the previous cell value successfully evaluates is kept by updating the path condition with $\sexp_\textit{old} \in \Val$.\footnote{This is not the only way to ensure this. It is also possible to, e.g., maintain an invariant saying that the path condition must include this type of information. The point being made here is that this needs to be ensured \emph{in some way}.} This is optional in OX-only models: our OX-only model is defined using rules that do not add evaluation information to the path condition, and we~have~still~been~able~to~prove~the~model~to~be~OX~sound.

\subsection{Fractional Ownership Memory Model}
\label{sec:mem-frac}

To illustrate that different ownership disciplines fit into our CSE theory, we have mechanised and proved OX and UX sound a linear memory model with fractional ownership~\cite{Boyland03,Bornat05} rather than exclusive ownership, as utilised in the memory models discussed up to this point. A memory model similar to the fractional ownership memory model we discuss here has previously been implemented in an experimental branch of Gillian and tested on a small set~of~hand-written~examples.

\paragraph{Model description} The memory model is best explained in terms of resources (i.e., IDef.~\ref{idef:res-srel}). Points-to assertions for the model are of the form $n \overset{q}{\mapsto} v$, where $q \in (0, 1] \subset \rationals$ specifies the amount of ownership. {\em Less-than-1 ownership} ($q < 1$) gives read permission to the location $n$, while {\em full ownership} ($q = 1$) gives both read and write permissions. The other IDefs. of the model are relatively straightforward extensions of our running example memory model. In particular, the concrete memory data type of the model is $\Nat \rightharpoonup_{\mathit{fin}} ((\Val, \rationals) \uplus \{ \cfreed \})$ and the symbolic memory data type is derived from this data type by symbolic lifting, i.e., $\LExp \rightharpoonup_{\mathit{fin}} ((\LExp, \LExp) \uplus \{ \cfreed \})$. The implementation of memory actions, $\resconsume$, and $\resproduce$ are, as one would expect, also similar but with additional ownership checks. For illustration, we show the two successful rules of $\resconsume$ for points-to assertions:\footnote{The ``$\sexp_l' \notin \dom(\smem_f)$'' expression in the first rule is needed for UX soundness, to not drop disjointedness information.}
\begin{mathparpagebreakable}
\small
\inferrule{\smem = \smem_f \uplus \{\sexp_l' \mapsto (\sexp_v, \sexp_q') \}}
          {\resconsume(m, O, \mapsto, [\sexp_l, \sexp_q], \smem)
           \rightsquigarrow \\\\ (ok, O, {[\sexp_v]},(\smem_f, \true, \sexp_l=\sexp_l' \land \sexp_q = \sexp_q' \land \sexp_l' \notin \dom(\smem_f)))}
\and
\inferrule{\smem = \smem_f \uplus \{\sexp_l' \mapsto (\sexp_v, \sexp_q') \} \\\\
           \smem' = \smem_f\uplus \{\sexp_l' \mapsto (\sexp_v, \sexp_q' - \sexp_q)\}}
          {\resconsume(m, O, \mapsto,[\sexp_l, \sexp_q],\smem)
           \rightsquigarrow \\\\\ (ok, O, {[\sexp_v]},(\smem', \true, \sexp_l=\sexp_l' \land \sexp_q < \sexp_q'))}
\end{mathparpagebreakable}

\subsection{Block-offset Memory Model for C}%
\label{sec:mem-block-offset}

Our CSE theory is not limited to different variants of the linear memory model. To illustrate this, we have mechanised and proved OX and UX sound a block-offset memory model for C. Originally inspired by the memory model of the verified CompCert C compiler~\cite{Leroy09}, the model has previously been implemented in Gillian and has been used in Gillian-based teaching, but no detailed definition or soundness results have previously been given.

\begin{figure}[t]
\begin{minipage}[t]{0.40\textwidth}
\small
\centering
\[
\begin{array}{r c l}
    \\
    \vspace*{0.5em}
    \cmem &\in& \cmemss\\
    \cmem(n_1) &=& (\{0 \mapsto 3, 1 \mapsto 5\}, \Some(2))\\
    \cmem(n_2) &=& (\{1 \mapsto 2, 2 \mapsto \false\}, \None)\\
    \cmem(n_3) &=& \cfreed \\
\end{array}
\]
\end{minipage}
\hfill
\begin{minipage}[t]{0.36\textwidth}
\vspace{0pt}
\begin{tikzpicture}[scale=0.6, every node/.style={scale=0.8}]
  \draw (0,0) rectangle (1,3);

  \draw(2,2.5) rectangle (7.8,3.5);

  \draw[dashed] (2,1) rectangle (8.3,2);

  \draw (2,-0.5) rectangle (3,.5);

  \draw  (2.2,2.7) rectangle (4.2,3.3);
  \draw[dashed] (2.2,1.2) rectangle (5.2,1.8);
  \draw (3.2,1.2) rectangle (5.2,1.8);

  \draw (3.2, 2.7) -- (3.2, 3.3);
  \draw (4.2, 1.2) -- (4.2, 1.8);
  
  \foreach \y in {1,2} {
    \draw (0,\y cm) -- (1cm,\y cm);
  }

    \node (n3) at (0.5cm, 0.5cm) {$n_3$};
    \node (n2) at (0.5cm, 1.5cm) {$n_2$};
    \node (n1) at (0.5cm, 2.5cm) {$n_1$};
    
    \node (bound) at (6,3) {\small  $bound: \Some(2)$};
    \node (uthree) at (2.7,3) {\small $3$};
    \node (ufive) at (3.7,3) {\small $5$};

    \node (bound) at (6.7,1.5) {\small  $bound: \None$};
    \node (mthree) at (3.7,1.5) {\small $3$};
    \node (mfalse) at (4.7,1.5) {\footnotesize false};

    \node (bot) at (2.5,0) {$\varnothing$};

    \draw[-{Stealth[length=2mm]},shorten <=1mm]  (n1.east) -- (2,2.9);
     \draw[-{Stealth[length=2mm]},shorten <=1mm]  (n2.east) -- (2,1.5);
      \draw[-{Stealth[length=2mm]},shorten <=1mm]  (n3.east) -- (2,0);
\end{tikzpicture}
\end{minipage}
\hfill
\begin{minipage}[t]{0.22\textwidth}
\small
\[
\begin{array}{l}
(n_1, 0) \mapsto 3 \lstar \\ (n_1, 1) \mapsto 5 \lstar \\ \mathsf{Bound}(n_1; 2) \lstar \vspace*{0.5em}\\ (n_2, 1) \mapsto 3 \lstar\\ (n_2, 2) \mapsto \false \lstar \vspace{0.5em}\\ n_3 \mapsto \cfreed
\end{array}
\]
\vspace{-1em}
\end{minipage}
\caption{Example block-offset memory instance $\mu$ expressed formally (left), visually (centre), and as a composition of resource assertions (right).}
\label{fig:block_diagram}
\end{figure}
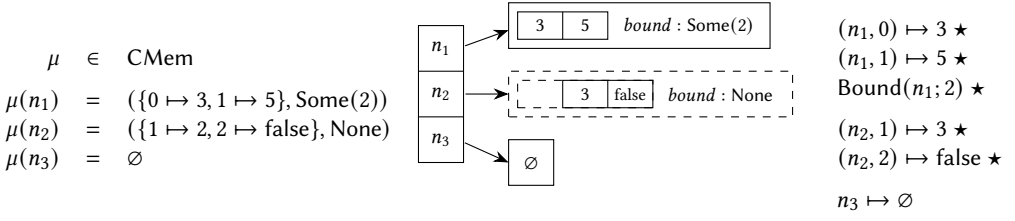

We describe, component by component, the concrete-memory instance data $(\cmemss, \cwf, \cmem_\emptyset, \cdot)$ for IDef.~\ref{idef:cmm}. The memory data type is as follows:
\[
    \cmemss \defeq \Nat \rightharpoonup_{\mathit{fin}} (\cmemss_\text{B} \uplus \{\cfreed\})\quad\text{where}\quad
        \cmemss_\text{B}\ \defeq (\Nat \rightharpoonup_{\mathit{fin}} \Val, \Nat?) 
\]
using notation  $t?$ to denote the option type for type $t$, with constructors $\None$ and $\Some$.
The concrete memory comprises two parts: $\cmemss$ is a mapping from block identifiers to {blocks};
$\cmemss_{\text{B}}$ is a {block} comprising a linear array and a bound indicating the fixed size of the array. (In C terms, the block identifiers are essentially pointers returned by \texttt{malloc()}, and the blocks describe the contents and size of the corresponding allocated memory.) This data structure allows us to represent \emph{partial blocks}, required to be able to define $\resmodels$, i.e., IDef.~\ref{idef:res-srel}, which we discuss shortly. 
Fig.~\ref{fig:block_diagram} gives an example of a partial concrete memory given by map $\cmem$ with domain of block identifiers $\{n_1, n_2, n_3\}$.
The mapping $\cmem(n_1)$ is a \emph{complete} block
as the bound is 2 and both cells are present in the block. 
The mapping $\cmem(n_2)$ is a \emph{partial} block due to both the missing bound and the missing map at offset~$0$. The mapping 
$\cmem(n_3)$ is a deallocated block, denoted by $\cfreed$. 

The well-formedness condition $\cwf$ provides constraints on the formation of blocks. Block-offset memories may not be well-formed for two reasons: first, a memory such as $\{ 3 \mapsto (\{ 0 \mapsto 1, 1 \mapsto 0 \}, \Some(1)) \}$ is not well-formed because its cells do not respect the bound; second, and more interestingly, a memory such as $\{ 3 \mapsto ( \emptyset, \None) \}$, which comprises an \emph{empty block} is not well-formed since such blocks break the frame properties of the model~(IProp.~\ref{iprop:cmm-frame}). We elaborate on this aspect more when we introduce concrete memory actions later (IDef.~\ref{idef:cmm-semantics}).

The empty memory is the empty mapping $\cmem_\emptyset = \emptyset$. Note that this trivially satisfies $\cwf$.

The definition of $\cdot$ is slightly complex. We exemplify using Fig.~\ref{fig:block_diagram}. Given another memory $ \cmem' = \{n_2 \mapsto (\{0 \mapsto 1, 3 \mapsto 0\}, \text{Some}(4))\}$, we have the composition
$\cmem \cdot \cmem'$ which is a mapping that gives the same results as $\cmem$ for block identifiers $n_1$ and $n_3$ and, for $n_2$,  gives:
\[
(\cmem \cdot \cmem')(n_2) =  (\{0 \mapsto 1, 1 \mapsto 3, 2 \mapsto \false, 3 \mapsto 0\}, \text{Some}(4))\}.
\]
We use the memory $\cmem'' = \{n_2 \mapsto (\{1 \mapsto 12\}, \text{Some}(2))\}$ to illustrate the two reasons why blocks may fail to compose. Indeed, the composition $\cmem \cdot \cmem''$ is not defined for two reasons: first, the addresses of the blocks at $n_2$ overlap (i.e., $\dom(\mathsf{fst}(\cmem(n_2))) \cap \dom(\mathsf{fst}(\cmem''(n_2))) \neq \emptyset$); second, the addresses of the block $\cmem(n_2)$ are not contained within the bound of the block $\cmem''(n_2)$, meaning that if the composition would be defined, then the result would not satisfy $\cwf$.

We now cover memory actions (IDef.~\ref{idef:cmm-semantics}).
The successful rules of $\mathtt{new}$ and $\mathtt{free}$ are given below:
\begin{mathpar}
\small
\infer{\cmem.\mathtt{new}([n]) \rightsquigarrow \oxok : (\cmem[n_b \mapsto \hp_b], [n_b])}
      {n_b \notin \dom(\cmem) \quad \hp_b = (\{ 0 \mapsto \nil, \dots, n - 1 \mapsto \nil \}, \Some(n))}
\quad
\infer{\cmem.\mathtt{free}([n_b]) \rightsquigarrow \oxok : (\cmem[ n_b \mapsto \cfreed ], [])}
      {\cmem(n_b) = (\hp_b, \Some(n)) \quad |\hp_b| = n}
\end{mathpar}
Because we are now working with blocks, allocation returns a fresh, complete block, and freeing a block deallocates an entire complete block given a block identifier. Load and store operations now also take in the offset as input in addition to the block identifier (and also value for store), and these operations only require components necessary to load or store, i.e., requiring the relevant partial block and not the complete block.

Going back to why empty blocks are not allowed by $\cwf$, note that UX frame breaks when allocation returns a block identifier $n_b$ pointing to a fresh, complete block, but the frame contains the same $n_b$ pointing to an empty block. OX frame instead breaks when you free a block identifier $n_b$ pointing to a complete block, but the frame contains the same $n_b$ pointing to~an~empty~block.

We now define the resource assertions and their satisfaction relation (IDef.~\ref{idef:res-srel}):
\[
\begin{array}{@{}l@{~}l@{~}c@{~\ }l}
 \hp \resmodels &
 (n_b, n_o) \mapsto v & \Leftrightarrow & \hp = \{n_b \mapsto (\{ n_o \mapsto v \}, \None)\} \\
 \hp \resmodels & \mathsf{Bound}(n_b; n) & \Leftrightarrow & \hp = \{n_b \mapsto (\emptyset,\Some(n)\} \\
 \hp \resmodels & n_b \mapsto \cfreed & \Leftrightarrow & \hp = \{n_b \mapsto \cfreed \}
 \end{array}
\]
The resource assertions consist of: the cell assertion, the bound assertion, and the freed cell assertion. Note that these resource assertions represent the smallest unit of memory from which to build larger memory using the separating conjunction. For example, the $\cmem$ memory model from Fig.~\ref{fig:block_diagram} can be represented by the assertion given on the right of the figure. Note that when we defined $\cmemss_\text{B}$ (for IDef.~\ref{idef:cmm}), we used $(\Nat \rightharpoonup_{\mathit{fin}} \Val, \Nat?)$ instead of the simpler $[\Val]$. This is to ensure we can define resource assertions for each unit of memory. If the concrete memory model used lists instead of finite maps, then defining $\resmodels$ would become impossible since the relation must define the entire memory in the relevant block.

The instance data $(\smemss, \smem_\emptyset)$ for IDef.~\ref{idef:smm} is as follows. The memory data type $\smemss$ is a simple symbolic lifting of $\cmemss$:
\[
\smemss \defeq \LExp \rightharpoonup_{\mathit{fin}} (\smemss_\text{B} \uplus \{ \cfreed \})\quad \text{where}\quad \smemss_\text{B} \defeq (\LExp \rightharpoonup_{\mathit{fin}} \LExp, \LExp?)
\]
The empty symbolic memory is, unsurprisingly, $\smem_\emptyset \defeq \emptyset$. Because the memory data type is a symbolic lifting, $\models_\text{Mem}$ (IDef.~\ref{idef:smm-srel}) is straightforward to define. The symbolic action semantics (IDef.~\ref{idef:smm-semantics}) is also symbolically lifted from its concrete counterpart and similarly straightforward. Lastly, the implementation of $\resconsume$ and $\resproduce$ (IDef.~\ref{idef:smm-cp}) are also straightforward.

\subsection{Memory Model for Object-oriented Languages}%
\label{sec:mem-oop}

Our CSE theory is not limited to low-level languages such as C, it is also compatible with high-level languages such as object-oriented languages like JavaScript and Python. In fact, the Gillian project that inspired our theory has a strong history of JavaScript support, starting from its predecessor JaVerT~\cite{javert,javert2} (an analysis tool specific to JavaScript). For example, Gillian has been used to test the data structure library Buckets.js~\cite{gillianpldi,buckets} and to verify a JavaScript implementation of a message header deserialisation module in the~AWS~Encryption~SDK~\cite{gilliancav}.

We now briefly describe the JavaScript memory model implemented in Gillian to show that it fits our theory. The model is a variant of the block-offset memory model introduced in the previous section; because of the large overlap between the models, we have not mechanised~this~JavaScript~model.

\paragraph{Model description} The concrete and symbolic memory data types of the JavaScript memory model implemented in Gillian are as follows (i.e., the data types of IDefs.~\ref{idef:cmm}~and~\ref{idef:smm}):\footnote{The memory model in Gillian additionally includes object metadata, which we do not discuss here.}
\[
\begin{array}{l@{~}l}
\cmemss &\defeq \nats \rightharpoonup_{\mathit{fin}} (\cmemss_\text{B}, \{ \strings \}?) \quad \text{where} \quad \cmemss_\text{B} \defeq \strings \rightharpoonup_{\mathit{fin}} (\Val \uplus \{ \cfreed \}) \\
\smemss &\defeq \LExp \rightharpoonup_{\mathit{fin}} (\smemss_\text{B}, \{ \LExp \}?) \quad \text{where} \quad \smemss_\text{B} \defeq \LExp \rightharpoonup_{\mathit{fin}} (\LExp \uplus \{ \cfreed \})
\end{array}
\]
where $\{ \strings \}$ and $\{ \LExp \}$ denote sets of $\strings$s and $\LExp$s, respectively, and $\cmemss_\text{B}$ and $\smemss_\text{B}$ represent JavaScript objects. The reader should compare these memory data types with the memory data types of the block-offset memory model and note the following differences. First, note that offsets (natural numbers) here have been replaced by property names (strings). Because of this, the bound from the block-offset memory model has been replaced by a set of strings, representing the ``domain'' of the object.\footnote{This change also leads to a slightly different well-formedness condition. Informally, a memory $\cmem \in \cmemss$ is only well-formed if $\forall (o, \mathsf{Just}(d)) \in \codom(\cmem). \dom(o) \subseteq d$ (see heap-domain invariant~\cite{Naudziuniene2018Infrastructure}). The rest of the well-formedness condition is similar to the block-offset memory model.} Second, in JavaScript, objects cannot be deallocated, but $\cfreed$-annotations for negative information are needed for a different reason. In JavaScript, reading fields that \emph{have not been set} or \emph{have been deleted} evaluates to \texttt{undefined} (see \ifarxiv App.~\ref{app:memory-models}\else the extended paper~\cite{arxiv}\fi{} for an example JavaScript REPL session where properties are added and deleted). To not break the frame properties of the model, such ``unset'' properties must be annotated with $\cfreed$.

The memory model has the following memory actions (IDef.~\ref{idef:cmm-semantics}): 
$$\passign{\pvar x}{\mathtt{newObj}()}, \mathtt{deleteField}(\pexp, \pexp), \passign{\pvar x}{\mathtt{lookup}(\pexp, \pexp)}, \mathtt{mutate}(\pexp, \pexp, \pexp).$$
E.g., $\mathtt{deleteField}(o, f)$ deletes field $f$ from the object at address $o$ and $\mathtt{mutate}(o, f, v)$ sets field $f$ in the object at address $o$ to $v$.
While the semantics of JavaScript is complex, this simple memory model is enough to capture its basic operations: in the JavaScript instantiation of Gillian, JavaScript programs are compiled to GIL, its intermediate representation, where complex operations (such as looking up an object field by following the ``prototype chain'') are compiled to a sequence of lower-level operations that are either side-effect free, or one of the actions provided above.

Lastly, the semantics of the actions of the memory model are obtained by applying minor modifications to the actions of the block-offset memory model. For instance, out-of-bound accesses happen when a memory lookup is outside the object domain instead of when an offset is outside the bound of the block, and so on.

\subsection{CHERI-assembly Memory Model}%
\label{sec:mem-cheri}

To show that our CSE theory can fit novel memory models beyond the usual suspects, we have designed and mechanised a new symbolic memory model for a CHERI-enabled idealised assembly language. CHERI~\cite{woodruff:2014:cheri} is a recently introduced memory-model-based capability protection model: it guarantees runtime spatial memory safety via hardware, and this is achieved using \textit{capabilities}: fat pointers that carry spatial metadata such as bound, permission, and a tag bit stating the validity of the capability, in addition to the memory address. Additional capability-aware instructions are added to the instruction set, where the \textit{monotonic property} is preserved: valid capabilities cannot gain more bounds or permissions than what they originally had.

We have proved our memory model OX sound, and plan to prove UX soundness and implement the memory model in Gillian in future work. To our best knowledge, our memory model is the first memory model for CHERI that supports SL-based symbolic execution and also comes with a soundness theorem. Details about related work are given in \S\ref{sec:related-work}.

We first discuss the definition of our new memory model. The CHERI-assembly model is the most substantial instantiation in this work: the CHERI-assembly model has roughly 19K lines of code, about 5 times larger than the block-offset model, the second most substantial instantiation. Afterwards, we discuss our design process. The design process is interesting because the memory model is the first memory model we have designed for our CSE theory without the guidance of an existing implementation. We explain how our CSE theory guided us to obtain the appropriate design: we made two failed design attempts before finally arriving at our current design.

\begin{figure}
\begin{minipage}[t]{0.45\textwidth}
\small
\centering
\[
\begin{array}{r c l}
    \mathsf{Cap} &\defeq& \{blo : \Nat;\ \mathit{off} : \Nat;\ base : \Nat;\ \\
    & & \ len : \Nat; \vec{perm}_{\mathsf{x}} : \vec{\bools};\ tag : \bools\}\\
    & & \quad \text{where } x \in \{\mathsf{load, store, ...}\} \\
    \mathsf{Cap}_{\mathsf{frag}} & \defeq & \{cap : \mathsf{Cap}, nth : \nats\} \\
\end{array}
\]
\end{minipage}
\hfill
\begin{minipage}[t]{0.53\textwidth}
\small
\centering
\[
\begin{array}{r c l}
    \cmemss_\text{B-CH} &\defeq& (\Nat \rightharpoonup_{fin} \Val + \mathsf{Cap}_{\mathsf{frag}}, \Nat?) \\
    \cmemss_\text{BO-CH}\ &\defeq& \Nat \rightharpoonup_{\mathit{fin}} (\cmemss_\text{B-CH} \uplus \cfreed)\\
    \cmemss_\text{CReg} &\defeq& \Nat \rightharpoonup_{\mathit{fin}} \mathsf{Cap} \\
    \cmemss &\defeq& (\cmemss_\text{CReg}, \cmemss_\text{BO-CH})
\end{array}
\]
\end{minipage}
%
\caption{The concrete memory data type.}%
\label{fig:cheri_cmm}
\begin{minipage}[t]{0.40\textwidth}
\footnotesize
\centering
\[
\begin{array}{r c l}
    \cmem &=& (\cmem_\text{R}, \cmem_{\text{BO}}) \\
    \vspace*{0.5em}
    (\cmem_\text{R}, \cmem_{\text{BO}}) &\in& \cmemss\\
    
    \cmem_{\text{BO}}(n_1) &=& (\{0 \mapsto 3, 1 \mapsto c_1\}, \Some(2))\\
    \cmem_{\text{BO}}(n_2) &=& (\{i \mapsto c_i\}, \Some(|\mathsf{Cap}|)\\
    & & \ \text{for }i \in \{0, 1, ..., |\mathsf{Cap} - 1|\}\\
    \vspace*{0.5cm}
    \cmem_\text{R}(r_1) &=& c\\
\end{array}
\]
\end{minipage}
\hfill
\begin{minipage}[t]{0.40\textwidth}
\vspace{0pt}
\begin{tikzpicture}[scale=0.6, every node/.style={scale=0.8}]
  \draw (0,1.25) rectangle (1, 3.25);
  \draw (0, -0.5) rectangle (1, 0.5);

  \draw(2,2.5) rectangle (6.7,3.5);

  \draw (2,1) rectangle (9.1,2);

  \draw (2,-0.5) rectangle (3,.5);

  \draw (2.2,2.7) rectangle (3.4,3.3);
  \draw (2.2,1.2) rectangle (5.1,1.8);

  \draw (2.8, 2.7) -- (2.8, 3.3);
  \draw (2.8, 1.2) -- (2.8, 1.8);
  \draw (3.4, 1.2) -- (3.4, 1.8);

  \draw[dashed] (-0.1, 0.75) -- (9.2, 0.75);
  
  \foreach \y in {2.25} {
    \draw (0,\y cm) -- (1cm,\y cm);
  }

    \node (r1) at (0.5cm, 0cm) {$r_1$};
    \node (n2) at (0.5cm, 1.75cm) {$n_2$};
    \node (n1) at (0.5cm, 2.75cm) {$n_1$};
    
    \node (bound) at (5.1,3) {\small  $bound: \Some(2)$};
    \node (uthree) at (2.5,3) {\small $3$};
    \node (ufive) at (3.1,3) {\small $c_1$};

    \node (bound) at (7.1,1.5) {\small  $bound: \Some(|\mathsf{Cap}|)$};
    \node (mone) at (2.5,1.5) {\small $c_0$};
    \node (mtwo) at (3.1,1.5) {\small $...$};
    \node (mfalse) at (4.3,1.45) {\small $c_{|\mathsf{Cap} - 1|}$};

    \node (bot) at (2.5,0) {$c$};

    \draw[-{Stealth[length=2mm]},shorten <=1mm]  (n1.east) -- (2,2.9);
     \draw[-{Stealth[length=2mm]},shorten <=1mm]  (n2.east) -- (2,1.5);
      \draw[-{Stealth[length=2mm]},shorten <=1mm]  (r1.east) -- (2,0);
\end{tikzpicture}
\end{minipage}
\hfill
\begin{minipage}[t]{0.18\textwidth}
\footnotesize
\[
\begin{array}{l}
(n_1, 0) \mapsto 3 \lstar\\ (n_1, 1) \mapsto_{\mathit{cf}} c_1 \lstar\\
\mathsf{Bound}(n_1; 2) \lstar \vspace{0.5em}\\ (n_2, 0) \mapsto_{\text{cap}} c \lstar \\ \mathsf{Bound}(n_2, |\mathsf{Cap}|) \lstar \vspace*{0.5em}\\
\mathsf{Reg}(r_1; c)
\end{array}
\]
\end{minipage}
\vspace{-1.5em}
\caption{Example CHERI memory instance $\mu$ expressed formally (left), visually (centre), and as a composition of resource assertions (right).}%
\label{fig:cheri_diagram}
\end{figure}

\paragraph{Model description} 

Our CHERI memory model extends the block-offset memory model of~\S\ref{sec:mem-block-offset} to be capability-aware. While the bit-level layout of capabilities may differ between architectures, even if the metadata is mostly similar; in this work, we work with a CHERI-assembly model with an abstract and architecture-agnostic design.

We now give the instance data $(\cmemss, \cwf, \mu_\emptyset, \cdot)$ conforming to IDef.~\ref{idef:cmm}; we first discuss $\cmemss$. Fig.~\ref{fig:cheri_cmm} shows the structure of the CHERI-assembly concrete memory model, and Fig.~\ref{fig:cheri_diagram} gives as an instance example $\cmem$. There are two main differences with the block-offset memory model: a separate mapping for capability registers (i.e. $\cmemss_{\text{CReg}}$) is added, and the main memory (i.e. $\cmemss_{\text{BO-CH}}$) is extended to be capability-aware. For capability registers, we use the abstract capability $\mathsf{Cap}$, which contains spatial metadata of capabilities. For the main memory $\cmemss_{\text{BO-CH}}$, we extend the block-offset model by also storing \emph{capability byte fragments}, represented as $\mathsf{Cap_{frag}}$, in addition to standard values $\vals$. In Fig.~\ref{fig:cheri_diagram}, we observe $\cmem_{\text{BO}}(n_1)$ contains the capability byte fragment $c_1$, which is the second byte fragment of some capability $c$. When a capability is stored in memory, the capability is stored as a sequence of abstract, contiguous, well-formed capability byte fragments -- and we can observe this in $\cmem_\text{BO}(n_2)$, where $|\mathsf{Cap}|$ is the size of a capability for a given architecture. We note that each capability byte fragment also stores a \emph{tag fragment}, and a capability in memory only has its tag bit set to true if and only if the tag fragment of all the capability fragments is set to $\true$. Usually, CHERI architectures store tags in the \emph{tagged memory}, separate from the main memory; in our model the two memories are merged, and tags are split into tag fragments -- the motivation behind this is explained when we discuss the design process below.

The well-formedness condition $\cwf$ also extends from that of the block-offset memory model. The additional constraints relate to the well-formedness of capability fragments in the memory. One obvious property is that the fragment value of $\mathsf{Cap_{frag}}$ should be between 0 and $|\mathsf{Cap} - 1|$. Another property is capability fragments whose tag fragment bit is set to true must be stored in the appropriate capability-offset-aligned position (the formal description is in \ifarxiv App.~\ref{app:memory-models}\else the extended paper~\cite{arxiv}\fi). This conforms to the specification that valid capabilities in memory are stored in a capability-aligned position~\cite{watson:2023:cheri}. All the capability fragments in $\cmem_{\text{BO}}(n_1)$ and $\cmem_{\text{BO}}(n_2)$ are stored in a capability-off-aligned position, which makes the overall memory well-formed; but note that if we instead have $\cmem_{\text{BO}}(n_1)(1) = c_4$, then the memory is well-formed only if $c_4.cap.tag = \false$. As we will see below, $\resmodels$ must account for this too, and part of $\cwf$ is expressed in $\resmodels$.  

The empty memory $\cmem_\emptyset$ can be straightforwardly defined as $(\emptyset, \emptyset)$ (which satisfies $\cwf$), and $\cdot$ is also a straightforward extension of that of the block-offset memory model.

\begin{wrapfigure}{r}{0.46\textwidth}
\centering
\begin{mathparpagebreakable}
\small
\inferrule 
{
\mathsf{fst}(\cmem)(r_s) = c_s \and c_s.tag = \mathtt{true} \and \\\\ c_s.perm_\mathsf{store} = \mathtt{true} \\\\
r_d \text{ is a capability register} \and \mathsf{fst}(\cmem)(r_d) = c_d \\\\
(c_s.perm_\mathsf{storecap} = \mathtt{true} \lor c_d.tag = \mathtt{false}) \\\\
c_s.\textit{off} + |\mathsf{Cap}| \leq c_s.base + c_s.len \\\\
c_s.\textit{off} \geq c_s.base \and c_s.\textit{off}\ \%\ |\mathsf{Cap}| = 0 \\\\
\mathsf{snd}(\mu)(c_s.blo) = (\mu_b, \Some(m)) \\\\
c_s.\textit{off} + |\mathsf{Cap}| \leq m \\\\
\{c_s.\textit{off}, ..., c_s.\textit{off} + |\mathsf{Cap}| - 1\} \subseteq \mathsf{dom}(\mu_b) \\\\
\mathsf{store\_capability}(c_s, c_d, \mu_b) = \mu_b' \\\\
\mu' = (\mathsf{fst}(\mu), \mathsf{snd}(\mu)[c_s.blo \mapsto (\mu_b', \Some(m))])
}
{\cmem.\mathtt{store}([r_s, r_d]) \rightsquigarrow \oxok : (\mu', [])}

\end{mathparpagebreakable}
\caption{Semantics of the capability store action.}%
\label{fig:storecap}
\vspace{-1em}
\end{wrapfigure}

We now discuss memory actions (IDef.~\ref{idef:cmm-semantics}). There are more than 100 memory-action rules, with relatively complex definitions. To exemplify, we discuss the successful case of the capability store action, shown in Fig.~\ref{fig:storecap}. 
The capability store action takes in $r_s$ and $r_d$, which are capability register numbers pointing to capability registers $c_s$ and $c_d$, respectively. The idea is that we store $c_d$ in the location pointed by $c_s$. The action then performs necessary spatial checks and throws relevant errors when a spatial safety property is violated, e.g. $c_s$ must have the tag bit set to true, and the offset of $c_s$ must be within bound and is capability-offset-aligned, etc. Afterwards, the $\mathsf{store\_capability}$ function stores $c_d$ as a sequence of well-formed contiguous capability byte fragments in the main memory.

We now discuss the resources of this model and their satisfaction relation $\resmodels$ (IDef.~\ref{idef:res-srel}). The three resources used in the block-offset memory model are directly ported. Additionally, we have two new resources: $\mathsf{Reg}(r_n; c)$, which describes that at register $r_n$ the capability $c$ is stored, and $(n_b, n_o) \mapsto_{cf} c_{n}$, which states that at block $n_b$ and offset $n_o$, the capability byte fragment $c_n$ is stored, where $n$ denotes the $n$th byte fragment. The resource satisfaction relation $\resmodels$~is~given~below:
\[
\begin{array}{@{}l@{~}l@{~}c@{~\ }l@{~\ }l}
  \hp \resmodels &
 \mathsf{Reg}(r_n; c) & \Leftrightarrow & \hp = (\{r_n \mapsto c\}, \emptyset) & \\
 \hp \resmodels & (n_b, n_o) \mapsto_{\mathit{cf}} c_n & \Leftrightarrow & \hp = (\emptyset, \{n_b \mapsto (\{n_o \mapsto c_n\}, \None)\}) \\
 & & & \land\ (c_n.cap.tag = \true \Longrightarrow n_o\ \%\ |\mathsf{Cap}| = n) \land\ n < |\mathsf{Cap}| \\ 
 \end{array} 
\]
Whereas defining $\resmodels$ for the block-offset memory was straightforward, defining $\resmodels$ here is slightly more involved. Due to $\cwf$ of CHERI, we cannot allow capability byte fragments whose tag fragment bit is true to be stored anywhere, and we require the fragment value to be valid. The assertion $(n_1, 1) \mapsto_{cf} c_1$ in Fig.~\ref{fig:cheri_diagram} is satisfiable, but $(n_1, 1) \mapsto_{cf} c_4$ is not if $c_4.cap.tag = \true$.

Note that one can define a full, valid capability resource assertion as a user-defined predicate assertion $(n_b, n_o) \mapsto_{\mathsf{cap}} c$ as follows:
\[
\begin{array}{@{}l@{~}l@{~}l}
    (n_b, n_o) \mapsto_{\mathsf{cap}} c & \defeq & \ilstar_{i=0}^{|\mathsf{Cap}| - 1}\ (n_b, n_o + i) \mapsto_{\mathit{cf}} c_i
\end{array}
\]

In Fig.~\ref{fig:cheri_diagram}, we can see the capability register mapping is represented using the register assertion, the capability fragment in $\cmem_{\text{BO}}(n_1)$ is represented using the capability fragment assertion, and the full capability in $\cmem_{\text{BO}}(n_2)$ is represented using the user-defined capability predicate assertion.

The symbolic memory (IDef.~\ref{idef:smm}) is a direct lifting of its concrete counterpart. The satisfaction relation $\memmodels$ (IDef.~\ref{idef:smm-srel}) extends that of the block-offset model by additionally relating symbolic capability registers to concrete ones. The symbolic memory actions (IDef.~\ref{idef:smm-semantics}) are a symbolic lifting of the concrete actions. The implementations of $\resconsume$ and $\resproduce$ (IDef.~\ref{idef:smm-cp}) extend those of the block-offset model: there is now an additional case when consuming or producing capability fragments depending on the tag fragment value of the capability fragment due to $\cwf$.

\paragraph{Design process}

Our first attempt at designing the memory model was based on that of~\citet{park:2023:gilliancheric}. That work separated the main memory into two: the (tagless) main memory, and the \emph{tagged} memory, where the tagged bit of a capability was stored in the tagged memory. While this model closely represented the CHERI hardware, the separation made it difficult to reason about the complex inter-dependency between the two memories, making it difficult to formalise resource assertions well-formed with respect to the concrete memory and also prove the OX frame property.

In our second attempt, to address the aforementioned issue, we introduced the notion of a ``chunk'' of memory, where a chunk is either a capability or a sequence of values and capability fragments of size $|\mathsf{Cap}|$. We removed the tagged memory and made capability tags implicitly defined depending on whether the chunk was a capability or not. Because this model no longer separates the main memory, there are no inter-dependencies between memories and no troubles proving the OX frame property or formalising well-formed resource assertions. However, we discovered writing function specification had limitations, e.g. when the precondition requires a capability fragment, but the memory comprises the full capability instead, which made the model not truly compositional.

Our third and final attempt introduced the notion of tag fragments in capability fragments. This ensured true compositionality, unlike the previous attempt, where there are no limitations on how to write specifications, whilst avoiding complex inter-dependencies between memories.

The structure of the concrete memory model naturally guided us to define the current resource assertions and their satisfaction relation. Indeed, this formalisation gave us confidence that our parametric CSE theory is well designed: while this work was done independently from the recently published Iris-MSWasm work~\cite{legoupil:2024:irismswasm}, we ended up with resource assertions essentially similar to those used in the Iris-MSWasm work.

\subsection{VeriFast-and-Viper-inspired Memory Model for C}%
\label{sec:mem-verifast}

We have mechanised a memory model for C inspired by the OX verification platforms VeriFast and Viper. Specifically, we have ported the memory model of Featherweight VeriFast~\cite{Jacobs15} (FVF), a formalisation of (a simplified version of) VeriFast, to our CSE theory and proved it OX sound. The motivation for this work is as follows. Beyond Gillian, VeriFast and Viper are the most well-known consume-and-produce-based CSE platforms. VeriFast and Viper have similar memory models: they both maintain a flat collection of ``heap chunks'' (explained below). The memory model we discuss in this section shows that our CSE theory can fit such memory models.

\paragraph{Model description} FVF analyses a simplified C language with the same memory actions as our running example memory model. In FVF, the concrete memory model (IDef.~\ref{idef:cmm}) is a multiset of concrete heap chunks. A concrete heap chunk is either a {\em points-to chunk} $l \mapsto v$ denoting that there is an allocated memory cell at address $l$ whose current value is $v$, or a {\em malloc-block chunk} $mb(l, \textit{size})$ denoting that a memory block of size $\textit{size}$ was allocated at address $l$ by {\tt malloc}, i.e., that the memory cells at addresses $l$ through $l+\textit{size}-1$ are part of a single block, which will be freed as one unit when {\tt free} is called with argument $l$. The memory composition operator is multiset union, and all heap chunks are disjoint. In our instantiation, we represent concrete heap~chunks~as~follows:
\[
\mathsf{cchunk} \defeq \mathsf{CCPointsTo}(\nats, \vals) \mid \mathsf{CCMB}(\nats, \nats)
\]

The symbolic memory model (IDef.~\ref{idef:smm}) used in FVF is a multiset of symbolic heap chunks. A symbolic heap chunk is either a points-to chunk, a malloc-block chunk, or a user-defined-predicate chunk. Since user-defined predicates are handled independently of the memory model in our CSE theory, the symbolic chunks of our memory-model instance are as follows:
\[
\mathsf{schunk} \defeq \mathsf{SCPointsTo}(\lexps, \lexps) \mid \mathsf{SCMB}(\lexps, \lexps)
\]

The memory actions of the concrete and symbolic memory models (IDefs.~\ref{idef:cmm-semantics} and~\ref{idef:smm-semantics} respectively) and $\resconsume$ and $\resproduce$ operations of the symbolic memory model (IDef.~\ref{idef:smm-cp}) are straightforward. In FVF, the semantics of the memory actions are simply defined in terms of the consume and produce operations (concrete consume and produce operations are defined for the concrete memory model) -- we therefore do not include any memory actions in our instance. The $\resconsume$ and $\resproduce$ operations (IDef.~\ref{idef:smm-cp}) of FVF implement unique-match branching (as discussed in \S\ref{sec:mem-linear}) -- our ported implementations therefore~do~the~same.

\section{Related Work}%
\label{sec:related-work}

\paragraph{Program logics} Multiple program logics -- such as abstract separation logic~\cite{calcagno:lics:2007}, views~\cite{dyoung:popl:2013,cisl}, and Iris~\cite{jung:jfunc:2018} -- are parametric on different PCM-like structures describing memory state and ghost state. Some of these program logics also feature other types of parametricity, such as programming-language parametricity. In contrast to CSE, program logics only describe sound inferences rather than a way to automate reasoning. For memory-model parametricity, the parameters we introduce in this paper show what is sufficient to animate reasoning and ensuring soundness of this animation.

\paragraph{Compositional symbolic execution} Since we have already discussed the previous work on foundations of memory-model-parametric CSE~\cite{gillianpldi,gilliancav} in the introduction and overview of this paper, we only discuss memory-model-monomorphic foundations here.

\citet{cse1} is the only previous work on CSE theory that treats both OX and UX soundness. The work is inspired by Gillian but monomorphised to the memory model we use as a running example in this paper. The work is similar to ours in scope in terms of engine features covered (function calls, user-defined predicates, etc.). Although the work is not mechanised, it is the monomorphic work that has influenced us the most; in particular, our consume and produce properties are inspired by the consume and produce properties they introduce, which they like us use to ensure interoperability of CSE analysis results with program logics and analysis tools built on top of program logics.

OX-only CSE is the most well-explored variant of CSE. We list the most significant projects in chronological order: \citet{Appel11} mechanises a subset of Smallfoot; \citet{Jacobs15} mechanise a subset of VeriFast; \citet{Keuchel22} argue for the use of Kripke specification monads in mechanising CSE and illustrate their techniques on small CSE case studies; \citet{Zimmerman24} formalise on paper a subset of Viper as part of larger work to enable sound gradual verification in Viper; \citet{Dardinier25} mechanise a soundness framework for translational verifiers, including CSE inspired by Viper. These projects have either smaller or similar coverage of engine features compared to us. Most closely related is the work by \citet{Dardinier25}, which like us ensure interoperability of CSE analysis results. Whereas our approach to interoperability forms a semantic connection to program logics, through the satisfaction relation for assertions $\subst, (\sto, \hp) \models A$, Dardinier et al. instead connect up syntactically by proof reconstruction. Larger case studies, for both approaches, are needed to better evaluate the trade-offs between~the~two~approaches.

Lastly, the Infer-Pulse work~\cite{Le22} treats only UX soundness for bi-abduction and is not mechanised. Since their engine is specialised to bi-abduction, their coverage of engine features is small.

\paragraph{Compilation to intermediate verification language} An alternative to symbolic execution is compilation to intermediate verification languages (IVLs) such as Boogie~\cite{Barnett06} and Why3~\cite{why3}, which turns the problem of automating reasoning into a compilation problem. We know of no such work addressing memory-model parametricity. Other IVL topics have received formal treatments: e.g., \citet{Parthasarathy24} mechanise proof-producing compilation to IVLs and targets Boogie in one case study, and \citet{Cohen24} mechanise the satisfaction relation of the logic fragment of Why3 and verify two compilation transformations inspired by Why3.

\paragraph{CHERI memory models} The most closely related previous work on CHERI have targeted CHERI-C, which extends the C language to support CHERI capabilities. There exist mechanised CHERI-C memory models formalised in Isabelle/HOL~\cite{park:2023:gilliancheric,park:2022:cheriafp} and Rocq~\cite{zaliva:2024:cheric,zaliva:2025:cheric,legoupil:2024:irismswasm}. The work of~\citet{park:2023:gilliancheric} provides an extractable CHERI-C model usable for concrete execution in Gillian, and the work of~\citet{legoupil:2024:irismswasm} extends the Iris-Wasm work~\cite{rao:2023:iriswasm} to incorporate \emph{handles}, a synonym for capabilities, and introduces resource assertions for handles. None of these works, however, cover symbolic execution. ESBMC-CHERI~\cite{brau_sse:2022:esbmccheri} is the only tool that supports symbolic execution of CHERI-C programs; however, the tool lacks a formal memory model and soundness proof and does not support compositional reasoning.

\paragraph{Additional interesting memory models} Gillian has two more memory-model instances which we have not discussed in this work. First, there is an optimised ``block-of-trees'' memory model for C which has been used in C verification case studies~\cite{gilliancav} (but not described in detail in previous publications). We will instantiate our CSE theory with this model in the future. Second, in work parallel to ours, \citet{Ayoun2025} have instantiated Gillian to Rust. Gillian's Rust memory model comprises several components, including a core heap model that extends the block-of-trees model for C with support for polymorphism and unknown layouts required by Rust. The memory model also includes ghost state for lifetime and prophecy reasoning. The model should be expressible using our theory; with the minor exception of the model's novel automation for reasoning about mutable borrows, for which a small generalisation of how Gillian handles user-defined predicates was required. The handling of user-defined predicates in our theory can easily be generalised by moving it from the memory-model-independent part of our theory into each memory-model instance. A bigger obstacle to overcome is the fact that the soundness justification of the memory model relies on results from RustBelt~\cite{Jung17} and RustHornBelt~\cite{Yusuke22}, requiring a formal connection between our theory and Iris to leverage~these~results.

Other interesting targets for future memory instantiations include well-validated formal memory models from, e.g., the Cerberus project~\cite{Memarian19,Lepigre22} or the WebAssembly project~\cite{wasm}.


\section{Conclusion}

In this paper, we have introduced a formal foundation for memory-model-parametric CSE platforms for verification and/or bug-finding. Multiple research groups have in recent years turned their attention to formally defining and proving sound CSE tools and platforms; despite this flurry of activity, the analysis platform Gillian is today the only CSE platform that supports memory-model parametricity. We hope this paper will inspire and help other CSE projects to also implement memory-model parametricity. We have also discussed a series of memory-model instantiations of our CSE theory, some based on or inspired by instantiations developed for Gillian.

Looking forward, now having in place a formal definition of memory model for CSE, in particular, sufficient memory-model requirements for memory-model instantiations of our CSE engine to be sound, we are now in the process of developing a combinator library for memory models, as defined in this paper, to make it easy to develop and prove sound large and complex memory models by composing smaller memory-model components together.

\section*{Data-availability Statement}

The Rocq mechanisation of our CSE theory and its instantiations are available in our artefact~\cite{artefact}.

\begin{acks}
We thank the anonymous reviewers for their constructive feedback. We also thank Dawid Lachowicz for careful comments on earlier drafts of this paper.

This work was supported by EPSRC Fellowship ``VetSpec: Verified Trustworthy Software Specification'' (EP/R034567/1) (Lööw, Nantes-Sobrinho, and Gardner), Gardner's faculty gift from Meta (Lööw and Nantes-Sobrinho), the Amazon Research Award ``Gillian-Rust: Unbounded Verification for Unsafe Rust Code'' (Ayoun), and an Imperial College London Department of Computing PhD Scholarship (Park).
\end{acks}

\bibliography{cse2.bib}


\begin{thebibliography}{58}


\ifx \showCODEN    \undefined \def \showCODEN     #1{\unskip}     \fi
\ifx \showISBNx    \undefined \def \showISBNx     #1{\unskip}     \fi
\ifx \showISBNxiii \undefined \def \showISBNxiii  #1{\unskip}     \fi
\ifx \showISSN     \undefined \def \showISSN      #1{\unskip}     \fi
\ifx \showLCCN     \undefined \def \showLCCN      #1{\unskip}     \fi
\ifx \shownote     \undefined \def \shownote      #1{#1}          \fi
\ifx \showarticletitle \undefined \def \showarticletitle #1{#1}   \fi
\ifx \showURL      \undefined \def \showURL       {\relax}        \fi
\providecommand\bibfield[2]{#2}
\providecommand\bibinfo[2]{#2}
\providecommand\natexlab[1]{#1}
\providecommand\showeprint[2][]{arXiv:#2}

\bibitem[Appel(2011)]%
        {Appel11}
\bibfield{author}{\bibinfo{person}{Andrew~W. Appel}.}
  \bibinfo{year}{2011}\natexlab{}.
\newblock \showarticletitle{{VeriSmall}: Verified Smallfoot Shape Analysis}.
\newblock In \bibinfo{booktitle}{\emph{Certified Programs and Proofs}}.
\newblock
\href{https://doi.org/10.1007/978-3-642-25379-9_18}{doi:\nolinkurl{10.1007/978-3-642-25379-9_18}}


\bibitem[Ayoun et~al\mbox{.}(2025)]%
        {Ayoun2025}
\bibfield{author}{\bibinfo{person}{Sacha-Élie Ayoun}, \bibinfo{person}{Xavier
  Denis}, \bibinfo{person}{Petar Maksimović}, {and} \bibinfo{person}{Philippa
  Gardner}.} \bibinfo{year}{2025}\natexlab{}.
\newblock \showarticletitle{A Hybrid Approach to Semi-automated Rust
  Verification}.
\newblock \bibinfo{journal}{\emph{Proceedings of the ACM on Programming
  Languages}} \bibinfo{volume}{9}, \bibinfo{number}{PLDI}
  (\bibinfo{year}{2025}).
\newblock
\href{https://doi.org/10.1145/3729289}{doi:\nolinkurl{10.1145/3729289}}


\bibitem[Baldoni et~al\mbox{.}(2018)]%
        {symb:exec:survey}
\bibfield{author}{\bibinfo{person}{Roberto Baldoni}, \bibinfo{person}{Emilio
  Coppa}, \bibinfo{person}{Daniele~Cono D’elia}, \bibinfo{person}{Camil
  Demetrescu}, {and} \bibinfo{person}{Irene Finocchi}.}
  \bibinfo{year}{2018}\natexlab{}.
\newblock \showarticletitle{A Survey of Symbolic Execution Techniques}.
\newblock \bibinfo{journal}{\emph{Comput. Surveys}} \bibinfo{volume}{51},
  \bibinfo{number}{3} (\bibinfo{year}{2018}).
\newblock
\href{https://doi.org/10.1145/3182657}{doi:\nolinkurl{10.1145/3182657}}


\bibitem[Barnett et~al\mbox{.}(2006)]%
        {Barnett06}
\bibfield{author}{\bibinfo{person}{Mike Barnett}, \bibinfo{person}{Bor-Yuh~Evan
  Chang}, \bibinfo{person}{Robert DeLine}, \bibinfo{person}{Bart Jacobs}, {and}
  \bibinfo{person}{K.~Rustan~M. Leino}.} \bibinfo{year}{2006}\natexlab{}.
\newblock \showarticletitle{Boogie: A Modular Reusable Verifier for
  Object-Oriented Programs}. In \bibinfo{booktitle}{\emph{Formal Methods for
  Components and Objects}}.
\newblock
\href{https://doi.org/10.1007/11804192_17}{doi:\nolinkurl{10.1007/11804192_17}}


\bibitem[Bornat et~al\mbox{.}(2005)]%
        {Bornat05}
\bibfield{author}{\bibinfo{person}{Richard Bornat}, \bibinfo{person}{Cristiano
  Calcagno}, \bibinfo{person}{Peter O'Hearn}, {and} \bibinfo{person}{Matthew
  Parkinson}.} \bibinfo{year}{2005}\natexlab{}.
\newblock \showarticletitle{Permission accounting in separation logic}. In
  \bibinfo{booktitle}{\emph{Symposium on Principles of Programming Languages}}.
\newblock
\href{https://doi.org/10.1145/1040305.1040327}{doi:\nolinkurl{10.1145/1040305.1040327}}


\bibitem[Boyland(2003)]%
        {Boyland03}
\bibfield{author}{\bibinfo{person}{John Boyland}.}
  \bibinfo{year}{2003}\natexlab{}.
\newblock \showarticletitle{Checking Interference with Fractional Permissions}.
  In \bibinfo{booktitle}{\emph{Static Analysis}}.
\newblock
\href{https://doi.org/10.1007/3-540-44898-5_4}{doi:\nolinkurl{10.1007/3-540-44898-5_4}}


\bibitem[Brau\ss{}e et~al\mbox{.}(2022)]%
        {brau_sse:2022:esbmccheri}
\bibfield{author}{\bibinfo{person}{Franz Brau\ss{}e}, \bibinfo{person}{Fedor
  Shmarov}, \bibinfo{person}{Rafael Menezes}, \bibinfo{person}{Mikhail~R.
  Gadelha}, \bibinfo{person}{Konstantin Korovin}, \bibinfo{person}{Giles
  Reger}, {and} \bibinfo{person}{Lucas~C. Cordeiro}.}
  \bibinfo{year}{2022}\natexlab{}.
\newblock \showarticletitle{ESBMC-CHERI: towards verification of C programs for
  CHERI platforms with ESBMC}. In \bibinfo{booktitle}{\emph{International
  Symposium on Software Testing and Analysis}}.
\newblock
\href{https://doi.org/10.1145/3533767.3543289}{doi:\nolinkurl{10.1145/3533767.3543289}}


\bibitem[Calcagno and Distefano(2011)]%
        {calcagno:nasa:2011}
\bibfield{author}{\bibinfo{person}{Cristiano Calcagno} {and}
  \bibinfo{person}{Dino Distefano}.} \bibinfo{year}{2011}\natexlab{}.
\newblock \showarticletitle{Infer: An Automatic Program Verifier for Memory
  Safety of {C} Programs}. In \bibinfo{booktitle}{\emph{NASA Formal Methods
  Symposium}}.
\newblock
\href{https://doi.org/10.1007/978-3-642-20398-5_33}{doi:\nolinkurl{10.1007/978-3-642-20398-5_33}}


\bibitem[Calcagno et~al\mbox{.}(2009)]%
        {calcagno:popl:2009}
\bibfield{author}{\bibinfo{person}{Cristiano Calcagno}, \bibinfo{person}{Dino
  Distefano}, \bibinfo{person}{Peter O'Hearn}, {and} \bibinfo{person}{Hongseok
  Yang}.} \bibinfo{year}{2009}\natexlab{}.
\newblock \showarticletitle{Compositional Shape Analysis by Means of
  Bi-Abduction}. In \bibinfo{booktitle}{\emph{Principles of Programming
  Languages}}.
\newblock
\href{https://doi.org/10.1145/1480881.1480917}{doi:\nolinkurl{10.1145/1480881.1480917}}


\bibitem[Calcagno et~al\mbox{.}(2011)]%
        {calcagno:jacm:2011}
\bibfield{author}{\bibinfo{person}{Cristiano Calcagno}, \bibinfo{person}{Dino
  Distefano}, \bibinfo{person}{Peter~W. O’Hearn}, {and}
  \bibinfo{person}{Hongseok Yang}.} \bibinfo{year}{2011}\natexlab{}.
\newblock \showarticletitle{Compositional Shape Analysis by Means of
  Bi-Abduction}.
\newblock \bibinfo{journal}{\emph{Journal of the {ACM}}} \bibinfo{volume}{58},
  \bibinfo{number}{6} (\bibinfo{year}{2011}).
\newblock
\href{https://doi.org/10.1145/2049697.2049700}{doi:\nolinkurl{10.1145/2049697.2049700}}


\bibitem[Calcagno et~al\mbox{.}(2007)]%
        {calcagno:lics:2007}
\bibfield{author}{\bibinfo{person}{Cristiano Calcagno},
  \bibinfo{person}{Peter~W. O'Hearn}, {and} \bibinfo{person}{Hongseok Yang}.}
  \bibinfo{year}{2007}\natexlab{}.
\newblock \showarticletitle{Local Action and Abstract Separation Logic}. In
  \bibinfo{booktitle}{\emph{Symposium on Logic in Computer Science}}.
\newblock
\href{https://doi.org/10.1109/LICS.2007.30}{doi:\nolinkurl{10.1109/LICS.2007.30}}


\bibitem[Cohen and Johnson-Freyd(2024)]%
        {Cohen24}
\bibfield{author}{\bibinfo{person}{Joshua~M. Cohen} {and}
  \bibinfo{person}{Philip Johnson-Freyd}.} \bibinfo{year}{2024}\natexlab{}.
\newblock \showarticletitle{A Formalization of Core Why3 in Coq}.
\newblock \bibinfo{journal}{\emph{Proc. ACM Program. Lang.}}
  \bibinfo{volume}{8}, \bibinfo{number}{POPL} (\bibinfo{year}{2024}).
\newblock
\href{https://doi.org/10.1145/3632902}{doi:\nolinkurl{10.1145/3632902}}


\bibitem[Dardinier et~al\mbox{.}(2025)]%
        {Dardinier25}
\bibfield{author}{\bibinfo{person}{Thibault Dardinier},
  \bibinfo{person}{Michael Sammler}, \bibinfo{person}{Gaurav Parthasarathy},
  \bibinfo{person}{Alexander~J. Summers}, {and} \bibinfo{person}{Peter
  M\"{u}ller}.} \bibinfo{year}{2025}\natexlab{}.
\newblock \showarticletitle{Formal Foundations for Translational Separation
  Logic Verifiers}.
\newblock \bibinfo{journal}{\emph{Proc. ACM Program. Lang.}}
  \bibinfo{volume}{9}, \bibinfo{number}{POPL} (\bibinfo{year}{2025}).
\newblock
\href{https://doi.org/10.1145/3704856}{doi:\nolinkurl{10.1145/3704856}}


\bibitem[De~Moura and Bj{\o}rner(2008)]%
        {z3}
\bibfield{author}{\bibinfo{person}{Leonardo De~Moura} {and}
  \bibinfo{person}{Nikolaj Bj{\o}rner}.} \bibinfo{year}{2008}\natexlab{}.
\newblock \showarticletitle{{Z3}: An Efficient {SMT} Solver}. In
  \bibinfo{booktitle}{\emph{Tools and Algorithms for the Construction and
  Analysis of Systems}}.
\newblock
\href{https://doi.org/10.1007/978-3-540-78800-3_24}{doi:\nolinkurl{10.1007/978-3-540-78800-3_24}}


\bibitem[Dinsdale-Young et~al\mbox{.}(2013)]%
        {dyoung:popl:2013}
\bibfield{author}{\bibinfo{person}{Thomas Dinsdale-Young},
  \bibinfo{person}{Lars Birkedal}, \bibinfo{person}{Philippa Gardner},
  \bibinfo{person}{Matthew Parkinson}, {and} \bibinfo{person}{Hongseok Yang}.}
  \bibinfo{year}{2013}\natexlab{}.
\newblock \showarticletitle{Views: compositional reasoning for concurrent
  programs}. In \bibinfo{booktitle}{\emph{Symposium on Principles of
  Programming Languages}}.
\newblock
\href{https://doi.org/10.1145/2429069.2429104}{doi:\nolinkurl{10.1145/2429069.2429104}}


\bibitem[Filli{\^a}tre and Paskevich(2013)]%
        {why3}
\bibfield{author}{\bibinfo{person}{Jean-Christophe Filli{\^a}tre} {and}
  \bibinfo{person}{Andrei Paskevich}.} \bibinfo{year}{2013}\natexlab{}.
\newblock \showarticletitle{Why3 --- Where Programs Meet Provers}. In
  \bibinfo{booktitle}{\emph{European Symposium on Programming}}.
\newblock
\href{https://doi.org/10.1007/978-3-642-37036-6_8}{doi:\nolinkurl{10.1007/978-3-642-37036-6_8}}


\bibitem[Fragoso~Santos et~al\mbox{.}(2018)]%
        {javert}
\bibfield{author}{\bibinfo{person}{Jos\'{e} Fragoso~Santos},
  \bibinfo{person}{Petar Maksimovi\'{c}}, \bibinfo{person}{Daiva
  Naudžiūnienė}, \bibinfo{person}{Thomas Wood}, {and}
  \bibinfo{person}{Philippa Gardner}.} \bibinfo{year}{2018}\natexlab{}.
\newblock \showarticletitle{{JaVerT}: JavaScript Verification Toolchain}.
\newblock \bibinfo{journal}{\emph{Proceedings of the ACM on Programming
  Languages}} \bibinfo{volume}{2}, \bibinfo{number}{{POPL}}
  (\bibinfo{year}{2018}).
\newblock
\href{https://doi.org/10.1145/3158138}{doi:\nolinkurl{10.1145/3158138}}


\bibitem[Fragoso~Santos et~al\mbox{.}(2019)]%
        {javert2}
\bibfield{author}{\bibinfo{person}{Jos\'{e} Fragoso~Santos},
  \bibinfo{person}{Petar Maksimovi\'{c}}, \bibinfo{person}{Gabriela Sampaio},
  {and} \bibinfo{person}{Philippa Gardner}.} \bibinfo{year}{2019}\natexlab{}.
\newblock \showarticletitle{{JaVerT} 2.0: Compositional Symbolic Execution for
  {JavaScript}}.
\newblock \bibinfo{journal}{\emph{Proceedings of the ACM on Programming
  Languages}} \bibinfo{volume}{3}, \bibinfo{number}{{POPL}}
  (\bibinfo{year}{2019}).
\newblock
\href{https://doi.org/10.1145/3290379}{doi:\nolinkurl{10.1145/3290379}}


\bibitem[{Fragoso Santos} et~al\mbox{.}(2020)]%
        {gillianpldi}
\bibfield{author}{\bibinfo{person}{Jos{\'{e}} {Fragoso Santos}},
  \bibinfo{person}{Petar Maksimović}, \bibinfo{person}{Sacha{-}{\'{E}}lie
  Ayoun}, {and} \bibinfo{person}{Philippa Gardner}.}
  \bibinfo{year}{2020}\natexlab{}.
\newblock \showarticletitle{Gillian, Part {I}: A Multi-language Platform for
  Symbolic Execution}. In \bibinfo{booktitle}{\emph{Conference on Programming
  Language Design and Implementation}}.
\newblock
\href{https://doi.org/10.1145/3385412.3386014}{doi:\nolinkurl{10.1145/3385412.3386014}}


\bibitem[Haas et~al\mbox{.}(2017)]%
        {wasm}
\bibfield{author}{\bibinfo{person}{Andreas Haas}, \bibinfo{person}{Andreas
  Rossberg}, \bibinfo{person}{Derek~L. Schuff}, \bibinfo{person}{Ben~L.
  Titzer}, \bibinfo{person}{Michael Holman}, \bibinfo{person}{Dan Gohman},
  \bibinfo{person}{Luke Wagner}, \bibinfo{person}{Alon Zakai}, {and}
  \bibinfo{person}{JF Bastien}.} \bibinfo{year}{2017}\natexlab{}.
\newblock \showarticletitle{Bringing the web up to speed with {WebAssembly}}.
  In \bibinfo{booktitle}{\emph{Conference on Programming Language Design and
  Implementation}}.
\newblock
\href{https://doi.org/10.1145/3062341.3062363}{doi:\nolinkurl{10.1145/3062341.3062363}}


\bibitem[Jacobs et~al\mbox{.}(2011)]%
        {verifast}
\bibfield{author}{\bibinfo{person}{Bart Jacobs}, \bibinfo{person}{Jan Smans},
  \bibinfo{person}{Pieter Philippaerts}, \bibinfo{person}{Fr{\'{e}}d{\'{e}}ric
  Vogels}, \bibinfo{person}{Willem Penninckx}, {and} \bibinfo{person}{Frank
  Piessens}.} \bibinfo{year}{2011}\natexlab{}.
\newblock \showarticletitle{{VeriFast}: A Powerful, Sound, Predictable, Fast
  Verifier for {C} and {Java}}. In \bibinfo{booktitle}{\emph{NASA Formal
  Methods Symposium}}.
\newblock
\href{https://doi.org/10.1007/978-3-642-20398-5_4}{doi:\nolinkurl{10.1007/978-3-642-20398-5_4}}


\bibitem[Jacobs et~al\mbox{.}(2015)]%
        {Jacobs15}
\bibfield{author}{\bibinfo{person}{Bart Jacobs}, \bibinfo{person}{Frédéric
  Vogels}, {and} \bibinfo{person}{Frank Piessens}.}
  \bibinfo{year}{2015}\natexlab{}.
\newblock \showarticletitle{Featherweight {VeriFast}}.
\newblock \bibinfo{journal}{\emph{Logical Methods in Computer Science}}
  \bibinfo{volume}{11} (\bibinfo{year}{2015}).
\newblock
Issue 3.
\href{https://doi.org/10.2168/LMCS-11(3:19)2015}{doi:\nolinkurl{10.2168/LMCS-11(3:19)2015}}


\bibitem[Jung et~al\mbox{.}(2017)]%
        {Jung17}
\bibfield{author}{\bibinfo{person}{Ralf Jung}, \bibinfo{person}{Jacques-Henri
  Jourdan}, \bibinfo{person}{Robbert Krebbers}, {and} \bibinfo{person}{Derek
  Dreyer}.} \bibinfo{year}{2017}\natexlab{}.
\newblock \showarticletitle{RustBelt: securing the foundations of the Rust
  programming language}.
\newblock \bibinfo{journal}{\emph{Proc. ACM Program. Lang.}}
  \bibinfo{volume}{2}, \bibinfo{number}{POPL} (\bibinfo{year}{2017}).
\newblock
\href{https://doi.org/10.1145/3158154}{doi:\nolinkurl{10.1145/3158154}}


\bibitem[Jung et~al\mbox{.}(2018)]%
        {jung:jfunc:2018}
\bibfield{author}{\bibinfo{person}{Ralf Jung}, \bibinfo{person}{Robbert
  Krebbers}, \bibinfo{person}{Jacques-Henri Jourdan}, \bibinfo{person}{Aleš
  Bizjak}, \bibinfo{person}{Lars Birkedal}, {and} \bibinfo{person}{Derek
  Dreyer}.} \bibinfo{year}{2018}\natexlab{}.
\newblock \showarticletitle{{Iris from the ground up: {A} modular foundation
  for higher-order concurrent separation logic}}.
\newblock \bibinfo{journal}{\emph{Journal of Functional Programming (JFP)}}
  \bibinfo{volume}{28} (\bibinfo{year}{2018}).
\newblock
\href{https://doi.org/10.1017/S0956796818000151}{doi:\nolinkurl{10.1017/S0956796818000151}}


\bibitem[Keuchel et~al\mbox{.}(2022)]%
        {Keuchel22}
\bibfield{author}{\bibinfo{person}{Steven Keuchel}, \bibinfo{person}{Sander
  Huyghebaert}, \bibinfo{person}{Georgy Lukyanov}, {and}
  \bibinfo{person}{Dominique Devriese}.} \bibinfo{year}{2022}\natexlab{}.
\newblock \showarticletitle{Verified symbolic execution with Kripke
  specification monads (and no meta-programming)}.
\newblock \bibinfo{journal}{\emph{Proc. ACM Program. Lang.}}
  \bibinfo{volume}{6}, \bibinfo{number}{ICFP} (\bibinfo{year}{2022}).
\newblock
\href{https://doi.org/10.1145/3547628}{doi:\nolinkurl{10.1145/3547628}}


\bibitem[Le et~al\mbox{.}(2022)]%
        {Le22}
\bibfield{author}{\bibinfo{person}{Quang~Loc Le}, \bibinfo{person}{Azalea
  Raad}, \bibinfo{person}{Jules Villard}, \bibinfo{person}{Josh Berdine},
  \bibinfo{person}{Derek Dreyer}, {and} \bibinfo{person}{Peter~W. O'Hearn}.}
  \bibinfo{year}{2022}\natexlab{}.
\newblock \showarticletitle{Finding Real Bugs in Big Programs with
  Incorrectness Logic}.
\newblock \bibinfo{journal}{\emph{Proceedings of the ACM on Programming
  Languages}} \bibinfo{volume}{6}, \bibinfo{number}{OOPSLA1}
  (\bibinfo{year}{2022}).
\newblock
\href{https://doi.org/10.1145/3527325}{doi:\nolinkurl{10.1145/3527325}}


\bibitem[Legoupil et~al\mbox{.}(2024)]%
        {legoupil:2024:irismswasm}
\bibfield{author}{\bibinfo{person}{Maxime Legoupil}, \bibinfo{person}{June
  Rousseau}, \bibinfo{person}{A\"{\i}na~Linn Georges}, \bibinfo{person}{Jean
  Pichon-Pharabod}, {and} \bibinfo{person}{Lars Birkedal}.}
  \bibinfo{year}{2024}\natexlab{}.
\newblock \showarticletitle{Iris-MSWasm: Elucidating and Mechanising the
  Security Invariants of Memory-Safe WebAssembly}.
\newblock \bibinfo{journal}{\emph{Proc. ACM Program. Lang.}}
  \bibinfo{volume}{8}, \bibinfo{number}{OOPSLA2} (\bibinfo{year}{2024}).
\newblock
\href{https://doi.org/10.1145/3689722}{doi:\nolinkurl{10.1145/3689722}}


\bibitem[Lepigre et~al\mbox{.}(2022)]%
        {Lepigre22}
\bibfield{author}{\bibinfo{person}{Rodolphe Lepigre}, \bibinfo{person}{Michael
  Sammler}, \bibinfo{person}{Kayvan Memarian}, \bibinfo{person}{Robbert
  Krebbers}, \bibinfo{person}{Derek Dreyer}, {and} \bibinfo{person}{Peter
  Sewell}.} \bibinfo{year}{2022}\natexlab{}.
\newblock \showarticletitle{VIP: Verifying Real-World C Idioms with
  Integer-Pointer Casts}.
\newblock \bibinfo{journal}{\emph{Proc. ACM Program. Lang.}}
  \bibinfo{volume}{6}, \bibinfo{number}{POPL} (\bibinfo{year}{2022}).
\newblock
\href{https://doi.org/10.1145/3498681}{doi:\nolinkurl{10.1145/3498681}}


\bibitem[Leroy(2009)]%
        {Leroy09}
\bibfield{author}{\bibinfo{person}{Xavier Leroy}.}
  \bibinfo{year}{2009}\natexlab{}.
\newblock \showarticletitle{Formal verification of a realistic compiler}.
\newblock \bibinfo{journal}{\emph{Commun. ACM}} \bibinfo{volume}{52},
  \bibinfo{number}{7} (\bibinfo{year}{2009}).
\newblock
\href{https://doi.org/10.1145/1538788.1538814}{doi:\nolinkurl{10.1145/1538788.1538814}}


\bibitem[L\"{o}\"{o}w et~al\mbox{.}(2025)]%
        {artefact}
\bibfield{author}{\bibinfo{person}{Andreas L\"{o}\"{o}w},
  \bibinfo{person}{Seung~Hoon Park}, \bibinfo{person}{Daniele Nantes-Sobrinho},
  \bibinfo{person}{Sacha-Élie Ayoun}, \bibinfo{person}{Opale Sj\"{o}stedt},
  {and} \bibinfo{person}{Philippa Gardner}.} \bibinfo{year}{2025}\natexlab{}.
\newblock \bibinfo{title}{Compositional Symbolic Execution for the Next 700
  Memory Models (Artefact)}.
\newblock
\href{https://doi.org/10.5281/ZENODO.16909361}{doi:\nolinkurl{10.5281/ZENODO.16909361}}


\bibitem[Lööw et~al\mbox{.}(2024a)]%
        {cse1}
\bibfield{author}{\bibinfo{person}{Andreas Lööw}, \bibinfo{person}{Daniele
  Nantes-Sobrinho}, \bibinfo{person}{Sacha{-}Élie Ayoun},
  \bibinfo{person}{Caroline Cronjäger}, \bibinfo{person}{Petar Maksimović},
  {and} \bibinfo{person}{Philippa Gardner}.} \bibinfo{year}{2024}\natexlab{a}.
\newblock \showarticletitle{Compositional Symbolic Execution for Correctness
  and Incorrectness Reasoning}. In \bibinfo{booktitle}{\emph{European
  Conference on Object-Oriented Programming}}.
\newblock
\href{https://doi.org/10.4230/LIPIcs.ECOOP.2024.25}{doi:\nolinkurl{10.4230/LIPIcs.ECOOP.2024.25}}


\bibitem[Lööw et~al\mbox{.}(2024b)]%
        {matchplanning}
\bibfield{author}{\bibinfo{person}{Andreas Lööw}, \bibinfo{person}{Daniele
  Nantes-Sobrinho}, \bibinfo{person}{Sacha{-}Élie Ayoun},
  \bibinfo{person}{Petar Maksimović}, {and} \bibinfo{person}{Philippa
  Gardner}.} \bibinfo{year}{2024}\natexlab{b}.
\newblock \showarticletitle{Matching Plans for Frame Inference in Compositional
  Reasoning}. In \bibinfo{booktitle}{\emph{European Conference on
  Object-Oriented Programming}}.
\newblock
\href{https://doi.org/10.4230/LIPIcs.ECOOP.2024.26}{doi:\nolinkurl{10.4230/LIPIcs.ECOOP.2024.26}}


\bibitem[Maksimovi\'{c} et~al\mbox{.}(2023)]%
        {esl}
\bibfield{author}{\bibinfo{person}{Petar Maksimovi\'{c}},
  \bibinfo{person}{Caroline Cronj\"{a}ger}, \bibinfo{person}{Andreas
  L\"{o}\"{o}w}, \bibinfo{person}{Julian Sutherland}, {and}
  \bibinfo{person}{Philippa Gardner}.} \bibinfo{year}{2023}\natexlab{}.
\newblock \showarticletitle{Exact Separation Logic}. In
  \bibinfo{booktitle}{\emph{European Conference on Object-Oriented
  Programming}}.
\newblock
\href{https://doi.org/10.4230/LIPIcs.ECOOP.2023.19}{doi:\nolinkurl{10.4230/LIPIcs.ECOOP.2023.19}}


\bibitem[Maksimović et~al\mbox{.}(2021)]%
        {gilliancav}
\bibfield{author}{\bibinfo{person}{Petar Maksimović},
  \bibinfo{person}{Sacha{-}{\'{E}}lie Ayoun},
  \bibinfo{person}{Jos{\'{e}}~Fragoso Santos}, {and} \bibinfo{person}{Philippa
  Gardner}.} \bibinfo{year}{2021}\natexlab{}.
\newblock \showarticletitle{Gillian, Part {II}: Real-World Verification for
  {JavaScript} and {C}}. In \bibinfo{booktitle}{\emph{Computer Aided
  Verification}}.
\newblock
\href{https://doi.org/10.1007/978-3-030-81688-9_38}{doi:\nolinkurl{10.1007/978-3-030-81688-9_38}}


\bibitem[Matsushita et~al\mbox{.}(2022)]%
        {Yusuke22}
\bibfield{author}{\bibinfo{person}{Yusuke Matsushita}, \bibinfo{person}{Xavier
  Denis}, \bibinfo{person}{Jacques-Henri Jourdan}, {and} \bibinfo{person}{Derek
  Dreyer}.} \bibinfo{year}{2022}\natexlab{}.
\newblock \showarticletitle{RustHornBelt: a semantic foundation for functional
  verification of Rust programs with unsafe code}. In
  \bibinfo{booktitle}{\emph{Conference on Programming Language Design and
  Implementation}}.
\newblock
\href{https://doi.org/10.1145/3519939.3523704}{doi:\nolinkurl{10.1145/3519939.3523704}}


\bibitem[Memarian et~al\mbox{.}(2019)]%
        {Memarian19}
\bibfield{author}{\bibinfo{person}{Kayvan Memarian}, \bibinfo{person}{Victor
  B.~F. Gomes}, \bibinfo{person}{Brooks Davis}, \bibinfo{person}{Stephen Kell},
  \bibinfo{person}{Alexander Richardson}, \bibinfo{person}{Robert N.~M.
  Watson}, {and} \bibinfo{person}{Peter Sewell}.}
  \bibinfo{year}{2019}\natexlab{}.
\newblock \showarticletitle{Exploring {C} semantics and pointer provenance}.
\newblock \bibinfo{journal}{\emph{Proc. ACM Program. Lang.}}
  \bibinfo{volume}{3}, \bibinfo{number}{POPL} (\bibinfo{year}{2019}).
\newblock
\href{https://doi.org/10.1145/3290380}{doi:\nolinkurl{10.1145/3290380}}


\bibitem[M{\"u}ller et~al\mbox{.}(2016)]%
        {Muller16}
\bibfield{author}{\bibinfo{person}{Peter M{\"u}ller}, \bibinfo{person}{Malte
  Schwerhoff}, {and} \bibinfo{person}{Alexander~J. Summers}.}
  \bibinfo{year}{2016}\natexlab{}.
\newblock \showarticletitle{Viper: A Verification Infrastructure for
  Permission-Based Reasoning}. In \bibinfo{booktitle}{\emph{Verification, Model
  Checking, and Abstract Interpretation}}.
\newblock
\href{https://doi.org/10.1007/978-3-662-49122-5_2}{doi:\nolinkurl{10.1007/978-3-662-49122-5_2}}


\bibitem[Naudziuniene(2018)]%
        {Naudziuniene2018Infrastructure}
\bibfield{author}{\bibinfo{person}{Daiva Naudziuniene}.}
  \bibinfo{year}{2018}\natexlab{}.
\newblock \emph{\bibinfo{title}{An Infrastructure for Tractable Verification of
  JavaScript Programs}}.
\newblock \bibinfo{thesistype}{Ph.\,D. Dissertation}. \bibinfo{school}{Imperial
  College London}.
\newblock


\bibitem[O'Hearn et~al\mbox{.}(2001)]%
        {seplogic}
\bibfield{author}{\bibinfo{person}{Peter~W. O'Hearn}, \bibinfo{person}{John~C.
  Reynolds}, {and} \bibinfo{person}{Hongseok Yang}.}
  \bibinfo{year}{2001}\natexlab{}.
\newblock \showarticletitle{Local Reasoning about Programs that Alter Data
  Structures}. In \bibinfo{booktitle}{\emph{Computer Science Logic}}.
\newblock
\href{https://doi.org/10.1007/3-540-44802-0_1}{doi:\nolinkurl{10.1007/3-540-44802-0_1}}


\bibitem[Owens et~al\mbox{.}(2016)]%
        {Owens16}
\bibfield{author}{\bibinfo{person}{Scott Owens}, \bibinfo{person}{Magnus~O.
  Myreen}, \bibinfo{person}{Ramana Kumar}, {and} \bibinfo{person}{Yong~Kiam
  Tan}.} \bibinfo{year}{2016}\natexlab{}.
\newblock \showarticletitle{Functional Big-Step Semantics}. In
  \bibinfo{booktitle}{\emph{European Symposium on Programming Languages and
  Systems}}.
\newblock
\href{https://doi.org/10.1007/978-3-662-49498-1_23}{doi:\nolinkurl{10.1007/978-3-662-49498-1_23}}


\bibitem[Park(2022)]%
        {park:2022:cheriafp}
\bibfield{author}{\bibinfo{person}{Seung~Hoon Park}.}
  \bibinfo{year}{2022}\natexlab{}.
\newblock \showarticletitle{A Formal CHERI-C Memory Model}.
\newblock \bibinfo{journal}{\emph{Archive of Formal Proofs}}
  (\bibinfo{date}{November} \bibinfo{year}{2022}).
\newblock
\showISSN{2150-914x}
\newblock
\shownote{\url{https://isa-afp.org/entries/CHERI-C_Memory_Model.html}, Formal
  proof development}.


\bibitem[Park et~al\mbox{.}(2023)]%
        {park:2023:gilliancheric}
\bibfield{author}{\bibinfo{person}{Seung~Hoon Park}, \bibinfo{person}{Rekha
  Pai}, {and} \bibinfo{person}{Tom Melham}.} \bibinfo{year}{2023}\natexlab{}.
\newblock \showarticletitle{A Formal CHERI-C Semantics for Verification}. In
  \bibinfo{booktitle}{\emph{Tools and Algorithms for the Construction and
  Analysis of Systems}}.
\newblock
\href{https://doi.org/10.1007/978-3-031-30823-9_28}{doi:\nolinkurl{10.1007/978-3-031-30823-9_28}}


\bibitem[Parthasarathy et~al\mbox{.}(2024)]%
        {Parthasarathy24}
\bibfield{author}{\bibinfo{person}{Gaurav Parthasarathy},
  \bibinfo{person}{Thibault Dardinier}, \bibinfo{person}{Benjamin Bonneau},
  \bibinfo{person}{Peter M\"{u}ller}, {and} \bibinfo{person}{Alexander~J.
  Summers}.} \bibinfo{year}{2024}\natexlab{}.
\newblock \showarticletitle{Towards Trustworthy Automated Program Verifiers:
  Formally Validating Translations into an Intermediate Verification Language}.
\newblock \bibinfo{journal}{\emph{Proc. ACM Program. Lang.}}
  \bibinfo{volume}{8}, \bibinfo{number}{PLDI} (\bibinfo{year}{2024}).
\newblock
\href{https://doi.org/10.1145/3656438}{doi:\nolinkurl{10.1145/3656438}}


\bibitem[Pulte et~al\mbox{.}(2023)]%
        {Pulte23}
\bibfield{author}{\bibinfo{person}{Christopher Pulte},
  \bibinfo{person}{Dhruv~C. Makwana}, \bibinfo{person}{Thomas Sewell},
  \bibinfo{person}{Kayvan Memarian}, \bibinfo{person}{Peter Sewell}, {and}
  \bibinfo{person}{Neel Krishnaswami}.} \bibinfo{year}{2023}\natexlab{}.
\newblock \showarticletitle{{CN}: Verifying Systems {C} Code with
  Separation-Logic Refinement Types}.
\newblock \bibinfo{journal}{\emph{Proceedings of the ACM on Programming
  Languages}} \bibinfo{volume}{7}, \bibinfo{number}{POPL}
  (\bibinfo{year}{2023}).
\newblock
\href{https://doi.org/10.1145/3571194}{doi:\nolinkurl{10.1145/3571194}}


\bibitem[Raad et~al\mbox{.}(2020)]%
        {isl}
\bibfield{author}{\bibinfo{person}{Azalea Raad}, \bibinfo{person}{Josh
  Berdine}, \bibinfo{person}{Hoang-Hai Dang}, \bibinfo{person}{Derek Dreyer},
  \bibinfo{person}{Peter O'Hearn}, {and} \bibinfo{person}{Jules Villard}.}
  \bibinfo{year}{2020}\natexlab{}.
\newblock \showarticletitle{Local Reasoning About the Presence of Bugs:
  Incorrectness Separation Logic}. In \bibinfo{booktitle}{\emph{Computer Aided
  Verification}}.
\newblock
\href{https://doi.org/10.1007/978-3-030-53291-8_14}{doi:\nolinkurl{10.1007/978-3-030-53291-8_14}}


\bibitem[Raad et~al\mbox{.}(2022)]%
        {cisl}
\bibfield{author}{\bibinfo{person}{Azalea Raad}, \bibinfo{person}{Josh
  Berdine}, \bibinfo{person}{Derek Dreyer}, {and} \bibinfo{person}{Peter~W.
  O'Hearn}.} \bibinfo{year}{2022}\natexlab{}.
\newblock \showarticletitle{Concurrent Incorrectness Separation Logic}.
\newblock \bibinfo{journal}{\emph{Proceedings of the ACM on Programming
  Languages}} \bibinfo{volume}{6}, \bibinfo{number}{POPL}
  (\bibinfo{year}{2022}).
\newblock
\href{https://doi.org/10.1145/3498695}{doi:\nolinkurl{10.1145/3498695}}


\bibitem[Rao et~al\mbox{.}(2023)]%
        {rao:2023:iriswasm}
\bibfield{author}{\bibinfo{person}{Xiaojia Rao},
  \bibinfo{person}{A\"{\i}na~Linn Georges}, \bibinfo{person}{Maxime Legoupil},
  \bibinfo{person}{Conrad Watt}, \bibinfo{person}{Jean Pichon-Pharabod},
  \bibinfo{person}{Philippa Gardner}, {and} \bibinfo{person}{Lars Birkedal}.}
  \bibinfo{year}{2023}\natexlab{}.
\newblock \showarticletitle{Iris-Wasm: Robust and Modular Verification of
  WebAssembly Programs}.
\newblock \bibinfo{journal}{\emph{Proc. ACM Program. Lang.}}
  \bibinfo{volume}{7}, \bibinfo{number}{PLDI} (\bibinfo{year}{2023}).
\newblock
\href{https://doi.org/10.1145/3591265}{doi:\nolinkurl{10.1145/3591265}}


\bibitem[Reynolds(2002)]%
        {reyseplogic}
\bibfield{author}{\bibinfo{person}{John~C. Reynolds}.}
  \bibinfo{year}{2002}\natexlab{}.
\newblock \showarticletitle{Separation Logic: A Logic for Shared Mutable Data
  Structures}. In \bibinfo{booktitle}{\emph{Logic in Computer Science}}.
\newblock
\href{https://doi.org/10.1109/LICS.2002.1029817}{doi:\nolinkurl{10.1109/LICS.2002.1029817}}


\bibitem[Rocq({[n.\,d.]})]%
        {Rocq}
\bibfield{author}{\bibinfo{person}{Rocq}.} \bibinfo{year}{[n.\,d.]}\natexlab{}.
\newblock \bibinfo{title}{The {Rocq} Prover}.
\newblock \bibinfo{howpublished}{\url{https://rocq-prover.org}}.
\newblock


\bibitem[Santos({[n.\,d.]})]%
        {buckets}
\bibfield{author}{\bibinfo{person}{Mauricio Santos}.}
  \bibinfo{year}{[n.\,d.]}\natexlab{}.
\newblock \bibinfo{title}{Buckets-JS: A JavaScript Data Structure Library}.
\newblock
  \bibinfo{howpublished}{\url{https://github.com/mauriciosantos/Buckets-JS}}.
\newblock


\bibitem[Schwerhoff(2016)]%
        {Schwerhoff16}
\bibfield{author}{\bibinfo{person}{Malte~Hermann Schwerhoff}.}
  \bibinfo{year}{2016}\natexlab{}.
\newblock \emph{\bibinfo{title}{Advancing Automated, Permission-Based Program
  Verification Using Symbolic Execution}}.
\newblock \bibinfo{thesistype}{Ph.\,D. Dissertation}. \bibinfo{school}{ETH
  Zürich}.
\newblock


\bibitem[Watson et~al\mbox{.}(2023)]%
        {watson:2023:cheri}
\bibfield{author}{\bibinfo{person}{Robert N.~M. Watson},
  \bibinfo{person}{Peter~G. Neumann}, \bibinfo{person}{Jonathan Woodruff},
  \bibinfo{person}{Michael Roe}, \bibinfo{person}{Hesham Almatary},
  \bibinfo{person}{Jonathan Anderson}, \bibinfo{person}{John Baldwin},
  \bibinfo{person}{Graeme Barnes}, \bibinfo{person}{David Chisnall},
  \bibinfo{person}{Jessica Clarke}, \bibinfo{person}{Brooks Davis},
  \bibinfo{person}{Lee Eisen}, \bibinfo{person}{Nathaniel~Wesley Filardo},
  \bibinfo{person}{Franz~A. Fuchs}, \bibinfo{person}{Richard Grisenthwaite},
  \bibinfo{person}{Alexandre Joannou}, \bibinfo{person}{Ben Laurie},
  \bibinfo{person}{A.~Theodore Markettos}, \bibinfo{person}{Simon~W. Moore},
  \bibinfo{person}{Steven~J. Murdoch}, \bibinfo{person}{Kyndylan Nienhuis},
  \bibinfo{person}{Robert Norton}, \bibinfo{person}{Alexander Richardson},
  \bibinfo{person}{Peter Rugg}, \bibinfo{person}{Peter Sewell},
  \bibinfo{person}{Stacey Son}, {and} \bibinfo{person}{Hongyan Xia}.}
  \bibinfo{year}{2023}\natexlab{}.
\newblock \bibinfo{booktitle}{\emph{{Capability Hardware Enhanced RISC
  Instructions: CHERI Instruction-Set Architecture (Version 9)}}}.
\newblock \bibinfo{type}{{T}echnical {R}eport} UCAM-CL-TR-987.
  \bibinfo{institution}{University of Cambridge, Computer Laboratory}.
\newblock
\href{https://doi.org/10.48456/tr-987}{doi:\nolinkurl{10.48456/tr-987}}


\bibitem[Woodruff et~al\mbox{.}(2014)]%
        {woodruff:2014:cheri}
\bibfield{author}{\bibinfo{person}{Jonathan Woodruff}, \bibinfo{person}{Robert
  N.~M. Watson}, \bibinfo{person}{David Chisnall}, \bibinfo{person}{Simon~W.
  Moore}, \bibinfo{person}{Jonathan Anderson}, \bibinfo{person}{Brooks Davis},
  \bibinfo{person}{Ben Laurie}, \bibinfo{person}{Peter~G. Neumann},
  \bibinfo{person}{Robert Norton}, {and} \bibinfo{person}{Michael Roe}.}
  \bibinfo{year}{2014}\natexlab{}.
\newblock \showarticletitle{The CHERI capability model: Revisiting RISC in an
  age of risk}. In \bibinfo{booktitle}{\emph{International Symposium on
  Computer Architecture (ISCA)}}.
\newblock
\href{https://doi.org/10.1109/ISCA.2014.6853201}{doi:\nolinkurl{10.1109/ISCA.2014.6853201}}


\bibitem[Yang(2001)]%
        {Yangphd}
\bibfield{author}{\bibinfo{person}{Hongseok Yang}.}
  \bibinfo{year}{2001}\natexlab{}.
\newblock \emph{\bibinfo{title}{Local Reasoning for Stateful Programs}}.
\newblock \bibinfo{thesistype}{Ph.\,D. Dissertation}.
  \bibinfo{school}{University of Illinois Urbana-Champaign}.
\newblock


\bibitem[Yang and O'Hearn(2002)]%
        {Yang02}
\bibfield{author}{\bibinfo{person}{Hongseok Yang} {and} \bibinfo{person}{Peter
  O'Hearn}.} \bibinfo{year}{2002}\natexlab{}.
\newblock \showarticletitle{A Semantic Basis for Local Reasoning}. In
  \bibinfo{booktitle}{\emph{Foundations of Software Science and Computation
  Structures}}.
\newblock
\showISBNx{978-3-540-45931-6}
\href{https://doi.org/10.1007/3-540-45931-6_28}{doi:\nolinkurl{10.1007/3-540-45931-6_28}}


\bibitem[Zaliva et~al\mbox{.}(2024)]%
        {zaliva:2024:cheric}
\bibfield{author}{\bibinfo{person}{Vadim Zaliva}, \bibinfo{person}{Kayvan
  Memarian}, \bibinfo{person}{Ricardo Almeida}, \bibinfo{person}{Jessica
  Clarke}, \bibinfo{person}{Brooks Davis}, \bibinfo{person}{Alexander
  Richardson}, \bibinfo{person}{David Chisnall}, \bibinfo{person}{Brian
  Campbell}, \bibinfo{person}{Ian Stark}, \bibinfo{person}{Robert N.~M.
  Watson}, {and} \bibinfo{person}{Peter Sewell}.}
  \bibinfo{year}{2024}\natexlab{}.
\newblock \showarticletitle{Formal Mechanised Semantics of CHERI C:
  Capabilities, Undefined Behaviour, and Provenance}. In
  \bibinfo{booktitle}{\emph{International Conference on Architectural Support
  for Programming Languages and Operating Systems}}.
\newblock
\href{https://doi.org/10.1145/3617232.3624859}{doi:\nolinkurl{10.1145/3617232.3624859}}


\bibitem[Zaliva et~al\mbox{.}(2025)]%
        {zaliva:2025:cheric}
\bibfield{author}{\bibinfo{person}{Vadim Zaliva}, \bibinfo{person}{Kayvan
  Memarian}, \bibinfo{person}{Brian Campbell}, \bibinfo{person}{Ricardo
  Almeida}, \bibinfo{person}{Nathaniel Filardo}, \bibinfo{person}{Ian Stark},
  {and} \bibinfo{person}{Peter Sewell}.} \bibinfo{year}{2025}\natexlab{}.
\newblock \showarticletitle{A CHERI C Memory Model for Verified Temporal
  Safety}. In \bibinfo{booktitle}{\emph{International Conference on Certified
  Programs and Proofs}}.
\newblock
\href{https://doi.org/10.1145/3703595.3705878}{doi:\nolinkurl{10.1145/3703595.3705878}}


\bibitem[Zimmerman et~al\mbox{.}(2024)]%
        {Zimmerman24}
\bibfield{author}{\bibinfo{person}{Conrad Zimmerman}, \bibinfo{person}{Jenna
  DiVincenzo}, {and} \bibinfo{person}{Jonathan Aldrich}.}
  \bibinfo{year}{2024}\natexlab{}.
\newblock \showarticletitle{Sound Gradual Verification with Symbolic
  Execution}.
\newblock \bibinfo{journal}{\emph{Proceedings of the ACM on Programming
  Languages}} \bibinfo{volume}{8}, \bibinfo{number}{POPL}
  (\bibinfo{year}{2024}).
\newblock
\href{https://doi.org/10.1145/3632927}{doi:\nolinkurl{10.1145/3632927}}


\end{thebibliography}


\ifarxiv
\appendix
\newpage
\section{Concrete Definitions and Semantics}%
\label{app:concrete}

This section provides additional concrete definitions and an exhastive enumeration of the rules of the concrete semantics for our programming language.

\subsection{Additional Definitions}

The sets of program variables for expressions and commands, denoted by $\pv{\pexp}$ and $\pv{\cmd}$ respectively, are defined in the standard way.

\subsection{Conventions}

The following two conventions help simplify some of our definitions and proofs. First, we only consider executions where no program variables are missing, that is, for a given program $\scmd$ and a program store $\sto$ we must have $\dom(\sto) \subseteq \pv{\scmd}$. Second, we consider the program variables $\pvar{err}$ (used for error reporting) and $\pvar{ret}$ (used in the definition of function specifications) to be reserved and assume they do not occur in programs or stores during execution.

\subsection{Expression Evaluation}

Expression evaluation is trivially defined, we provide some illustrative cases:
\[
\begin{array}{rcl}
\esem{\gv}{\sto} & = &\gv \\
\esem{\pvar{x}}{\sto}& = &\sto(\pvar{x})\\
\esem{\pexp_1 ~{+}~ \pexp_2}{\sto} & = &
	\begin{cases}
		\esem{\pexp_1}{\sto} + \esem{\pexp_2}{\sto}, &
		\esem{\pexp_1}{\sto} \in \nats, \esem{\pexp_2}{\sto} \in \nats\\
		\undefd, & \mbox{otherwise} 
	\end{cases}\\
\esem{\pexp_1 ~{/}~ \pexp_2}{\sto} & = &
	\begin{cases}
		\esem{\pexp_1}{\sto} / \esem{\pexp_2}{\sto}, &
		\esem{\pexp_1}{\sto} \in \nats, \esem{\pexp_2}{\sto} \in \nats,\esem{\pexp_2}{\sto} \neq 0 \\
		\undefd, & \mbox{otherwise} 
	\end{cases}\\
\end{array}
\]

\subsection{Functions and Function Implementation Contexts}

A function, $\pfunction{\fid}{\lst{\pvar x}}{\cmd; \preturn{\pexp}}$, comprises:
an identifier, $\fid \in \fids \subseteq \strings$;  the
parameters, $\lst{\pvar x}$, given by a list of distinct program variables; a  function
body, $\cmd \in \Cmd$; and a {return expression}, $\pexp \in \PExp$, with  $\pv{\pexp} \subseteq
\{\lst{\pvar x}\} \cup \pv{\cmd}$.
A function implementation context, $\fictx$,  is a finite partial function from function identifiers to their parameters, bodies and return expressions:
$\fictx : \fids \rightharpoonup_{\mathit{fin}} ([\pvars], \cmds, \pexps)$, where
for $\fictx(\fid)=(\lst{\pvar{x}}, \cmd, {\pexp})$,  we also write
$\fid(\lst{\pvar{x}})\{\cmd; \preturn{\pexp}\} \in \fictx$.

\subsection{Concrete Semantics}

The rules of the semantics is as follows:
\begin{mathparpagebreakable}
\small
       \inferrule[\textsc{Skip}]{ \ }{
       \st, \pskip \baction_{\fictx} \oxok: \st
       } \and
       \inferrule[\textsc{Assign}]{
       \esem{\pexp}{\sto} = v \\ \sto' = \sto[\pvar{x} \storearrow v]
       }{
          \sthreadp{ \sto }{ \hp }, \passign{\pvar{x}}{\pexp} \baction_{\fictx} {\oxok} : \sthreadp{ \sto' }{ \hp }
       } \and
       \inferrule[\textsc{Assign-Err-Eval}]{
       \esem{\pexp}{\sto} = \undefd \\ \sto' = \sto[\pvar{err} \mapsto ``\mathsf{ExprEval}"]
       }{
          \sthreadp{ \sto }{ \hp }, \passign{\pvar x}{\pexp} \baction_{\fictx} {\oxerr} : \sthreadp{ \sto' }{ \hp }
       } \and
%
\inferrule[\textsc{Memory-Action}]{
    \esem{\vec{\pexp}}{\sto} = \vec{v} \\ \cmem.\act(\vec{v}) \rightsquigarrow \oxok : (\cmem', \vec{v}') \\ |\vec{v}'| = |\vec{\pvar{x}}| \\\\ \sto' = \sto [\vec{\pvar{x}} \storearrow \vec{v}' ]
  }{
    \csesemtrans{\sto, \cmem}{\vec{\pvar{x}} := \act(\lst\pexp)}{\sto', \cmem'}{\fictx}{\oxok}
  }
\and
\inferrule[\textsc{Memory-Action-Err-Length}]{
    \esem{\vec{\pexp}}{\sto} = \vec{v} \\ \cmem.\act(\vec{v}) \rightsquigarrow \oxok : (\cmem', \vec{v}') \\ |\vec{v}'| \neq |\vec{\pvar{x}}| \\\\
    \sto' = \sto[\pvar{err} \storearrow ``\mathsf{ActionArgsLength}"]
  }{\csesemtrans{\sto, \cmem}{\vec{\pvar{x}} := \act(\lst\pexp)}{\sto', \cmem'}{\fictx}{\oxerr}}
\and
\inferrule[\textsc{Memory-Action-Err}]{
    \esem{\vec{\pexp}}{\sto} = \vec{v} \\ \cmem.\act(\vec{v}) \rightsquigarrow \outcome : (\cmem', \vec{v}') \\ \outcome \not= \oxok \\\\ \sto' = \sto[\pvar{err} \storearrow \vec{v}']
  }{
    \csesemtrans{\sto, \cmem}{\vec{\pvar{x}} := \act(\lst\pexp)}{\sto', \cmem'}{\fictx}{\outcome}
  }
\and
\inferrule[\textsc{Memory-Action-Err-Eval}]{\esem{\vec{\pexp}}{\sto} = \undefd \\ \sto' = \sto[\pvar{err} \storearrow ``\mathsf{ExprEval}"]}
          {\csesemtrans{\sto, \cmem}{\vec{\pvar{x}} := \act(\lst\pexp)}{\sto', \cmem}{\fictx}{\oxerr}}

\and
%
       \inferrule[\textsc{If-Then}]{
       \esem{ \pexp }{ \sto } = \true \quad  (\sto, \cmem), \scmd_1 \baction_{\fictx} \outcome: \st'
       }{
          (\sto, \cmem), \pifelse{\pexp}{\scmd_1}{\scmd_2} \baction_{\fictx} \outcome: \st'
       }
       \and
       \inferrule[\textsc{If-Else}]{
          \esem{ \pexp }{ \sto } = \false \quad  (\sto, \cmem), \scmd_2 \baction_{\fictx} \outcome: \st'
       }{
          (\sto, \cmem), \pifelse{\pexp}{\scmd_1}{\scmd_2} \baction_{\fictx} \outcome: \st'
       } \and
       \inferrule[\textsc{If-Err-Eval}]{
       \esem{\pexp}{\sto} = \undefd \\ \sto' = \sto[\pvar{err} \storearrow ``\mathsf{ExprEval}"]
       }{
          \sthreadp{ \sto }{ \hp }, \pifelse{\pexp}{\scmd_1}{\scmd_2} \baction_{\fictx} {\oxerr} : \sthreadp{ \sto' }{ \hp }
       }
       \and
       \inferrule[\textsc{If-Err-Type}]{
       \esem{\pexp}{\sto} = v \notin \bools \quad \sto' = \sto[\pvar{err} \storearrow ``\mathsf{Type}"]
       }{
          \sthreadp{ \sto }{ \hp },  \pifelse{\pexp}{\scmd_1}{\scmd_2}
          \baction_{\fictx} {\oxerr} : \sthreadp{ \sto' }{ \hp }
       } \and
       \inferrule[\textsc{Seq}]{
       \st, \scmd_1 \baction_{\fictx} \oxok : \st' \\  \st', \scmd_2 \baction_{\fictx} \outcome:  \st''
       }{
    \st, \scmd_1 ; \scmd_2 \baction_{\fictx} \outcome : \st''
       } \and
       \inferrule[\textsc{Seq-Err}]{
       \st, \scmd_1 \baction_{\fictx} \outcome:   \st' \\  \outcome \neq \oxok
       }{
    \st, \scmd_1 ; \scmd_2 \baction_{\fictx} \outcome : \st'
       } \and
 \inferrule[\textsc{Func}]{
    \pfunction{\procname}{\vec{\pvar{x}}}{\scmd; \preturn{\pexp'}} \in \scontext
    \\\\
    |\vec{\pexp}| = |\vec{\pvar{x}}| \\ \esem{\vec{\pexp}}{\sto} = \vec{v} 
     \quad \vec{\pvar{z}} = (\pv{\scmd} \setminus \{\vec{\pvar{x}}\})
    \\\\
 \sto_p  =  \emptyset [ \vec{\pvar{x}} \storearrow \vec{v}] [ \vec{\pvar{z}} \storearrow \nil]
     \\  \esem{\pexp'}{\sto_q} =v' \\\\ (\sto_p, \hp), \scmd
      \baction_{\fictx} \oxok : \sthreadp{ \sto_q }{ \hp' }  \\
      \sto' = \sto[\pvar{y} \storearrow v']
  }{
     \sthreadp{ \sto }{ \hp },
     \pfuncall{\pvar{y}}{\procname}{\vec{\pexp}} \baction_{\fictx}
     \oxok : \sthreadp{\sto' }{ \hp' }
  }
  \and
  \inferrule[\textsc{Func-Err-Missing}]
      {\fid \notin \dom(\scontext) \\ \sto' = \sto[\pvar{err} \storearrow ``\mathsf{FuncMissing}"]}
      {\sthreadp{ \sto }{ \hp }, \pfuncall{\pvar{x}}{\procname}{\vec{\pexp}} \baction_{\fictx} {\oxerr} :  (\sto', \hp)}
  \and
  \inferrule[\textsc{Func-Err-Args-Length}]{
    \pfunction{\procname}{\vec{\pvar{x}}}{\scmd; \preturn{\pexp'}} \in \scontext
    \\
    |\vec{\pexp}| \neq |\vec{\pvar x}| \\\\ \sto' = \sto[\pvar{err} \storearrow ``\mathsf{FuncArgsLength}"]
  }{
     \sthreadp{ \sto }{ \hp },
     \pfuncall{\pvar{y}}{\procname}{\vec{\pexp}} \baction_{\fictx}
     \oxerr: \sthreadp{\sto' }{ \hp }
  }
  \and
  \inferrule[\textsc{Func-Err-Args-Eval}]{
    \pfunction{\procname}{\vec{\pvar{x}}}{\scmd; \preturn{\pexp'}} \in \scontext 
    \\
    |\vec{\pexp}| = |\vec{\pvar x}| \\\\ \esem{\vec{\pexp}}{\sto} = \undefd \\ \sto' = \sto[\pvar{err} \storearrow ``\mathsf{ExprEval}"]
  }{
     \sthreadp{ \sto }{ \hp },
     \pfuncall{\pvar{y}}{\procname}{\vec{\pexp}} \baction_{\fictx}
     \oxerr: \sthreadp{\sto' }{ \hp }
  }
  \and
  \inferrule[\textsc{Func-Err-Body}]{
    \pfunction{\procname}{\vec{\pvar{x}}}{\scmd; \preturn{\pexp'}} \in \scontext
    \\\\
    |\vec{\pexp}| = |\vec{\pvar x}| \\ \esem{\vec{\pexp}}{\sto} = \vec{v} 
     \\ \pvar{z} = (\pv{\scmd} \setminus \{\vec{\pvar{x}}\})
    \\\\
 \sto_p  =  \emptyset [ \vec{\pvar{x}} \storearrow \vec{v}] [ \vec{\pvar{z}} \storearrow \nil] \\ \outcome \neq \oxok
     \\\\ (\sto_p, \hp), \scmd
      \baction_{\fictx} \outcome: \sthreadp{ \sto_q }{ \hp' } \\ \sto' = \sto[\pvar{err} \storearrow \sto_q(\pvar{err}) ]
  }{
     \sthreadp{ \sto }{ \hp },
     \pfuncall{\pvar{y}}{\procname}{\vec{\pexp}} \baction_{\fictx}
     \outcome: \sthreadp{\sto' }{ \hp' }
  }
  \and
  \inferrule[\textsc{Func-Err-Eval-Ret}]{
    \pfunction{\procname}{\vec{\pvar{x}}}{\scmd; \preturn{\pexp'}} \in \scontext
    \\\\
    |\vec{\pexp}| = |\vec{\pvar x}| \\ \esem{\vec{\pexp}}{\sto} = \vec{v} 
     \\ \vec{\pvar{z}} = (\pv{\scmd} \setminus \{\vec{\pvar{x}}\})
    \\\\
 \sto_p  =  \emptyset [ \vec{\pvar{x}} \storearrow \vec{v}] [ \vec{\pvar{z}} \storearrow \nil]
     \\\\ (\sto_p, \hp), \scmd
      \baction_{\fictx} \oxok : \sthreadp{ \sto_q }{ \hp' }  \\
      \esem{\pexp'}{\sto_q} = \undefd \\\\ \sto' = \sto[\pvar{err} \storearrow ``\mathsf{ExprEval}"]
  }{
     \sthreadp{ \sto }{ \hp },
     \pfuncall{\pvar{y}}{\procname}{\vec{\pexp}} \baction_{\fictx}
     \oxerr: \sthreadp{\sto' }{ \hp' }
  }
\end{mathparpagebreakable}

\subsection{OX and UX Frame Properties}

For completeness, we state the OX and UX frame properties (in the familiar style of SL and ISL) for the full semantics:
\begin{theorem}[OX and UX Frame]\label{thm:frame} For $\cwf(\cmem)$ and $\cwf(\cmem_f)$:
\[
\begin{array}{l}
\csesemtrans{\sto, \cmem \memcomp \cmem_f}{\cmd}{\sto', \cmem'}{\fictx}{\result} \implies \\
\quad \exists \sto'', \cmem'', \result'.~
\csesemtrans{\sto, \cmem}{\cmd}{\sto'', \cmem''}{\fictx}{\result'} \land
 (\result' \neq \omiss \implies (\result' = \result \land \sto'' = \sto' \land \cmem' = \cmem'' \memcomp \cmem_f))
\\[1mm]
\csesemtrans{\sto, \cmem}{\cmd}{\sto', \cmem'}{\fictx}{\result} \land
\result \neq \omiss \land
\cmem' \memcomp \cmem_f \text{ defined} \implies \\
\quad \csesemtrans{\sto, \cmem \memcomp \cmem_f}{\cmd}{\sto', \cmem' \memcomp \cmem_f}{\fictx}{\result}
\end{array}
\]
\end{theorem}

\newpage
\section{Assertions and Function Specifications: Extended Definitions}\label{app:specifications}

In this section, we define all the necessary constructs required for defining the relevant program logics. In particular, we formally define the function specification for SL, ISL, and ESL, including the internalisation of function specifications.

\subsection{Additional Definitions}

The expression $\pv{A}$ denotes the set of program variables of $A$, and $\lv{A}$ denotes the set of logical variables of $A$.

\subsection{Logical Expression Evaluation}

We also define a logical expression evaluation function; some cases are provided below:
\[
\begin{array}{rcl}
\esem{\gv}{\subst,\sto} & = &\gv \\
\esem{\pvar{x}}{\subst,\sto} & = &\sto(\pvar{x})\\
\esem{\lvar{x}}{\subst,\sto} & = &
	\begin{cases}
		\subst(\lvar{x}), & \lvar{x} \in \dom(\subst) \\
		\undefd, &\mbox{otherwise}
	\end{cases}\\
\esem{\lexp \in \tau}{\subst,\sto} & = &
	\begin{cases}
		\true, & \esem{\lexp}{\subst,\sto} \in \tau \\
		\false, &\mbox{otherwise}
	\end{cases}\\
\esem{\lexp \in \Val}{\subst,\sto} & = &
	\begin{cases}
		\true, & \esem{\lexp}{\subst,\sto} \neq \undefd \\
		\false, &\mbox{otherwise}
	\end{cases}\\
\end{array}
\]
We sometimes omit the store and write $\esem{\lexp}{\subst}$ instead of $\esem{\lexp}{\subst,\sto}$ when the store irrelevant to the outcome of evaluating the logical expression $\lexp$.

\subsection{Assertion Satisfaction Relation}

To define the assertion satisfaction relation, we assume we are given a set $\preds$ containing predicate definitions. The elements of the set $(\pred, \predin, \predout, A) \in \preds$, denoted $\pred(\predin; \predout)~\{ A \} \in \preds$, have type $(\strings, \vec{\LVar}, \vec{\LVar}, \Assert)$, where $\pred$ is the predicate name, $\predin$ the input parameters, $\predout$ the output parameters, and $A$ the predicate body. Predicate definitions satisfy the following:

\begin{itemize}
\item No duplicates in $\predin$ or $\predout$

\item $\predin \cap \predout = \emptyset$

\item $\pv{A} = \emptyset$

\item $\lv{A} \subseteq \predin \cup \predout$

\item $A$ is of the form $A = \bigvee_i A_i$ where for each $A_i$ we have $\predout \subseteq \lv{A_i}$
\end{itemize}

We now define the assertion satisfaction relation. Most of the below cases are defined straightforwardly, but we note that for the logical expression case, we assume the heap is empty. The memory resource case depends on the resource satisfaction relation introduced in \S\ref{sec:specifications}. 
\[
\begin{array}{@{}l@{~}l@{~}c@{~\ }l}
  \subst, (\sto, \hp)  \models &
 \lexp & \Leftrightarrow & \esem{{\lexp}}{\subst,\sto} = \true \wedge \hp = \emptycmem \\
 & \AssTrue &\Leftrightarrow& \cwf(\hp) \\
 & A_1 \Rightarrow  A_2 &\Leftrightarrow& \subst, (\sto, \hp)  \models A_1 \Rightarrow  \subst, (\sto, \hp) \models A_2 \\
 & A_1 \lor A_2 &\Leftrightarrow& \subst, (\sto, \hp)  \models A_1 \lor  \subst, (\sto, \hp) \models A_2 \\
 & \exists \lvar{x} \ldotp A &\Leftrightarrow& \exists v \in \Val \ldotp \subst[\lvar{x} \mapsto v], (\sto, \hp) \models A \\
 & {\emp} &\Leftrightarrow& {\hp = \emptycmem } \\
 & A_1 \lstar A_2 &\Leftrightarrow& \exists \hp_1, \hp_2 \ldotp \hp = \hp_1 \memcomp \hp_2 \land \subst, (\sto, \hp_1) \models A_1 \wedge \subst, (\sto, \hp_2) \models A_2 \\
 & \resource(\vec{\lexp_1}; \vec{\lexp_2}) &\Leftrightarrow&

 \esem{\vec{\lexp_1}}{\subst,\sto} = \vec{v_1} \land \esem{\vec{\lexp_2}}{\subst,\sto} = \vec{v_2} \land \hp \models_{\resources} \resource(\vec{v_1}; \vec{v_2}) \\
 & \pred(\vec{\lexp_1}; \vec{\lexp_2}) &\Leftrightarrow& \subst, (\sto, \hp) \models A[\vec{\lexp_1}/\predin][\vec{\lexp_2}/\predout] \text{ for } \pred(\predin; \predout)~\{ A \} \in \preds
 \end{array}
\]

An assertion $A$ is {\it valid}, denoted $\models A$, iff $\forall \subst, \sto, \hp.~\subst, (\sto, \hp) \models A$.

\subsection{Function Specifications}

\paragraph{SL, ISL, and ESL quadruples} We use the following notations for SL, ISL, and ESL command quadruples:
\[
\quadruple{P}{\cmd}{\Qok}{\Qerr},
\quad \islquadruple{P}{\cmd}{\Qok}{\Qerr},
\quad \eslquadruple{P}{\cmd}{\Qok}{\Qerr},
\]
for command $\cmd$, precondition $P \in \Assert$, successful postcondition $\Qok \in \Assert$, and faulting postcondition~$\Qerr \in \Assert$. The corresponding triples are defined as degenerate cases of the above quadruples, with $\Qok = \false$ or $\Qerr = \false$.

The validity of SL, ISL, and ESL quadruples are defined as follows:
%
\[
  \begin{array}{l}
\fictx \models \islquadruple{P}{\cmd}{\Qok}{\Qerr} \eqdef (\forall \subst, \sto', \hp', \outcome.\\
\qquad \subst, (\sto', \hp') \models Q_\outcome \implies (\exsts{\sto,\hp}~\subst, (\sto, \hp) \models
    P~\land~(\sto, \hp), \cmd \baction_\fictx \outcome: (\sto', \hp')))\\[1mm]
\fictx \models \quadruple{P}{\cmd}{\Qok}{\Qerr} \eqdef (\forall \subst, \sto, \hp, \outcome, \sto', \hp'. \\
\qquad \subst, (\sto, \hp) \models P \land (\sto, \hp), \cmd \baction_{\fictx} \outcome: (\sto', \hp') \implies 
(\outcome \neq \oxm \land \subst, (\sto', \hp') \models Q_\outcome)) \\[1mm]
\fictx \models \eslquadruple{P}{\cmd}{\Qok}{\Qerr} \eqdef \\
\qquad \fictx \models \quadruple{P}{\cmd}{\Qok}{\Qerr} \land \fictx \models \islquadruple{P}{\cmd}{\Qok}{\Qerr}
\end{array}
\]
Informally, an SL quadruple is valid if all terminating executions from $P$ of $\cmd$ are contained in $\Qok$ and $\Qerr$; an ISL quadruple is valid if all models of $\Qok$ and $\Qerr$ are reachable from $P$ through $\cmd$; and an ESL quadruple is valid if both corresponding SL and ISL quadruples are valid.

We write $\genquadruple{P}{\!}{\Qok}{\Qerr}$ to denote the type of specification may be SL, ISL, or ESL.

\paragraph{SL, ISL, and ESL function specifications} Function
specifications comprise 
{\em external} specifications, which describe the interface
of the function toward its callers, 
and {\em internal} specifications, satisfied by
the function body.
An {external} specification has the form $\genquadruple{\vec{\pvar{x}} = \vec x \lstar P}{}{\Qok}{\Qerr}$~where:
\begin{itemize}
\item $\vec{\pvar{x}}$ are distinct, $\vec x$ are distinct, and $\pv{P} = \emptyset$; 
\item $\pv{\Qok} = \{\pvar{ret}\}$ or $\Qok = \false$;
\item and $\pv{\Qerr} = \{\pvar{err}\}$ or $\Qerr = \false$.
\end{itemize}
The set of external specifications is denoted by $\especs$. A function specification context (also: specification context),~$\fsctx$, is a finite partial function from function identifiers to a finite set of external specifications: $\fsctx \in \fids \rightharpoonup_{\mathit{fin}} \mathcal{P}(\especs).$

We define an {\em internalisation} function for external specifications, adapting the definition introduced in ESL~\cite{esl}. Given implementation context~$\fictx$ and function $\pfunction{\fid}{\lst{\pvar x}}{\cmd; \preturn{\pexp}} \in  \gamma$, an ISL specification internalisation, $\fext_{\fictx, \fid}^\macUX$, is defined by: 
\[
\begin{array}{l}
 \fext_{\fictx, \fid}^{\macUX} (\islquadruple{
 \mathit{\vec{\pvar x} = \vec x \lstar P}}{\!}{\Qok}{\Qerr}) \defeq \\ \quad
 \{~\islquadruple{\mathit{\vec{\pvar x} = \vec x \lstar P} \lstar \vec {\pvar z}\doteq\nil}{\!}{\Qok'}{\Qerr'} : \\ \qquad \quad
 \models \Qok' \Rightarrow \pexp \in \Val \lstar \AssTrue, \\ \qquad \quad
 \models \Qok \Rightarrow \exsts{\vec{p}} \Qok'[\vec{p} / \vec{\pvar p}]\lstar\pvar{ret}\doteq \pexp[\vec{p} / \vec{\pvar p}], \\ \qquad \quad
 \models \Qerr \Rightarrow \exsts{\vec{p}} \Qerr'[\vec{p} / \vec{\pvar p}]~\}
\end{array}
\]
where 
$\pvvar z=\pv{\cmd}\backslash \{\vec{\pvar{x}} \} $, $\vec{\pvar p} =
\{\vec{\pvar{x}} \} \uplus
\{\pvvar z\}$, and the logical variables $\vec p$ are fresh with respect to $\Qok'$ and $\Qerr'$.
The internal precondition simply extends the
external one with locals initialised to $\nil$. 
As UX reasoning cannot lose information, internal
postconditions will contain information about program variables, 
whereas external ones must not, given variable scoping. 
The transition from internal to external postconditions forgets
program variables by substituting them with fresh logical variables,
and uses backwards consequence. SL specification internalisation, denoted by
$\fext_{\fictx, \fid}^\macOX$, is defined analogously, with the two latter
implications reversed to use forward consequence (e.g.,
$\models \Qok \Leftarrow \exsts{\vec{p}} \Qok'[\vec{p} / \vec{\pvar
    p}]\lstar\pvar{ret}\doteq \pexp[\vec{p} / \vec{\pvar p}]$). ESL specification internalisation instead uses equivalence.
    
\paragraph{Function environments} A function environment, $(\fictx,\fsctx)$, is a pair comprising an implementation context $\fictx$ and a specification context $\fsctx$.
A function environment $(\fictx,\fsctx)$ is {valid}, written $\models (\fictx, \fsctx)$, iff
every function in $\fsctx$ has an implementation in $\fictx$ and for every external specification
in $\fsctx$ there exists a corresponding internal specification valid under $\fictx$. Lastly, a function specification is valid in a context $\fsctx$, denoted by $\fsctx \models \genquadruple{P}{\fid(\vec{\pvar{x}})}{\Qok}{\Qerr}$ iff for all $\fictx$ such that $\models (\fictx,\fsctx)$ it holds that
$\fictx \models \genquadruple{P}{\fid(\vec{\pvar{x}})}{\Qok}{\Qerr}$.

\newpage

\section{Symbolic Definitions and Semantics}%
\label{app:symbolic}

This section provides additional symbolic definitions and additional rules of the symbolic semantics $\cseenginetrans{}{\fsctx}{\oracle}{\sst}{\scmd}{\outcome}{\oracle'}{\sst'}$. For the rules, in order to keep clutter to a minimum, we omit mentioning the oracles for the rules unless the oracle plays a role, i.e., $\csesemtransabstract{\sst}{\scmd}{\sst'}{\fsctx}{\outcome}$ means $\cseenginetrans{}{\fsctx}{\oracle}{\sst}{\scmd}{\outcome}{\oracle}{\sst'}$.

We use the following notation: given a (countable) set $S$, $(s, \oracle') \in \oracle(S)$ denotes an angelic choice $s \in S$ by oracle $\oracle$ with resulting shifted oracle $\oracle' = \lambda{}n.~\oracle(n + 1)$.

\subsection{Additional Symbolic Satisfaction Relations}

The symbolic store satisfaction relation is as follows:
\[
\subst, \sto \models_\text{Sto} \ssto \iff \forall \pvar{x} \in \dom(\ssto).~ \esem{\ssto(\pvar{x})}{\subst} = \sto(\pvar{x})
\]
Let $\sps$ be a multiset of symbolic predicates $\{ \pred^1(\sexpin^1; \sexpout^1), \dots, \pred^n(\sexpin^n; \sexpout^n) \}$. Now, the symbolic predicate satisfaction relation is as follows:
\[
\subst, \cmem \models_\text{Pred} \sps \iff \exists \hp_1, \dots, \hp_n.~\hp=\hp_1 \memcomp \dots \memcomp \hp_n \land \forall i \in \{ 1, \dots, n \}.~ \subst, \hp_i \models \pred^i(\sexpin^i; \sexpout^i)
\]

\subsection{Symbolic Expression Evaluation}

We define the symbolic expression evaluation function. The role of this function is to return a logical expression where all the program variables are evaluated with respect to the symbolic store. Note that the symbolic expression evaluation function does not take in or produce a path condition and does not branch -- this obligation is, instead, handled by the symbolic semantics of commands, as we will see further below.
In the main text, symbolic expression evaluation was introduced as evaluating program expressions. Because of the fold and unfold commands introduced below (which have expressions that may contain logical variables), symbolic expression evaluation need to take in logical expressions rather than only program expressions. For other commands that take in standard expressions, these can be straightforwardly translated to logical expressions~without~issue.

We give a few sample rules below, other rules are similarly straightforward:
\begin{mathpar}
\small
\inferrule{}
{\cseeval{v}{\ssto}{}{v}{}}
\and
\inferrule{}
{\cseeval{\pvar{x}}{\ssto}{}{\ssto(\pvar x)}{}}
\and
\inferrule{}
{\cseeval{\lvar{x}}{\ssto}{}{\lvar{x}}{}}
\and
\inferrule{
	\cseeval{\lexp_1}{\ssto}{}{\sexp_1}{}\quad
	\cseeval{\lexp_2}{\ssto}{}{\sexp_2}{}
}
{\cseeval{\lexp_1~{+}~\lexp_2}{\ssto}{}{\sexp_1 + \sexp_2}{}}
\and
\inferrule{
	\cseeval{\lexp_1}{\ssto}{}{\sexp_1}{}\quad
	\cseeval{\lexp_2}{\ssto}{}{\sexp_2}{}
}
{\cseeval{\lexp_1~{/}~\lexp_2}{\ssto}{}{\sexp_1 / \sexp_2}{}}
\end{mathpar}

\subsection{Core Commands}

The rules for the core commands, that is, commands that are not memory-action commands or based on \mac and \produce, are as follows:

\begin{mathparpagebreakable}
\small
\inferrule[\textsc{Skip}]
{ }
{\csesemtransabstractm{\sst}{\pskip}{\sst}{\fsctx}{m}{\osucc}}
\and
\inferrule[\textsc{Assign}]
{\cseeval{\pexp}{\ssto}{}{\sexp}{} \\ \ssto' = \ssto[\pvar{x} \storearrow \sexp] \\\\
\spc' = (\ssto(\pvar{x}) \in \Val \land \spc)}
{\csesemtransm{\ssto, \smem, \spred, \spc}{\passign{\pvar{x}}{\pexp}}{\ssto', \smem, \spred, \spc'}{\fsctx}{m}{\osucc}}
\and
\inferrule[\textsc{Assign-Err-Eval}]
{\cseeval{\pexp}{\ssto}{}{\sexp}{} \and \ssto' = \ssto[\pvar{err} \storearrow \mathstr{\mathsf{ExprEval}}] \\\\
\spc' = (\sexp \not\in \Val \land \spc)}
{\csesemtransm{\ssto, \smem, \spred \spc}{\passign{\pvar{x}}{\pexp}}{\ssto', \smem, \spred, \spc'}{\fsctx}{m}{\oerr}}
\and
%
%



%
\inferrule[\textsc{If-Then}]
{\cseeval{\pexp}{\ssto}{}{\sexp}{} \and \spc' = (\sexp = \true \land \spc) \\\\
\cseenginetrans{m}{\fsctx}{\oracle}{(\ssto, \smem, \spred, \spc')}{\scmd_1}{\outcome}{\oracle'}{\sst'}
}
{
\cseenginetrans{m}{\fsctx}{\oracle}{(\ssto, \smem, \spred, \spc)}{\pifelse{\pexp}{\scmd_1}{\scmd_2}}{\outcome}{\oracle'}{\sst'}
}
\and
\inferrule[\textsc{If-Else}]
{\cseeval{\pexp}{\ssto}{}{\sexp}{} \and \spc' = (\sexp = \false \land \spc) \\\\
\cseenginetrans{m}{\fsctx}{\oracle}{(\ssto, \smem, \spred, \spc')}{\scmd_2}{\outcome}{\oracle'}{\sst'}
}
{
\cseenginetrans{m}{\fsctx}{\oracle}{(\ssto, \smem, \spred, \spc)}{\pifelse{\pexp}{\scmd_1}{\scmd_2}}{\outcome}{\oracle'}{\sst'}
}
\and
\inferrule[\textsc{If-Err-Eval}]
{\cseeval{\pexp}{\ssto}{}{\sexp}{} \and \spc' = (\sexp \not\in \Val \land \spc) \\\\
 \ssto' = \ssto[\pvar{err} \storearrow \mathstr{\mathsf{ExprEval}}]}
{\csesemtransm{\ssto, \smem, \spred, \spc}{\pifelse{\pexp}{\scmd_1}{\scmd_2}} {\ssto', \smem, \spred, \spc'}{\fsctx}{m}{\oerr}}
\and
\inferrule[\textsc{If-Err-Type}]
{\cseeval{\pexp}{\ssto}{}{\sexp}{} \and \spc' = (\sexp \not\in \bools \land \sexp \in \Val \land \spc) \\\\
\ssto' = \ssto[\pvar{err} \storearrow \mathstr{\mathsf{Type}}]}
{\csesemtransm{\ssto, \smem, \spred, \spc}{\pifelse{\pexp}{\scmd_1}{\scmd_2}} {\ssto', \smem, \spred, \spc'}{\fsctx}{m}{\oerr}}
\and
\inferrule[\textsc{Seq}]
{\cseenginetrans{m}{\fsctx}{\oracle}{\sst}{\scmd_1}{\oxok}{\oracle'}{\sst'} \\\\
\cseenginetrans{m}{\fsctx}{\oracle'}{\sst'}{\scmd_2}{\outcome}{\oracle''}{\sst''}}
{\cseenginetrans{m}{\fsctx}{\oracle}{\sst}{\scmd_1;\scmd_2}{\outcome}{\oracle''}{\sst''}}
\and
\inferrule[\textsc{Seq-Err}]
{\cseenginetrans{m}{\fsctx}{\oracle}{\sst}{\scmd_1}{\outcome}{\oracle'}{\sst'} \quad \result \neq \osucc}
{\cseenginetrans{m}{\fsctx}{\oracle}{\sst}{\scmd_1;\scmd_2}{\outcome}{\oracle'}{\sst'}}
\end{mathparpagebreakable}

\subsection{Memory-action Command}

The rules for the memory-action command are as follows:

\begin{mathparpagebreakable}
\small
  \inferrule[\textsc{Memory-Action}]
  {
    \cseeval{\vec{\pexp}}{\ssto}{}{\vec{\sexp}}{} \and
    \smem.\act(\vec{\sexp}) \rightsquigarrow \oxok : (\smem', \spc', \vec{\sexp}') \\\\
    |\vec{\sexp}'| = |\vec{\pvar{x}}| \and
    \spc'' = (\ssto(\vec{\pvar{x}}), \vec{\sexp}, \vec{\sexp}' \in \Val \land \spc' \land \spc)
  }
  {
     \csesemtransm{\ssto, \smem, \spred, \spc}{\vec{\pvar{x}} := \act(\lst\pexp)}{\ssto [\vec{\pvar{x}} \storearrow \vec{\sexp}' ], \smem', \spred, \spc''}{\fsctx}{m}{\oxok}
  }
\and
\inferrule[\textsc{Memory-Action-Err-Length}]
  {
    \cseeval{\vec{\pexp}}{\ssto}{}{\vec{\sexp}}{} \and
    \smem.\act(\vec{\sexp}) \rightsquigarrow \oxok : (\smem', \spc', \vec{\sexp}') \\\\
    |\vec{\sexp}'| \neq |\vec{\pvar{x}}| \and
    \spc'' = (\vec{\sexp}, \vec{\sexp}' \in \Val \land \spc' \land \spc) \\\\
    \ssto' = \ssto[\pvar{err} \storearrow \mathstr{\mathsf{ActionArgsLength}}]
  }
  {
     \csesemtransm{\ssto, \smem, \spred, \spc}{\vec{\pvar{x}} := \act(\lst\pexp)}{\ssto', \smem', \spred, \spc''}{\fsctx}{m}{\oxok}
  }
\and
  \inferrule[\textsc{Memory-Action-Err}]{
    \cseeval{\vec{\pexp}}{\ssto}{}{\vec{\sexp}}{} \and
    \smem.\act(\vec{\sexp}) \rightsquigarrow \outcome : (\smem', \spc', \vec{\sexp}') \\\\
    \outcome \neq \oxok \and
    \spc'' = (\vec{\sexp} \in \Val \land \spc' \land \spc) \\\\
    \ssto' = \ssto[\pvar{err} \storearrow \vec{\sexp}']
  }{
     \csesemtransm{\ssto, \smem, \spred, \spc}{\vec{\pvar{x}} := \act(\lst\pexp)}{\ssto', \smem', \spred, \spc''}{\fsctx}{m}{\outcome}
  }
\and
  \inferrule[\textsc{Memory-Action-Err-Eval}]
  {
    \cseeval{\vec{\pexp}}{\ssto}{}{\sexp}{} \and \spc' = (\vec{\sexp} \not\in \Val \land \spc) \\\\
    \ssto' = \ssto[\pvar{err} \storearrow \mathstr{\mathsf{ExprEval}}]
  }
  {
     \csesemtransm{\ssto, \smem, \spred, \spc}{\vec{\pvar{x}} := \act(\lst\pexp)}{\ssto', \smem, \spred, \spc'}{\fsctx}{m}{\oxerr}
  }
\end{mathparpagebreakable}

\noindent
We formally state the OX/UX soundness of memory actions for IProp.~\ref{iprop:mem-sound}:
\[
\begin{array}{lll}
 \text{(OX)} & \text{If } & \cmem.\act(\vec{v}) \rightsquigarrow \outcome : (\cmem', \vec{v}') \text{ and } \subst, \hp \memmodels \smem \text{ and } \esem{\vec{\sexp}}{\subst}=\vec{v}\ \text{and } \\
  & & (\forall \outcome, \smem', \spc', \vec{\sexp}'.\ \smem.\act(\vec{\sexp}) \rightsquigarrow \outcome : (\smem', \spc', \vec{\sexp}') \text{ and } \esem{\spc'}{\subst} = \true \text{ and } \esem{\vec{\sexp}'}{\subst} \neq \undefd \implies \outcome \neq \oxabort) \\
& \text{then } &    \exists \smem', \spc', \vec{\sexp}', \subst'.\ \smem.\act(\vec{\sexp}) \rightsquigarrow \outcome : (\smem', \spc', \vec{\sexp}') \text{ and } \subst'|_{\lv{\smem}} = \subst \\
& & \text{ and } \esem{\spc'}{\subst'}=\true \text{ and } \subst', \hp' \memmodels \smem' \text{ and } \esem{\vec{\sexp}'}{\subst'}=\vec{v}' \\[2mm]
\text{(UX)}&  \text{If } & \smem.\act(\vec{\sexp}) \rightsquigarrow \outcome : (\smem', \spc', \vec{\sexp}') \text{ and } \outcome \neq \oxabort \text{ and } \esem{\vec{\sexp}}{\subst}=\vec{v} \text{ and} \\
 & & \esem{\spc'}{\subst}=\true \text{ and } \subst, \hp' \memmodels \smem'\text{ and } \esem{\vec{\sexp}'}{\subst}=\vec{v}' \\
 & \text{then } & \exists \hp.\ \subst, \hp \memmodels \smem \text{ and } \cmem.\act(\vec{v}) \rightsquigarrow \outcome : (\cmem', \vec{v}')
\end{array}
\]

\subsection{Consume and Produce Operations Soundness Properties}

Below, we give the soundness properties for the \mac and \produce~operations. We assume $\pv{A} = \emptyset$ for all cases. We also formally state the OX/UX soundness of $\resconsume$ and $\resproduce$ for IProp.~\ref{iprop:cp-sound}.

\begin{align*}
\intertext{OX soundness of \mac:}
&\begin{tabular}{ll}
   If & \(m\in \{OX,EX\}\) and \\
    &  \(( \forall \outcome, \oracle', \ssubst', \sst_f. ~   \mac(m, O, A, \ssubst, \sst)\rightsquigarrow (o, O', \ssubst', \sst_f) \text{ and } \sat (\sst_f) \implies o\neq abort ) \) and \\
    &  \(\subst, \sto, \hp  \models \sst \) \\
   then & \( \exists \hp_A, \hp_f, O', \ssubst', \sst_f.~\mac(m, O, A, \ssubst, \sst) \rightsquigarrow  (\oxok, O', \ssubst', \sst_f) \text{ and } \hp=\hp_A\cdot \hp_f \) and\\
    & \qquad \(\subst, \sto, \hp_A \smodels \ssubst'(A) \text{ and } \ \subst,  \sto, \hp_f \models \sst_f\)
     \end{tabular}
\intertext{UX soundness of \mac:}
&\begin{tabular}{ll}
If & \(\mac(m, O, A, \ssubst, \sst) \rightsquigarrow  (ok, O', \ssubst', \sst_f)\) and \\
&  \(\subst, \sto, \hp_A \models \ssubst'(A)\) and \(\subst,  \sto, \hp_f \models \sst_f\) and \((\hp_A\cdot \hp_f)\) is defined \\
then &
     \(
   \subst, \sto, (\hp_A\cdot \hp_f)\models \sst\)
\end{tabular}
\intertext{OX soundness of \produce:}
&\begin{tabular}{ll}
 If &  $\lv{A}\subseteq \dom(\ssubst)$ and $\subst, \sto, \hp_A \smodels \ssubst(A)$ and $\subst, \sto,\hp_f \models \sst_f $ and $(\hp_A\cdot \hp_f)$ is defined \\
 then &  $\exists \subst',\sst.~\subst'|_{\lv{\sst_f}}=\subst \text{ and } \produce(A, \ssubst, \sst_f) \rightsquigarrow \sst \text{ and } \subst', \sto, (\hp_A \cdot \hp_f) \models \sst$
\end{tabular}
\intertext{UX soundness of \produce:}
&\begin{tabular}{ll}
  If &
     \(\produce(A, \ssubst, \sst_f) \rightsquigarrow \sst\)
      and~ \(  \subst, \sto,\hp \models \sst\)\\
  then &
  \(
   \exists \hp_A,\hp_f. ~\hp=\hp_A\cdot \hp_f
    \) and \( \subst, \sto, \hp_A \smodels \ssubst(A) \text{ and } \subst,  \sto, \hp_f
    \models \sst_f
  \)
\end{tabular}
\intertext{OX soundness of $\resconsume$ operation:}
&\begin{array}{ll}
   \text{ If } & m \in \{OX,EX\} \text{ and } \\
   & (\forall \outcome, \oracle', \sexpout, \smem_f,\spc_\textit{fi}, \spc_f.  ~ \resconsume(m, O, \resource, \sexpin, \smem) \rightsquigarrow (\outcome, O', \sexpout,(\smem_f,\spc_\textit{fi}, \spc_f)) \text{ and} \\
   & \quad (\esem{\spc_\textit{fi}}{\subst}=\true \implies (\exists \hp_f.~ \esem{\spc_f}{\subst}=\true  \text{ and }  \subst, \hp_f\memmodels \smem_f)) \implies \outcome\neq abort)
\text{ and } \\
& \subst, \hp \memmodels \smem \text{ and } \esem{\sexpin}{\subst} = \vin\\
\text{ then } & \\
& \left(
\begin{array}{l}
\exists O', \sexpout, \vout, \smem_f, \spc_{\mathit{fi}}, \spc_f.
~ \resconsume(m, O, \resource, \sexpin, \smem) \rightsquigarrow (ok, O',  \sexpout, (\smem_f,\spc_{\mathit{fi}}, \spc_f)) \text{ and} \\
\qquad (\esem{\spc_\mathit{fi}}{\subst} = \true \implies \\
\qquad\quad \exists \hp_\resource, \hp_f. ~ \esem{\sexpout}{\subst}=\vout \text{ and } \hp_r \resmodels \resource(\vin;\vout) \text{ and } \subst, \hp_f \memmodels \smem_f \text{ and} \\
\qquad\qquad \hp=\hp_\resource\cdot \hp_f \text{ and } \esem{\spc_f}{\subst} =\true)
\end{array}
\right)
\end{array}
\intertext{UX soundness of $\resconsume$ operation:}
&\begin{array}{ll}
\text{If}   &\resconsume(m, O, \resource, \sexpin, \smem)     \rightsquigarrow (ok, O', \sexpout,(\smem_f,\spc_{\mathit{fi}}, \spc_f)) \text{ and } \esem{\sexpin}{\subst} = \vin \text{ and }  \esem{\sexpout}{\subst}= \vout
   \\
& \text{and } \esem{\spc_f}{\subst}=\true \text{ and }  \subst, \hp_f\memmodels \smem_f \text{ and } \hp_r \resmodels \resource(\vin; \vout)\text{ and } (\hp_r\cdot \hp_f) \text{ is defined } \\
\text{then}&  \esem{\spc_{\mathit{fi}}}{\subst}=\true \text{ and }  \subst, (\hp_r\cdot \hp_f) \memmodels \smem
\end{array}
\intertext{OX soundness of $\resproduce$ operation:}
&\begin{array}{ll}
\text{If }& \subst,\hp_f \memmodels \smem_f \text{ and } \esem{\sexpin}{\subst}=\vin \text{ and } \esem{\sexpout}{\subst}=\vout \text{ and } \hp_r \resmodels \resource (\vin;\vout) \\ & \text{and } (\hp_r \cdot \hp_f) \text{ is defined} \\
  \text{then }&  \exists \smem, \spc. 
    ~\resproduce(\resource,\sexpin,\sexpout, \smem_f) \rightsquigarrow (\smem,\spc) \text{ and } \subst, (\hp_r\cdot \hp_f)\memmodels \smem \text{ and }\esem{\spc}{\subst} =\true\\
\end{array}
\intertext{UX soundness of $\resproduce$ operation:}
&\begin{array}{ll}
\text{If } & \resproduce(\resource,\sexpin,\sexpout, \smem_f)\rightsquigarrow (\smem,\spc) \text{ and }\esem{\sexpin}{\subst}=\vin\text{ and  } \esem{\sexpout}{\subst}=\vout  \text{ and } \\
& \esem{\spc}{\subst}=\true \text{ and } \subst, \hp\memmodels \smem\\ 
 \text{then } & \exists \hp_r, \hp_f. ~ \hp=\hp_r\cdot \hp_f \text{ and }  \hp_r \resmodels \resource(\vin;\vout) \text{ and } \subst, \hp_f\memmodels \smem_f
\end{array}
\end{align*}

\subsection{Function-call Command}

We implement the standard function-call rule for the consume-produce engine architecture; e.g., see \citet[Def.~4.8]{Jacobs15} for VeriFast, \citet[Fig.~3.6]{Schwerhoff16} for Viper, and \citet[Sec.~5]{cse1} for Gillian. In short, given a function call $\pfuncall{\pvar{y}}{\fid}{\lst{\pexp}}$ to execute, the function-call rule angelically picks a function specification from $\fsctx(f)$ compatible with the current execution mode $m$, to exemplify, say, $\quadruple{\vec{\pvar{x}} = \vec{x} \lstar P}{f(\vec{\pvar{x}})}{\Qok}{\Qerr} \in \fsctx(f)$, and first consumes the precondition $\vec{\pvar{x}} = \vec{x} \lstar P$ and then produces the postconditions $\Qok$ and $\Qerr$. The OX/UX soundness of the rule follows from the OX/UX soundness of \mac and \produce~(Lem.~\ref{lem:cp}).

We now give the successful rule for function calls, specialised to only produce $\Qok$ to simplify the presentation (the full rule produces both $\Qok$ and $\Qerr$). We use the following notation: given some function identifier $f$ and a function specification context $\fsctx$, $\fsctx(f)|_m$ returns a set of function specifications for the function identifier $f$ compatible with mode $m$ (e.g., ESL specifications are compatible with both OX and UX mode). The uninteresting parts of the rule, which simply involves updates and setup, are greyed out:

\begin{mathpar}
\small
\infer
{\cseenginetrans{m}{\fsctx}{\oracle}{(\ssto, \smem, \spred, \spc)}{\passign{\pvar y}{\fid(\vec{\pexp})}}{\osucc}{\oracle''}{(\ssto[\pvar{y} \mapsto \sym{r}], \smem', \spred', \spc''''')}}
{
\begin{array}{l@{\hspace*{+0.0cm}}r}
(\genquadruple{\vec{\pvar{x}} = \vec{x} \lstar P}{\fid(\vec{\pvar{x}})}{\Qok}{\Qerr}, \oracle') \in \oracle(\fsctx(f)|_m) & \text{get function specification} \\
\cseeval{\vec{\pexp}}{\ssto}{}{\vec{\sexp}}{} \text{~and~} \hat{\theta} = [\vec{x} \mapsto \vec{\sexp}] & \text{evaluate function parameters} \\
\textcolor{gray}{\spc' = (\vec{\sexp} \in \Val \land \spc)} & \textcolor{gray}{\text{successful evaluation in path condition}} \\
\mac(m, \oracle', P, \hat{\theta}, (\ssto, \smem, \spred, \spc')) \rightsquigarrow (\oxok, \oracle'', \hat{\theta}', (\ssto, \smem_f, \spred_f, \spc''))& \text{consume precondition} \\
\textcolor{gray}{r \text{ and } \sym{r} \text{ fresh}} & \textcolor{gray}{\text{fresh variables for return value}} \\
\textcolor{gray}{\Qok' = \Qok[r/\pvar{ret}] \text{~and~} \ssubst'' = \ssubst'[r \mapsto \sym{r}]} & \textcolor{gray}{\text{set up return value}} \\
\textcolor{gray}{\spc''' = (\sym{r} \in \Val \land \spc'')} & \textcolor{gray}{\text{include $\sym{r}$ in path condition}} \\
\produce(\Qok', \ssubst'', (\ssto,\smem_f, \spred_f, \spc''')) \rightsquigarrow  (\ssto, \smem', \spred', \spc'''') & \text{produce postcondition} \\
\textcolor{gray}{\spc''''' = (\ssto(\pvar{y}) \in \Val \land \spc'''')} & \textcolor{gray}{\text{include \textsf{y} in path condition}}
\end{array}
}
\end{mathpar}

\subsection{Fold and Unfold Commands}

We implement the standard consume-produce rules for folding and unfolding user-defined predicates; e.g., see \citet[Def.~4.8]{Jacobs15} for VeriFast, \citet[Fig.~3.6]{Schwerhoff16} for Viper, and \citet[App.~B in extended paper]{cse1} for Gillian.

Folding and unfolding commands are expressed in the following extended syntax:
\[
\cmd ::= \dots \mid \pfold{\pred}{\vec{\lexp}} \mid \punfold{\pred}{\vec{\lexp}}
\]
where $\vec{\lexp} \in \vec{\LExp}$.

In the concrete semantics, fold and unfold are no-ops. Formally, that is:
\begin{mathparpagebreakable}
\infer{\cst, \pfold{p}{\vec{\lexp}} \baction_{\fictx} {\oxok} : \cst}{}
\and
\infer{\cst, \punfold{p}{\vec{\lexp}} \baction_{\fictx} {\oxok} : \cst}{}
\end{mathparpagebreakable}

The symbolic semantics of the commands is defined using \mac and \produce. In short, given a predicate definition $\pred(\predin; \predout)~\{\bigvee_i A_i\}$, the folding rule angelically picks a disjunct $A_i$, consumes the disjunct, and produces a symbolic predicate for the predicate, whereas the unfolding rule angelically picks a symbolic predicate, consumes the symbolic predicate, and produces all of the disjuncts $A_i$ of the predicate.

We now give the formal symbolic rules. The notion $\sst.\consPure(m, \lexp) \rightsquigarrow \sst'$ is consume for Boolean assertions (also known as pure assertions). For the rules here, the role of $\consPure$ (in $\macOX$ mode) is to check whether the path condition of $\sst$ implies the given Boolean assertion and does not change the state. If the implication does not hold, $\consPure$ will not return. Now, let $\sst = (\ssto, \smem, \spred, \spc)$:

\begin{minipage}{0.5\textwidth}
\begin{prooftree}
    \AxiomC{\(
    \begin{array}{l}
      \pred(\predin; \predout)~\{\bigvee_i A_i\} \in \preds \\
      \cseeval{\vec{\lexp}}{\ssto}{}{\vec{\sexp}}{} \qquad \spc' = (\vec{\sexp} \in \Val \land \spc) \\
      (A_i, \oracle') \in \oracle(\bigvee_i A_i) \\
      \ssubst = [\predin \mapsto \vec{\sexp} ] \\
      \mac(\macOX, \oracle', A_i, \ssubst, (\ssto, \smem, \spred, \spc')) \rightsquigarrow \\ \quad (\oxok, \oracle'', \ssubst', (\ssto,\smem',\spred',\spc'')) \\
      \spred'' = \{\pred(\vec{\sexp}; \ssubst'(\predout))\} \cup \hat{\mathcal{P}}'\\
      \spc''' = (\ssubst'(\predout) \in \Val \land \spc'')
    \end{array}
    \)
    }
    \UnaryInfC{
    \(\cseenginetrans{\macOX}{\fsctx}{\oracle}{\sst}{\pfold{\pred}{\vec{\lexp}}}{\osucc}{\oracle''}{(\ssto, \smem', \spred'', \spc''')}\)}
\end{prooftree}
\end{minipage}
~
\begin{minipage}{0.5\textwidth}
\begin{prooftree}
\AxiomC{\(
\begin{array}{l}
\pred(\predin; \predout)~\{\bigvee_i A_i\}\in \preds \\
\cseeval{\vec{\pexp}}{\ssto}{}{\vec{\sexp}}{} \qquad
\spc' = (\vec{\sexp} \in \Val \land \spc) \\
\spred \neq \emptyset \qquad
(\pred(\sexpin; \sexpout), \oracle') \in \oracle(\spred) \\
\sst.\consPure(\macOX, \vec{\sexp} = \vec{\sexp}_{in}) \rightsquigarrow \sst \\
\spred' = \spred \setminus \{\pred(\sexpin; \sexpout)\} \\
\ssubst = [\predin \mapsto \vec{\sexp}_{in}, \predout \mapsto \sexpout] \\
\produce(A_i,\ssubst,(\ssto,\smem,\spred',\spc')) \rightsquigarrow \\ \quad (\ssto,\smem',\spred'',\spc'')
\end{array}
\)}
\UnaryInfC{
\(\cseenginetrans{\macOX}{\fsctx}{\oracle}{\sst}{\punfold{\pred}{\vec{\lexp}}}{\osucc}{\oracle'}{(\ssto, \smem', \spred'', \spc'')}\)}
\end{prooftree}
\end{minipage}

\vspace{0.15cm}
The OX soundness of the rules follows from the OX soundness of \mac and \produce~(Lem.~\ref{lem:cp}).

\subsection{Function-Specification Verification Procedure}

We have formalised the standard OX function verification procedure from the consume-produce literature on top of our CSE engine, which we here call $\texttt{verifyOX}$; e.g., see \citet[Def.~4.7]{Jacobs15} for VeriFast, \citet[Fig.~3.19]{Schwerhoff16} for Viper, and \citet[App.~G in extended paper]{cse1} for Gillian. In short, given a specification context $\fsctx$, a function $\pfunction{f}{\vec{\pvar{x}}}{\scmd; \preturn{\pexp}}$, and an SL specification $t_f = \tripleok{\vec{\pvar{x}} = \vec{x} \lstar P}{f(\vec{\pvar{x}})}{\Qok}$, if $\texttt{verifyOX}(\fsctx, f, t_f)$ terminates successfully, then $\fsctx$ can soundly be extend to $\fsctx[f \mapsto \fsctx(f) \cup \{ t_f \}]$. The procedure $\texttt{verifyOX}$ succeeds if it can produce the precondition $\vec{\pvar{x}} = \vec{x} \lstar P$, symbolically execute the function body $\cmd$ (and return expression $\pexp$), and consume the postcondition $\Qok$. We make one simplification orthogonal to the main topic of the paper, i.e., memory-model parametricity: we do not extend the specification context $\fsctx$ with the new specification $t_f$ before verification, meaning that for now our verification procedure only handles non-recursive functions. The soundness of $\texttt{verifyOX}$ follows from the OX soundness of our CSE engine~(Thm.~\ref{thm:ox-sound}) and \mac and \produce~(Lem.~\ref{lem:cp}).

\newpage
\section{Details of Memory-Model Instances}\label{app:memory-models}

This section contains additional definitions, information, and rules of the various memory models we discuss in the main text. To minimise clutter, we omit oracles and $\spc_i$ in the rules of resource consume if the oracle is not used and $\spc_i = \true$, i.e., $\resconsume(m, r, \sexpin, \smem) \rightsquigarrow (\outcome, \sexpout, (\smem', \spc))$ means $\resconsume(m, O, r, \sexpin, \smem) \rightsquigarrow (\outcome, O, \sexpout, (\smem', \true, \spc))$.

\subsection{Linear Memory Model (Running Example)}

Our running example linear memory model, for simplicity, use simplified new and free actions without arguments, i.e., we allocate and deallocate only one cell at a time. See, e.g., the block-offset memory model for C in \S\ref{sec:more-memory-models} for a memory model which can allocate and deallocate multiple cells at a time.

\subsubsection{Concrete Semantics}\hfill

\begin{mathparpagebreakable}
\footnotesize
\inferrule[\textsc{Lookup}]
    {\hp(n) = v}
    {\cmem.\mathtt{lookup}([n]) \rightsquigarrow \oxok : (\cmem, [v])}
\and
\inferrule[\textsc{Lookup-Err-Type}]
    {v \notin \nats}
    {\cmem.\mathtt{lookup}([v]) \rightsquigarrow \oxerr : (\cmem, [``\mathsf{Type}"])}
\and
\inferrule[\textsc{Lookup-Err-Missing}]
    {n \notin \dom(\hp)}
    {\cmem.\mathtt{lookup}([n]) \rightsquigarrow \oxm : (\cmem, [``\mathsf{MissingCell}", n])}
\and
\inferrule[\textsc{Lookup-Err-Use-After-Free}]
    {\hp(n) = \cfreed}
    {\cmem.\mathtt{lookup}([n]) \rightsquigarrow \oxerr : (\cmem, [``\mathsf{UseAfterFree}", n])}
\and
\inferrule[\textsc{Mutate}]
    {\hp(n) \in \Val}
    {\cmem.\mathtt{mutate}([n, v]) \rightsquigarrow \oxok : (\cmem[n \mapsto v], [])}
\and
\inferrule[\textsc{Mutate-Err-Type}]
    {v \notin \nats}
    {\cmem.\mathtt{mutate}([v, v']) \rightsquigarrow \oxerr : (\cmem, [``\mathsf{Type}"])}
\and
\inferrule[\textsc{Mutate-Err-Missing}]
    {n \notin \dom(\hp)}
    {\cmem.\mathtt{mutate}([n, v]) \rightsquigarrow \oxm : (\cmem, [``\mathsf{MissingCell}", n])}
\and
\inferrule[\textsc{Mutate-Err-Use-After-Free}]
    {\hp(n) = \cfreed}
    {\cmem.\mathtt{mutate}([n, v]) \rightsquigarrow \oxerr : (\cmem, [``\mathsf{UseAfterFree}", n])}
\and
\inferrule[\textsc{New}]
    {n \not\in \domain(\cmem) }
    {\cmem.\mathtt{new}([]) \rightsquigarrow \oxok : (\cmem[n \mapsto \texttt{null}], [n])}
\and
\inferrule[\textsc{Free}]
    {\hp(n) \in \Val}
    {\cmem.\mathtt{free}([n]) \rightsquigarrow \oxok : (\cmem[n \mapsto \cfreed], [])}
\and
\inferrule[\textsc{Free-Err-Type}]
    {v \notin \nats}
    {\cmem.\mathtt{free}([v]) \rightsquigarrow \oxerr : (\cmem, [``\mathsf{Type}"])}
\and
\inferrule[\textsc{Free-Err-Missing}]
    {n \notin \dom(\hp)}
    {\cmem.\mathtt{free}([n]) \rightsquigarrow \oxm : (\cmem, [``\mathsf{MissingCell}", n])}
\and
\inferrule[\textsc{Free-Err-Use-After-Free}]
    {\hp(n) = \cfreed}
    {\cmem.\mathtt{free}([n]) \rightsquigarrow \oxerr : (\cmem, [``\mathsf{UseAfterFree}", n])}
\end{mathparpagebreakable}

\subsubsection{Symbolic Semantics}\hfill

\begin{mathparpagebreakable}
\footnotesize
\inferrule[\textsc{Lookup}]
 {\smem(\sexp_l') = \sexp \and \spc' = (\sexp_l = \sexp_l')}
 {\acttrans{\smem}{lookup}{[\sexp_l]}{\oxok}{\smem, \spc', [\sexp]}}
\and
\inferrule[\textsc{Lookup-Err-Type}]
 {\spc' = (\sexp_l \not\in \nats)}
 {\acttrans{\smem}{lookup}{[\sexp_l]}{\oerr}{\smem, \spc', [\mathstr{\mathsf{Type}}]}}
\and
\inferrule[\textsc{Lookup-Err-Missing}]
 {\spc' = (\sexp_l \in \nats \land \sexp_l \not\in \domain(\smem))}
 {\acttrans{\smem}{lookup}{[\sexp_l]}{\omiss}{\smem, \spc', [\mathstr{\mathsf{MissingCell}}, \sexp_l]}}
\and
\inferrule[\textsc{Lookup-Err-Use-After-Free}]
 {\smem(\sexp_l') = \cfreed \and \spc' = (\sexp_l = \sexp_l') }
 {\acttrans{\smem}{lookup}{[\sexp_l]}{\oxerr}{\smem, \spc', [\mathstr{\mathsf{UseAfterFree}}, \sexp_l]}}
\and
\inferrule[\textsc{Mutate}]
 {\smem(\sexp_l') = \sexp_\textit{old} \and \smem' = \smem[\sexp_l' \mapsto \sexp] \\\\
  \spc' = (\sexp_l = \sexp_l' \land \sexp_\textit{old} \in \Val)}
 {\acttrans{\smem}{mutate}{[\sexp_l, \sexp]}{\oxok}{\smem', \spc', []}}
\and
\inferrule[\textsc{Mutate-Err-Type}]
 {\spc' = (\sexp_l \not\in \nats)}
 {\acttrans{\smem}{mutate}{[\sexp_l, \sexp]}{\oerr}{\smem, \spc', [\mathstr{\mathsf{Type}}]}}
\and
\inferrule[\textsc{Mutate-Err-Missing}]
 {\spc' = (\sexp_l \in \nats \land \sexp_l \not\in \domain(\smem))}
 {\acttrans{\smem}{mutate}{[\sexp_l, \sexp]}{\omiss}{\smem, \spc', [\mathstr{\mathsf{MissingCell}}, \sexp_l]}}
\and
\inferrule[\textsc{Mutate-Err-Use-After-Free}]
 {\smem(\sexp_l') = \cfreed \and \spc' = (\sexp_l = \sexp_l')}
 {\acttrans{\smem}{mutate}{[\sexp_l, \sexp]}{\oxerr}{\smem, \spc', [\mathstr{\mathsf{UseAfterFree}}, \sexp_l]}}
\and
\inferrule[\textsc{New}]
 {\sexp_l \text{ fresh} \and \spc' = (\sexp_l \in \Val)}
 {\acttrans{\smem}{new}{[]}{\oxok}{\smem[\sexp_l \mapsto \texttt{null}], \spc', [\sexp_l]}}
\and
\inferrule[\textsc{Free}]
 {\smem(\sexp_l') = \sexp_\textit{old} \and \smem' = \smem[\sexp_l' \mapsto \cfreed] \\\\ \spc' = (\sexp_l = \sexp_l' \land \sexp_\textit{old} \in \Val)}
 {\acttrans{\smem}{free}{[\sexp_l]}{\oxok}{\smem', \spc', []}}
\and
\inferrule[\textsc{Free-Err-Type}]
 {\spc' = (\sexp_l \not\in \nats)}
 {\acttrans{\smem}{free}{[\sexp_l]}{\oerr}{\smem, \spc', [\mathstr{\mathsf{Type}}]}}
\and
\inferrule[\textsc{Free-Err-Missing}]
 {\spc' = (\sexp_l \in \nats \land \sexp_l \not\in \domain(\smem))}
 {\acttrans{\smem}{free}{[\sexp_l]}{\omiss}{\smem, \spc', [\mathstr{\mathsf{MissingCell}}, \sexp_l]}}
\and
\inferrule[\textsc{Free-Err-Use-After-Free}]
 {\smem(\sexp_l') = \cfreed \and \spc' = (\sexp_l = \sexp_l')}
 {\acttrans{\smem}{free}{[\sexp_l]}{\oxerr}{\smem, \spc', [\mathstr{\mathsf{UseAfterFree}}, \sexp_l]}}
\end{mathparpagebreakable}

\subsubsection{Resources, Consume, and Produce}

Recall, there are two types resources: ${\mapsto}(\lexp_{in}; \lexp_{out})$ and ${\mapsto} \cfreed(\lexp_{in};)$; pretty-printed as, respectively, $\lexp_{in} \mapsto \lexp_{out}$ and $\lexp_{in} \mapsto \cfreed$. The former states that the location denoted by $\lexp_{in}$ points to a certain value denoted by $\lexp_{out}$. The latter states that the location denoted by $\lexp_{in}$ has been freed.

The full implementation of $\resconsume$ is as follows:
\begin{mathparpagebreakable}
\footnotesize
\inferrule{\smem=\smem_f \uplus \{\sexp_1 \mapsto \sexp_2\}}
{\resconsume(m, \mapsto,[\sexp],\smem)
\rightsquigarrow (ok, {[\sexp_2]},(\smem_f,\sexp=\sexp_1))}
\and
\inferrule{\smem = \smem' \uplus \{\sexp_1\mapsto \cfreed\}}
{\resconsume(m, {\mapsto}\cfreed,[\sexp],\smem)
\rightsquigarrow (ok, {[\sexp_2]},(\smem_f,\sexp=\sexp_1))}
\and
\inferrule{\smem(\sexp_1) = \cfreed}
{\resconsume(m, \mapsto,[\sexp],\smem)
\rightsquigarrow (\oxabort, [], (\smem, \sexp = \sexp_1))}
\and
\inferrule{\smem(\sexp_1) = \sexp_2}
{\resconsume(m, {\mapsto}\cfreed,[\sexp],\smem)
\rightsquigarrow (\oxabort, [], (\smem, \sexp = \sexp_1))}
\and
\inferrule{ }
{\resconsume(m, r,[\sexp],\smem)
\rightsquigarrow (\oxabort, [\mathstr{\mathsf{MissingCell}}, \sexp], (\smem, \sexp \not\in \domain(\smem)))}
\and
\inferrule{|\vec{\sexp}|\neq 1}
{\resconsume(m, r,\vec{\sexp},\smem)
\rightsquigarrow (\oxabort, [], (\smem, \true))}
\end{mathparpagebreakable}

The full implementation of $\resproduce$ is as follows:
\begin{mathparpagebreakable}
\footnotesize
\inferrule{\smem'=\smem \uplus \{\sexp_1\mapsto \sexp_2\}}
{\resproduce(\mapsto, [\sexp_1],[\sexp_2],\smem)\rightsquigarrow (\smem', \true)}
\and
\inferrule{\smem'=\smem \uplus \{\sexp_1\mapsto \cfreed\}}
{\resproduce({\mapsto}\cfreed, [\sexp_1],[],\smem)\rightsquigarrow (\smem', \true)}
\end{mathparpagebreakable}

\subsection{Block-offset Memory Model for C}

\subsubsection{Well-formedness and Composition}

We use the following definition in both the definition of well-formedness and composition:
\begin{align*}
\cwf_\text{B}^\text{size}((\cmem_b, \Some(n))) &\defeq (\forall i \in \dom(\cmem_b), i < n) \\
\cwf_\text{B}^\text{size}((\cmem_b, \None)) &\defeq \text{always} \\
\intertext{Well-formedness is defined as follows:}
\cwf_\text{B}((\cmem_b, n)) &\defeq \cwf_\text{B}^\text{size}((\cmem_b, n)) \land (\cmem_b \neq \emptyset \lor n \neq \None) \\
\cwf(\cmem) &\defeq \forall n_b, \cmem(n_b) \neq \cfreed \implies \cwf_\text{B}(\cmem(n_b))
\intertext{To define composition, we define disjoint union of two option values as follows:}
v_1 \uplus v_2 &\defeq
\begin{cases}
v_1 \text{ when } v_2 = \None \\
v_2 \text{ when } v_1 = \None \\
\text{undefined otherwise}
\end{cases} \\
\intertext{Now, composition is defined as follows (note that $n_1$ and $n_2$ are swapped in the condition):}
(\cmem_{b1}, n_1) \memcomp (\cmem_{b2}, n_2) &\defeq \begin{cases}
(\cmem_{b1} \uplus \cmem_{b2}, n_1 \uplus n_2) \text{ when } \cwf_\text{B}^\text{size}((\cmem_{b1}, n_2)) \land \cwf_\text{B}^\text{size}((\cmem_{b2}, n_1))\\
\text{undefined otherwise}
\end{cases} \\
(\cmem_1 \memcomp \cmem_2)(n_b) &\defeq
\begin{cases}
\cmem_1(n_b) \text{ when } n_b \notin \dom(\cmem_2) \\
\cmem_2(n_b) \text{ when } n_b \notin \dom(\cmem_1) \\
\cmem_1(n_b) \memcomp \cmem_2(n_b) \text{ when } \cmem_1(n_b) \neq \cfreed \land \cmem_2(n_b) \neq \cfreed \\
\text{undefined otherwise}
\end{cases}
\end{align*}

\subsubsection{Concrete Semantics}

We give the semantics of actions \texttt{lookup}, \texttt{new}, and \texttt{free}.

\begin{mathparpagebreakable}
\footnotesize
\inferrule[\textsc{Lookup}]
    {\cmem(n_b) = (\hp_b, \_) \and \hp_b(n_o) = v}
    {\cmem.\mathtt{lookup}([n_b, n_o]) \rightsquigarrow \oxok : (\cmem, [v])}
\and
\inferrule[\textsc{Lookup-Err-Type-Block}]
    {v_b \not\in \nats}
    {\cmem.\mathtt{lookup}([v_b, v_o]) \rightsquigarrow \oxerr : (\cmem, [``\mathsf{Type}"])}
\and
\inferrule[\textsc{Lookup-Err-Type-Offset}]
    {v_o \not\in \nats}
    {\cmem.\mathtt{lookup}([n_b, v_o]) \rightsquigarrow \oxerr : (\cmem, [``\mathsf{Type}"])}
\and
\inferrule[\textsc{Lookup-Err-Missing-Block}]
    {n_b \notin \dom(\cmem)}
    {\cmem.\mathtt{lookup}([n_b, n_o]) \rightsquigarrow \omiss : (\cmem, [``\mathsf{MissingBlock}", n_b])}
\and
\inferrule[\textsc{Lookup-Err-Use-After-Free}]
    {\hp(n_b) = \cfreed}
    {\cmem.\mathtt{lookup}([n_b, n_o]) \rightsquigarrow \oxerr : (\cmem, [``\mathsf{UseAfterFree}", n_b])}
\and
\inferrule[\textsc{Lookup-Err-Missing-Offset-No-Size}]
    {\cmem(n_b) = (\hp_b, \None) \and n_o \notin \dom(\hp_b)}
    {\cmem.\mathtt{lookup}([n_b, n_o]) \rightsquigarrow \omiss : (\cmem, [``\mathsf{MissingCell}", n_b, n_o])}
\and
\inferrule[\textsc{Lookup-Err-Missing-Offset-Size}]
    {\cmem(n_b) = (b, \Some(n)) \\ n_o \notin \dom(b) \\\\ n_o < n}
    {\cmem.\mathtt{lookup}([n_b, n_o]) \rightsquigarrow \omiss : (\cmem, [``\mathsf{MissingCell}", n_b, n_o])}
\quad
\inferrule[\textsc{Lookup-Err-Offset-Size}]
    {\cmem(n_b) = (\hp_b, \Some(n)) \\ n_o \notin \dom(\hp_b) \\\\ n_o \not< n}
    {\cmem.\mathtt{lookup}([n_b, n_o]) \rightsquigarrow \oxerr : (\cmem, [``\mathsf{OutOfBounds}", n_b, n_o])}
\and
\inferrule[\textsc{New}]
    {n_b \notin \dom(\cmem) \\ \hp_b = (\{ 0 \mapsto \nil, \dots, n - 1 \mapsto \nil \}, \Some(n))}
    {\cmem.\mathtt{new}([n]) \rightsquigarrow \oxok : (\cmem[n_b \mapsto \hp_b], [n_b])}
\and
\inferrule[\textsc{New-Err-Type}]
    {v \notin \nats}
    {\cmem.\mathtt{new}([v]) \rightsquigarrow \oxerr : (\cmem, [``\mathsf{Type}"])}
\and
\inferrule[\textsc{Free}]
    {\cmem(n_b) = (\hp_b, \Some(\textit{n})) \and |\hp_b| = n}
    {\cmem.\mathtt{free}([n_b]) \rightsquigarrow \oxok : (\cmem[ n_b \mapsto \cfreed ], [])}
\and
\inferrule[\textsc{Free-Err-Type}]
    {v \notin \nats}
    {\cmem.\mathtt{free}([v]) \rightsquigarrow \oxerr : (\cmem, [``\mathsf{Type}"])}
\and
\inferrule[\textsc{Free-Err-Miss-Block}]
    {n_b \notin \dom(\cmem)}
    {\cmem.\mathtt{free}([n_b]) \rightsquigarrow \omiss : (\cmem, [``\mathsf{MissingBlock}", n_b])}
\and
\inferrule[\textsc{Free-Err-Use-After-Free}]
    {\hp(n_b) = \cfreed}
    {\cmem.\mathtt{free}([n_b]) \rightsquigarrow \oxerr : (\cmem, [``\mathsf{UseAfterFree}", n_b])}
\and
\inferrule[\textsc{Free-Err-Miss-Size}]
    {\cmem(n_b) = (\hp_b, \None)}
    {\cmem.\mathtt{free}([n_b]) \rightsquigarrow \omiss : (\cmem, [``\mathsf{MissingBound}", n_b])}
\and
\inferrule[\textsc{Free-Err-Miss-Offset}]
    {\cmem(n_b) = (\mu_b, \Some(n)) \and |\mu_b| \neq n}
    {\cmem.\mathtt{free}([n_b]) \rightsquigarrow \omiss : (\cmem, [``\mathsf{MissingCells}", n_b])}
\end{mathparpagebreakable}

\subsubsection{Symbolic Semantics}

We give the successful rules for \texttt{lookup}, \texttt{new}, and \texttt{free}. Note that the input to \texttt{new} must be a constant; for nonconstant input the action must return an $\oxabort$ outcome because dynamically sized allocations cannot be represented using the underlying memory~data~type.
\begin{mathparpagebreakable}
\footnotesize
\inferrule[\textsc{Lookup}]
 {\smem(\sexp_b') = (\smem_b', \_) \and \smem_b'(\sexp_o') = \sexp \and \spc' = (\sexp_b = \sexp_b' \land \sexp_o = \sexp_o')}
 {\acttrans{\smem}{lookup}{[\sexp_b, \sexp_o]}{\oxok}{\smem, \spc', [\sexp]}}
\and
\inferrule[\textsc{New}]
 {\sexp_b \text{ fresh} \and \smem_b = (\{ 0 \mapsto \nil, \dots, n - 1 \mapsto \nil \}, \Some(n)) \and \spc' = (\sexp_b \in \Val)}
 {\acttrans{\smem}{new}{[n]}{\oxok}{\smem[\sexp_b \mapsto \smem_b], \spc', [\sexp_b]}}
\and
\inferrule[\textsc{Free}]
 {\smem(\sexp_b') = (\smem_b', \Some(\sexp_n)) \and \spc' = (\sexp_b = \sexp_b' \land |\smem_b'| = \sexp_n \land \smem_b' \in \Val))}
 {\acttrans{\smem}{free}{[\sexp_b]}{\oxok}{\smem[\sexp_b' \mapsto \cfreed], \spc', []}}
\end{mathparpagebreakable}

\subsubsection{Resources, Consume, and Produce}

The block-offset memory model comprises three resource assertions and are as follows:
\begin{itemize}
    \item The memory-cell resource assertion $(n_b, n_o) \mapsto v$ states that the memory value $v$ is stored in offset $n_o$ of block $n_b$
    \item The bounds resource assertion $\mathsf{Bound}(n_b; n)$ states that nothing is defined beyond offset $n$ of block $n_b$
    \item The freed resource assertion $n_b \mapsto \cfreed$ states that the block $n_b$ was freed
\end{itemize}

The satisfaction relation for resource assertions is defined below:
\[
\begin{array}{@{}l@{~}l@{~}c@{~\ }l}
 \hp \resmodels &
 (n_b, n_o) \mapsto v & \Leftrightarrow & \hp = \{n_b \mapsto (\{ n_o \mapsto v \}, \None)\} \\
 \hp \resmodels & \mathsf{Bound}(n_b; n) & \Leftrightarrow & \hp = \{n_b \mapsto (\emptyset,\Some(n))\} \\
 \hp \resmodels & n_b \mapsto \cfreed & \Leftrightarrow & \hp = \{n_b \mapsto \cfreed \}
 \end{array}
\]

Below, we show all of the successful cases for $\resconsume$ and $\resproduce$:

\begin{mathparpagebreakable}
\footnotesize
\inferrule{\smem(\sexp_b') = (\{\sexp_o' \mapsto \sexp_v\}, \None) \\\\
\smem' = \smem|_{\dom(\smem) - \{\sexp_b'\}} \\\\
\spc' = (\sexp_b=\sexp_b' \land \sexp_o=\sexp_o' \land \sexp_b' \notin \dom(\smem'))}
{\resconsume(m, \mapsto,[\sexp_b, \sexp_o],\smem)
\rightsquigarrow (ok, {[\sexp_v]},(\smem', \spc'))}
\and
\inferrule{\sexp_b\notin \dom(\smem)\\\\
\smem'=\smem [ \sexp_b \mapsto (\{\sexp_o \mapsto \sexp_v\}, \None) ] }
{\resproduce(\mapsto, [\sexp_b, \sexp_o],[\sexp_v],\smem)\rightsquigarrow (\smem', \true)}
\and
\inferrule{\smem(\sexp_b') = (\smem_b', \sexp_n) \\ \neg(|\smem_b'| = 1 \land \sexp_n = \None) \\\\
\smem_b'(\sexp_o') = \sexp_v \\ \smem_b'' = \smem_b'|_{\dom(\smem_b') - \{\sexp_o'\}} \\\\
\smem' = \smem [ \sexp_b' \mapsto (\smem_b'', \sexp_{n}) ] \\
\spc' = (\sexp_b=\sexp_b' \land \sexp_o=\sexp_o')}
{\resconsume(m, \mapsto,[\sexp_b, \sexp_o],\smem)
\rightsquigarrow (ok, {[\sexp_v]},(\smem', \spc'))}
\and
\inferrule{\smem(\sexp_b') = (\smem_b', \sexp_n) \\ (\smem_b' \neq \emptyset \lor \sexp_n \neq \None) \\\\
\sexp_o \notin \dom(\smem_b') \\ \smem_b'' = \smem_b'[ \sexp_o \mapsto \sexp_v ] \\\\
\smem'=\smem [ \sexp_b' \mapsto (\smem_b'', \sexp_n) ]}
{\resproduce(\mapsto, [\sexp_b, \sexp_o],[\sexp_v],\smem)\rightsquigarrow (\smem', \sexp_b = \sexp_b')}
\and
\inferrule{\smem(\sexp_b') = (\emptyset, \Some(\sexp_n)) \\\\
\smem' = \smem|_{\dom(\smem) - \{\sexp_b'\}} \\
\spc' = (\sexp_b=\sexp_b')}
{\resconsume(m, \mathsf{Bound},[\sexp_b],\smem)
\rightsquigarrow (ok, {[\sexp_n]},(\smem', \spc'))}
\and
\inferrule{\sexp_b\notin \dom(\smem)\\
\smem' = \smem [ \sexp_b \mapsto (\emptyset, \Some(\sexp_n)) ] }
{\resproduce(\mathsf{Bound}, [\sexp_b],[\sexp_n],\smem)\rightsquigarrow (\smem', \true)}
\and
\inferrule{\smem(\sexp_b') = (\smem_b', \Some(\sexp_n)) \\ \smem_b' \neq \emptyset \\\\
\smem' = \smem [ \sexp_b' \mapsto (\smem_b', \None) ] \\
\spc' = (\sexp_b=\sexp_b')}
{\resconsume(m, \mathsf{Bound},[\sexp_b],\smem)
\rightsquigarrow (ok, {[\sexp_n]},(\smem', \spc'))}
\and
\inferrule{\smem(\sexp_b') = (\smem_b', \None) \\ \smem_b' \neq \emptyset \\\\
\smem' = \smem [ \sexp_b' \mapsto (\smem_b', \Some(\sexp_n)) ] }
{\resproduce(\mathsf{Bound}, [\sexp_b],[\sexp_n],\smem)\rightsquigarrow (\smem', \sexp_b = \sexp_b')}
\and
\inferrule{\smem(\sexp_b') = \cfreed \\
\smem' = \smem|_{\dom(\smem) - \{ \sexp_b' \}}\\
\spc' = (\sexp_b = \sexp_b')}
{\resconsume(m, {\mapsto}\cfreed,[\sexp_b],\smem) 
\rightsquigarrow (ok, [], (\smem', \spc'))}

\and
\inferrule{\sexp_b \notin \dom(\smem)\\
\smem' = \smem [ \sexp_b \mapsto \cfreed ] }
{\resproduce({\mapsto}\cfreed, [\sexp_b],[],\smem)\rightsquigarrow (\smem', \true)}
\end{mathparpagebreakable}

\subsection{Memory Model for Object-oriented Languages}

Here, we illustrate adding and deleting properties of objects in JavaScript. The following JavaScript REPL session is from Node.js v23.10.0:
\begin{verbatim}
> var obj = {}
undefined
> obj.foo
undefined
> obj.foo = 5
5
> obj.bar = 1
1
> obj
{ foo: 5, bar: 1 }
> delete obj.foo
true
> obj.foo
undefined
> obj
{ bar: 1 }
\end{verbatim}

\subsection{CHERI-assembly Memory Model}

\subsubsection{Well-formedness Property}

Given that the CHERI-assembly memory model is an extension of the block-offset memory model, all well-formedness properties in the block-offset memory model were ported to CHERI-assembly's well-formedness property. The additional condition added was that all capability fragments with a valid capability tag fragment are in an appropriately aligned position, which is denoted below:
\begin{align*}
\cwf_\text{B-CH}^\text{size}((\cmem_b, \Some(n))) &\defeq (\forall i \in \dom(\cmem_b), i < n) \text{ and } \cwf_\text{B-CH}^\text{size}((\cmem_b, \None)) \defeq \text{always} \\
\cwf_\text{B-CH}((\cmem_b, n)) &\defeq \cwf_\text{B-CH}^\text{size}((\cmem_b, n)) \land (\cmem_n \neq \emptyset \lor n \neq \None)\ \land \\
&\quad\ \ \ (\forall (o, c_{n'}) \in \mu_b,\  (c_{n'}.cap.tag = \true \implies o\ \%\ |\mathsf{Cap}| = n') \land n' < |\mathsf{Cap}|) \\
\cwf(\cmem) &\defeq \forall n_b, \cmem(n_b) \neq \cfreed \implies \cwf_\text{B-CH}(\cmem(n_b)) \\
\end{align*}

\subsubsection{Concrete Semantics}

We give the semantics of some of the memory-related actions. There are 14 actions. Four relates to accessing/mutating the memory (i.e. \texttt{load}, \texttt{store}, \texttt{alloc}, and \texttt{free}), and ten relates to accessing/mutating the capability registers (i.e. \texttt{get\_addr}, \texttt{get\_base}, \texttt{get\_length}, \texttt{get\_offset}, \texttt{get\_perms}, \texttt{get\_tag}, \texttt{addr\_set}, \texttt{tag\_clear}, \texttt{perms\_clear}, and \texttt{bounds\_set}). For loading and storing, we have merged cases where one attempts to load a capability or a byte value. There are 92 cases within the action semantics -- many cases contribute to spatial checking, yielding an appropriate hardware exception as an error in cases where the check fails. For mutating capability registers, the actions were designed such that bounds can only be restricted, permissions can only be taken away, and the tag bit can only be cleared, which preserves the monotonic property. We highlight some of the main successful memory-related cases below:

\begin{mathparpagebreakable}
\footnotesize
\inferrule[\textsc{Free}]
{
\mathsf{fst}(\cmem)(r_s) = c_s \and c_s.tag = \true \\\\
c_s.\textit{off} = 0 \\\\
c_s.\textit{off} \leq c_s.base + c_s.len \\\\
\mathsf{snd}(\mu)(c_s.blo) = (\mu_b, \Some(m)) \\\\
|\mu_b| = m \\\\
\mu' = (\mathsf{fst}(\mu), \mathsf{snd}(\mu)[c_s.blo \mapsto \cfreed])
}
{\cmem.\mathtt{free}([r_s]) \rightsquigarrow \oxok : (\mu', [])}
\and
\inferrule[\textsc{Load}]
{
\mathsf{fst}(\cmem)(r_s) = c_s \and c_s.tag = \true \and c_s.perm_\mathsf{load} = \true \\\\
c_s.\textit{off} + \mathsf{type\_size}(b) \leq c_s.base + c_s.len \\\\
c_s.\textit{off} \geq c_s.base \and c_s.\textit{off}\ \%\ \mathsf{type\_size}(b) = 0 \\\\
\mathsf{snd}(\mu)(c_s.blo) = (\mu_b, \Some(m)) \\\\
c_s.\textit{off} + |\mathsf{Cap}| \leq m \\\\
\{c_s.\textit{off}, ... c_s.\textit{off} + \mathsf{type\_size}(b) - 1\} \subseteq \mathsf{dom}(\mu_b) \\\\
c_f \text{ fresh w.r.t. } \domain(\mathsf{fst}(\cmem)) \\\\
\mathsf{load}(c_s, b, \mu_b, \mathsf{fst}(\mu), c_f) = (\mu'_r, v)
}
{\cmem.\mathtt{load}([r_s, b]) \rightsquigarrow \oxok : ((\mu'_r, \mathsf{snd}(\mu)), [v])}
\and
\inferrule[\textsc{Store-NonCap}]
{
\mathsf{fst}(\cmem)(r_s) = c_s \and c_s.tag = \true \\\\ 
c_s.perm_\mathsf{store} = \true \and
v \text{ is not a capability register} \\\\
c_s.\textit{off} + \mathsf{val\_size}(v) \leq c_s.base + c_s.len \\\\
c_s.\textit{off} \geq c_s.base \and c_s.\textit{off}\ \%\ \mathsf{val\_size}(v) = 0 \\\\
\mathsf{snd}(\mu)(c_s.blo) = (\mu_b, \Some(m)) \\\\
c_s.\textit{off} + |\mathsf{Cap}| \leq m \and c_s.\textit{off} \in \mathsf{dom}(\mu_b) \\\\
\mathsf{store\_value}(c_s, v, \mu_b) = \mu_b' \\\\
\mu' = (\mathsf{fst}(\mu), \mathsf{snd}(\mu)[c_s.blo \mapsto (\mu_b', \Some(m))])
}
{\cmem.\mathtt{store}([r_s, v]) \rightsquigarrow \oxok : (\mu', [])}
\and
\inferrule[\textsc{Store-Cap}]
{
\mathsf{fst}(\cmem)(r_s) = c_s \and c_s.tag = \true \and c_s.perm_\mathsf{store} = \true \\\\
r_d \text{ is a capability register} \and \mathsf{fst}(\cmem)(r_d) = c_d \\\\
(c_s.perm_\mathsf{storecap} = \true \lor c_d.tag = \mathtt{false}) \\\\
c_s.\textit{off} + |\mathsf{Cap}| \leq c_s.base + c_s.len \\\\
c_s.\textit{off} \geq c_s.base \and c_s.\textit{off}\ \%\ |\mathsf{Cap}| = 0 \\\\
\mathsf{snd}(\mu)(c_s.blo) = (\mu_b, \Some(m)) \\\\
c_s.\textit{off} + |\mathsf{Cap}| \leq m \\\\
\{c_s.\textit{off}, ..., c_s.\textit{off} + |\mathsf{Cap}| - 1\} \subseteq \mathsf{dom}(\mu_b) \\\\
\mathsf{store\_capability}(c_s, c_d, \mu_b) = \mu_b' \\\\
\mu' = (\mathsf{fst}(\mu), \mathsf{snd}(\mu)[c_s.blo \mapsto (\mu_b', \Some(m))])
}
{\cmem.\mathtt{store}([r_s, r_d]) \rightsquigarrow \oxok : (\mu', [])}
\end{mathparpagebreakable}

There are a number of auxiliary functions used in the concrete semantics:
\begin{itemize}
    \item $\mathsf{type\_size}(b)$ returns \textsf{capsize}, the size of a capability, if $b$ is true and 1 (i.e. the size of a byte) otherwise. 
    \item $\mathsf{load}(c_s, b, \mu_b, \mathsf{fst}(\mu), c_f)$ attempts to read, at location $c_s.\textit{off}$ in memory $\mu_b$, the value or capability fragment byte and returns corresponding value or capability fragment byte if $b$ is false. If $b$ is true, then locations $c_s.\textit{off}$ to $c_s.\textit{off} + |\mathsf{Cap}| - 1$ is read, and if a valid capability is stored in a contiguous way, then the new value is saved to the fresh register $c_f$ and the updated register map and the fresh register $c_f$ are returned.
    \item $\mathsf{store\_value}(c_s, v, \mu_b)$ updates the memory $\mu_b$ at position $c_s.\textit{off}$ with value $v$, where $v$ may either be a value or a capability fragment.
    \item $\mathsf{store\_capability}(c_s, c_d, \mu_b)$ updates the memory $\mu_b$ at positions $c_s$.\textit{off} to $c_s$.\textit{off} + \\ $\mathsf{type\_size}(\true) - 1$ by splitting $c_d$ into capability fragments and storing them in a contiguous order. The tag fragment bits for each capability fragment is set to the value of $c_d.\textit{tag}$.
\end{itemize}

\subsubsection{Symbolic Memory Model Structure}

First, we define the symbolic memory below:

\[
\begin{array}{r c l}
    \mathsf{SCap} &\defeq& \{blo : \lexps;\ \mathit{off} : \lexps;\ base : \lexps;\ len : \lexps;\ \vec{perm}_{\mathsf{x}} : \vec{\lexps};\ tag : \lexps\}\\
    & & \quad \text{where } x \in \{\mathsf{load, store, loadcap, storecap}\} \\
    \mathsf{SCap_{frag}} &\defeq& \{cap : \mathsf{SCap}; nth : \lexps\} \\
    \smemss_\text{B-CH}\ &\defeq& \lexps \rightharpoonup_{\mathit{fin}} (\lexps + \mathsf{SCap_{frag}}, \lexps?)\\
    \smemss_\text{SCReg} &\defeq& \lexps \rightharpoonup_{\mathit{fin}} \mathsf{SCap}\\
    \smemss &\defeq& (\smemss_\text{SCReg}, \smemss_\text{B-CH} \uplus \{ \cfreed \})
\end{array}
\]

The symbolic semantics are directly lifted from the concrete versions, much like the block offset memory model. There are a total of 116 cases for the symbolic actions semantics -- whereas the concrete semantics merged the capability/non-capability cases for load, for simplicity, we wrote separate cases in the symbolic semantics. Capabilities and capability fragments have been lifted to symbolic capabilities and symbolic capability fragments, which are used in the memory.

\subsubsection{Resources, Consume, and Produce}

We define $\resmodels$, the resource satisfaction relation for CHERI-assembly:
\[
\begin{array}{@{}l@{~}l@{~}c@{~\ }l@{~\ }l}
  \hp \resmodels &
 \mathsf{Reg}(r_n; c) & \Leftrightarrow & \hp = (\{r_n \mapsto c\}, \emptyset) & \\
 & (n_b, n_o) \mapsto v & \Leftrightarrow & \hp = (\emptyset, \{n_b \mapsto (\{n_o \mapsto v\}, \None)\}) & \text{if } v \in \Val\\
 & (n_b, n_o) \mapsto_{\mathit{cf}} c_n & \Leftrightarrow & \hp = (\emptyset, \{n_b \mapsto (\{n_o \mapsto c_n\}, \None)\}) & \\
 & & & \land\ (c_n.cap.tag = \true \Longrightarrow n_o\ \%\ |\mathsf{Cap}| = n) & \\ 
 & & & \land\ n < |\mathsf{Cap}| & \text{if } c_n \in \mathsf{Cap_{frag}} \\ 
 & \mathsf{Bound}(n_b; n) & \Leftrightarrow & \hp = (\emptyset, \{n_b \mapsto (\emptyset, \Some(n)\}) & \\
 & n_b \mapsto \cfreed & \Leftrightarrow & \hp = (\emptyset, \{n_b \mapsto \cfreed \}) &
 \end{array} 
\]

We require a new resource assertion for capability registers, given that the memory model now comprises capability registers. As for the memory resource assertions, there are two cases, where one of them is identical to that of the block-offset memory model. The other one is specifically for capability fragments. Given $c_n \in \mathsf{Cap_{frag}}$, $n$ indicates that the fragment is the $n$th byte of a full capability $c \in \mathsf{Cap}$. When the tag fragment bit is true, which indicates that the particular capability byte fragment has not been modified by an non-capability-aware store, we require that $n_o\ \%\ |\mathsf{Cap}| = n$, given that $n_o$ is an offset in a block; this not only conforms with the CHERI-ISA specification~\cite{watson:2023:cheri}, but this also ensures each resource assertion forms a well-formed memory, and composing resource assertions also results in a well-formed memory. 

The more interesting memory capability resource assertion can be defined as a predicate assertion, i.e. $\mathsf{Cap}(n_b, n_o; c) \{A_{\text{body}}\} \in \mathsf{Preds}$, pretty-printed as $(n_b, n_o) \mapsto_{\mathsf{cap}} c$. Here, assume $c \in \mathsf{Cap}$ and $c_i \in \mathsf{Cap_{frag}}$ where $c_i.nth = i$, i.e. the $i^{th}$ byte of the capability $c$.
\[
\begin{array}{@{}l@{~}l@{~}l}
    A_{\text{body}} & \defeq & \ilstar_{i=0}^{|\mathsf{Cap}| - 1}\ (n_b, n_o + i) \mapsto_{\mathit{cf}} c_i
\end{array}
\]

Note that we do not require the assertion $n_o\ \%\ |\mathsf{Cap}| = 0$, i.e., the offset starts in a capability-aligned position (based on the CHERI-ISA specification~\cite{watson:2023:cheri}), since the resource satisfaction relation implicitly implies each resource is in an aligned position.

Below, we show some of the successful cases for $\resconsume$ and $\resproduce$:

\begin{mathparpagebreakable}
\footnotesize
\inferrule{\mathsf{fst}(\smem)(\sexp_r') = \hat{c} \\\\ 
\smem' = (\mathsf{fst}(\smem|_{\dom(\mathsf{fst}(\smem)) - \{ \sexp_r' \}}), \mathsf{snd}(\smem)) \\
\spc' = (\sexp_r = \sexp_r')}
{\resconsume(m, \mathsf{Reg},[\sexp_r],\smem)
\rightsquigarrow (ok, {[\hat{c}]},(\smem', \spc'))}
\and
\inferrule{\sexp_r \notin \dom(\mathsf{fst}(\smem))\\
\smem' = (\mathsf{fst}(\smem) [ \sexp_r \mapsto \hat{c} ] , \mathsf{snd}(\smem))}
{\resproduce(\mathsf{Reg}, [\sexp_r],[\hat{c}],\smem)\rightsquigarrow (\smem', \true)}
\and
\inferrule{\mathsf{snd}(\smem)(\sexp_b') = (\{\sexp_o' \mapsto \sexp_v\}, \None) \\\\
\smem' = (\mathsf{fst}(\smem), \mathsf{snd}(\smem)|_{\dom(\mathsf{snd}(\smem)) - \{ \sexp_b' \}}) \\\\
\spc' = (\sexp_b=\sexp_b' \land \sexp_o=\sexp_o' \land \sexp_b' \notin \dom(\mathsf{snd}(\smem')))}
{\resconsume(m, \mapsto,[\sexp_b, \sexp_o],\smem)
\rightsquigarrow (ok, {[\sexp_v]},(\smem', \spc'))}
\and
\inferrule{\sexp_b\notin \dom(\mathsf{snd}(\smem))\\\\
\smem' = (\mathsf{fst}(\smem), \mathsf{snd}(\smem) [ \sexp_b \mapsto (\{\sexp_o \mapsto \sexp_v\}, \None) ] )}
{\resproduce(\mapsto, [\sexp_b, \sexp_o],[\sexp_v],\smem)\rightsquigarrow (\smem', \true)}
\and
\inferrule{\mathsf{snd}(\smem)(\sexp_b') = (\{\sexp_o' \mapsto \hat{c}_{\sexp_n}\}, \None) \\\\
\smem' = (\mathsf{fst}(\smem), \mathsf{snd}(\smem)|_{\dom(\mathsf{snd}(\smem)) - \{ \sexp_b' \}}) \\\\
\spc' = (\sexp_b=\sexp_b' \land \sexp_o=\sexp_o' \land \sexp_b' \notin \dom(\mathsf{snd}(\smem')) \\\\
\land\ \hat{c}_{\sexp_n}.cap.tag \neq \true \land \sexp_n < |\mathsf{Cap}|) }
{\resconsume(m, \mapsto_{\mathit{cf}},[\sexp_b, \sexp_o],\smem)
\rightsquigarrow (ok, {[\hat{c}_{\sexp_n}]},(\smem', \spc'))}
\and
\inferrule{\sexp_b\notin \dom(\mathsf{snd}(\smem))\\\\
\smem' = (\mathsf{fst}(\smem), \mathsf{snd}(\smem) [ \sexp_b \mapsto (\{\sexp_o \mapsto \hat{c}_{\sexp_n}\}, \None) ] ) \\\\
\spc = (\hat{c}_{\sexp_n} \neq \true \land \sexp_n < |\mathsf{Cap}|)}
{\resproduce(\mapsto_{\mathit{cf}}, [\sexp_b, \sexp_o],[\hat{c}_{\sexp_n}],\smem)\rightsquigarrow (\smem', \spc)}
\and
\inferrule{\mathsf{snd}(\smem)(\sexp_b') = (\{\sexp_o' \mapsto \hat{c}_{\sexp_n}\}, \None) \\\\
\smem' = (\mathsf{fst}(\smem), \mathsf{snd}(\smem)|_{\dom(\mathsf{snd}(\smem)) - \{ \sexp_b' \}}) \\\\
\spc' = (\sexp_b=\sexp_b' \land \sexp_o=\sexp_o' \land \sexp_b' \notin \dom(\mathsf{snd}(\smem')) \\\\
\land\ \hat{c}_{\sexp_n}.cap.tag = \true \land \sexp_n < |\mathsf{Cap}| \\\\ \land\ \sexp_o\ \%\ |\mathsf{Cap}| = \sexp_n)\qquad\qquad\qquad\ \ }
{\resconsume(m, \mapsto_{\mathit{cf}},[\sexp_b, \sexp_o],\smem)
\rightsquigarrow (ok, {[\hat{c}_{\sexp_n}]},(\smem', \spc'))}
\and
\inferrule{\sexp_b\notin \dom(\mathsf{snd}(\smem))\\\\
\smem' = (\mathsf{fst}(\smem), \mathsf{snd}(\smem) [ \sexp_b \mapsto (\{\sexp_o \mapsto \hat{c}_{\sexp_n}\}, \None) ] ) \\\\
\spc = (\hat{c}_{\sexp_n} \neq \true \land \sexp_n < |\mathsf{Cap}| \land \sexp_o\ \%\ |\mathsf{Cap}| = \sexp_n)}
{\resproduce(\mapsto_{\mathit{cf}}, [\sexp_b, \sexp_o],[\hat{c}_{\sexp_n}],\smem)\rightsquigarrow (\smem', \spc)}
\and
\inferrule{\mathsf{snd}(\smem)(\sexp_b') = (\emptyset, \Some(\sexp_n)) \\\\
\smem' = (\mathsf{fst}(\smem), \mathsf{snd}(\smem)|_{\dom(\mathsf{snd}(\smem)) - \{ \sexp_b' \}}) \\
\spc' = (\sexp_b=\sexp_b')}
{\resconsume(m, \mathsf{Bound},[\sexp_b],\smem)
\rightsquigarrow (ok, {[\sexp_n]},(\smem', \spc'))}
\and
\inferrule{\sexp_b\notin \dom(\mathsf{snd}(\smem))\\\\
\smem' = (\mathsf{fst}(\smem), \mathsf{snd}(\smem) [ \sexp_b \mapsto (\emptyset, \Some(\sexp_n)) ] )}
{\resproduce(\mathsf{Bound}, [\sexp_b],[\sexp_n],\smem)\rightsquigarrow (\smem', \true)}
\and
\inferrule{\mathsf{snd}(\smem)(\sexp_b') = \cfreed \\ \spc' = (\sexp_b = \sexp_b') \\\\
\smem' = (\mathsf{fst}(\smem), \mathsf{snd}(\smem)|_{\dom(\mathsf{snd}(\smem)) - \{ \sexp_b' \}})}
{\resconsume(m, {\mapsto}\cfreed,[\sexp_b],\smem) 
\rightsquigarrow (ok, [], (\smem', \spc'))}

\and
\inferrule{\sexp_b \notin \dom(\mathsf{snd}(\smem))\\
\smem' = (\mathsf{fst}(\smem), \mathsf{snd}(\smem) [ \sexp_b \mapsto \cfreed ] )}
{\resproduce({\mapsto}\cfreed, [\sexp_b],[],\smem)\rightsquigarrow (\smem', \true)}
%
%
%
%
%
%
%
%
%
\end{mathparpagebreakable}

Note that for both consume and produce, we split the $\mapsto_{\mathit{cf}}$ case based on the value of the tag fragment bit of the capability fragment. Strictly speaking, the split is not necessary for the purpose of proving OX soundness, but the split is required for proving UX soundness, which is not part of the contribution of this paper. For future work, we plan on proving UX soundness with our existing formalism.

\else
\fi

\end{document}